\documentclass[aps,prd,twocolumn,superscriptaddress,showpacs,floatfix]{revtex4}

\usepackage[dvips]{graphicx}
\usepackage{xspace}
\usepackage{latexsym}
\usepackage{amssymb}
\usepackage{times}
\usepackage{mathptm}
 
\graphicspath{{./figs/}}

\begin{document}

\newcommand{\superk}    {Super-Kamiokande\xspace}       
\newcommand{\nue}       {$\nu_{e}$\xspace}
\newcommand{\numu}      {$\nu_{\mu}$\xspace}
\newcommand{\nutau}     {$\nu_{\tau}$\xspace}
\newcommand{\nusterile} {$\nu_{sterile}$\xspace}
\newcommand{\mutau}     {$\nu_\mu \rightarrow \nu_{\tau}$\xspace}
\newcommand{\musterile} {$\nu_\mu \rightarrow \nu_{sterile}$\xspace}
\newcommand{\dms}       {$\Delta m^2$\xspace}
\newcommand{\sstt}      {$\sin^2 2 \theta$\xspace}
\newcommand{\asymerr}[2]{\ooalign{{\scriptsize \raisebox{4pt}{+~#1}}\crcr
{\scriptsize \raisebox{-4pt}{--~#2}}}}

\title{A Measurement of Atmospheric Neutrino Oscillation Parameters 
by Super-Kamiokande I}

\date{\today}

\newcommand{\icrr}{\affiliation{Kamioka Observatory, Institute for Cosmic Ray Research, University of Tokyo, Kamioka, Gifu, 506-1205, Japan}}
\newcommand{\ncen}{\affiliation{Research Center for Cosmic Neutrinos, Institute for Cosmic Ray Research, University of Tokyo, Kashiwa, Chiba 277-8582, Japan}}
\newcommand{\bu}{\affiliation{Department of Physics, Boston University, Boston, MA 02215, USA}}
\newcommand{\bnl}{\affiliation{Physics Department, Brookhaven National Laboratory, Upton, NY 11973, USA}}
\newcommand{\uci}{\affiliation{Department of Physics and Astronomy, University of California, Irvine, Irvine, CA 92697-4575, USA}}
\newcommand{\csu}{\affiliation{Department of Physics, California State University, Dominguez Hills, Carson, CA 90747, USA}}
\newcommand{\cnu}{\affiliation{Department of Physics, Chonnam National University, Kwangju 500-757, Korea}}
\newcommand{\duke}{\affiliation{Department of Physics, Duke University, Durham, NC 27708, USA}}
\newcommand{\gmu}{\affiliation{Department of Physics, George Mason University, Fairfax, VA 22030, USA}}
\newcommand{\gifu}{\affiliation{Department of Physics, Gifu University, Gifu, Gifu 501-1193, Japan}}
\newcommand{\uh}{\affiliation{Department of Physics and Astronomy, University of Hawaii, Honolulu, HI 96822, USA}}
\newcommand{\ui}{\affiliation{Department of Physics, Indiana University, Bloomington,  IN 47405-7105, USA} }
\newcommand{\kek}{\affiliation{High Energy Accelerator Research Organization (KEK), Tsukuba, Ibaraki 305-0801, Japan}}
\newcommand{\kobe}{\affiliation{Department of Physics, Kobe University, Kobe, Hyogo 657-8501, Japan}}
\newcommand{\kyoto}{\affiliation{Department of Physics, Kyoto University, Kyoto 606-8502, Japan}}
\newcommand{\lanl}{\affiliation{Physics Division, P-23, Los Alamos National Laboratory, Los Alamos, NM 87544, USA}}
\newcommand{\lsu}{\affiliation{Department of Physics and Astronomy, Louisiana State University, Baton Rouge, LA 70803, USA}}
\newcommand{\umd}{\affiliation{Department of Physics, University of Maryland, College Park, MD 20742, USA}}
\newcommand{\duluth}{\affiliation{Department of Physics, University of Minnesota, Duluth, MN 55812-2496, USA}}
\newcommand{\miyagi}{\affiliation{Department of Physics, Miyagi University of Education, Sendai,Miyagi 980-0845, Japan}}
\newcommand{\suny}{\affiliation{Department of Physics and Astronomy, State University of New York, Stony Brook, NY 11794-3800, USA}}
\newcommand{\nagoya}{\affiliation{Department of Physics, Nagoya University, Nagoya, Aichi 464-8602, Japan}}
\newcommand{\niigata}{\affiliation{Department of Physics, Niigata University, Niigata, Niigata 950-2181, Japan}}
\newcommand{\osaka}{\affiliation{Department of Physics, Osaka University, Toyonaka, Osaka 560-0043, Japan}}
\newcommand{\seoul}{\affiliation{Department of Physics, Seoul National University, Seoul 151-742, Korea}}
\newcommand{\shizuokaseika}{\affiliation{International and Cultural Studies, Shizuoka Seika College, Yaizu, Shizuoka 425-8611, Japan}}
\newcommand{\shizuoka}{\affiliation{Department of Systems Engineering, Shizuoka University, Hamamatsu, Shizuoka 432-8561, Japan}}
\newcommand{\skku}{\affiliation{Department of Physics, Sungkyunkwan University, Suwon 440-746, Korea}}
\newcommand{\tohoku}{\affiliation{Research Center for Neutrino Science, Tohoku University, Sendai, Miyagi 980-8578, Japan}}
\newcommand{\tokyo}{\affiliation{University of Tokyo, Tokyo 113-0033, Japan}}
\newcommand{\tokai}{\affiliation{Department of Physics, Tokai University, Hiratsuka, Kanagawa 259-1292, Japan}}
\newcommand{\tit}{\affiliation{Department of Physics, Tokyo Institute for Technology, Meguro, Tokyo 152-8551, Japan}}
\newcommand{\warsaw}{\affiliation{Institute of Experimental Physics, Warsaw University, 00-681 Warsaw, Poland}}
\newcommand{\uw}{\affiliation{Department of Physics, University of Washington, Seattle, WA 98195-1560, USA}}
\newcommand{\tsukubanow}{\altaffiliation{ Present address: Department of Physics, Univ. of Tsukuba, Tsukuba, Ibaraki 305 8577, Japan}}
\newcommand{\okayamanow}{\altaffiliation{ Present address: Department of Physics, Okayama University, Okayama 700-8530, Japan}}
%
\author{Y.Ashie}\icrr
\author{J.Hosaka}\icrr
\author{K.Ishihara}\icrr
\author{Y.Itow}\icrr
\author{J.Kameda}\icrr
\author{Y.Koshio}\icrr
\author{A.Minamino}\icrr
\author{C.Mitsuda}\icrr
\author{M.Miura}\icrr
\author{S.Moriyama}\icrr
% ...
\collaboration{The Super-Kamiokande Collaboration}\noaffiliation

\begin{abstract}
  
  We present a combined analysis of fully-contained, partially-contained and
  upward-going muon atmospheric neutrino data from a 1489 day exposure of the
  Super--Kamiokande detector. The data samples span roughly five decades in
  neutrino energy, from 100~MeV to 10~TeV.  A detailed Monte Carlo comparison
  is described and presented.  The data is fit to the Monte Carlo
  expectation, and is found to be consistent with neutrino oscillations of
  $\nu_\mu \leftrightarrow \nu_\tau$ with $\sin^22\theta > 0.92$ and $1.5\times
  10^{-3} < \Delta m^2 < 3.4\times 10^{-3}{\rm eV}^2$ at 90\% confidence level.

\end{abstract}


\maketitle

\section{Introduction}
\label{sec:introduction}

Atmospheric neutrinos are produced from the decays of particles resulting
from interactions of cosmic rays with Earth's atmosphere.  We have previously
reported the results of a number of atmospheric neutrino observations
spanning energies from 100~MeV to
10~TeV~\cite{Fukuda:1998tw,Fukuda:1998ub,Fukuda:1998ah,Fukuda:1999pp}.  
In each case, a
significant zenith-angle dependent deficit of $\nu_\mu$ was observed. These
deficits have been interpreted as evidence for neutrinos
oscillations~\cite{Fukuda:1998mi}. If neutrinos have a non-zero mass, then the
probability that a neutrino of energy $E_\nu$ produced in a weak flavor
eigenstate $\nu_\alpha$ will be observed in eigenstate 
$\nu_\beta$ after traveling a distance
$L$ through the vacuum is:
\begin{equation}
\label{eqn:oscillation}
P(\nu_\alpha \rightarrow \nu_\beta) = \sin^2 2 \theta \sin^2{\Big(}\frac{1.27 
\Delta m^2(\textrm{eV}^2) L (\textrm{km})}{E_\nu(\textrm{GeV})}\Big{)},
\end{equation}
where $\theta$ is the mixing angle between the mass eigenstates and the weak
eigenstates and $\Delta m^2$ is the difference of the squared mass eigenvalues.
This equation is valid in the 2-flavor approximation.
The analysis reported in this paper is under the assumption of effective
2-flavor neutrino oscillations, $\nu_\mu \leftrightarrow \nu_\tau$,
which is considered to be dominant in atmospheric neutrino oscillations.
Equation~\ref{eqn:oscillation} is also true in matter for
$\nu_\mu \leftrightarrow \nu_\tau$, but may be modified for oscillation
involving $\nu_e$ which travel through matter.
The zenith angle dependence of the observed
deficits results from the variation of $L$ with the direction of the
neutrino. Neutrinos produced directly overhead travel roughly 15~km to
the detector while those produced directly below traverse the full
diameter of the Earth (13,000~km) before reaching the detector. By
measuring the neutrino event rate over these wide ranges of $E_\nu$
and $L$, we have measured the neutrino oscillation parameters $\Delta
m^2$ and $\sin^2 2\theta$.


Super-Kamiokande (also Super-K or SK) is a 50-kiloton water Cherenkov
detector located deep underground in Gifu Prefecture, Japan. Atmospheric
neutrinos are observed in Super--K in two ways. At the lowest energies,
100~MeV -- 10~GeV, atmospheric neutrinos are observed via their
charged-current interactions with nuclei in the 22.5~kiloton water fiducial
mass: $\nu + N \rightarrow l + X$. These interactions are classified as
fully-contained (FC) if all of the energy is deposited inside the inner
Super--K detector, or as partially-contained (PC) if a high energy muon
exits the inner detector, depositing energy in the outer veto region.
The neutrino energies that produce partially-contained events are typically
10 times higher than those that produce fully-contained events. The Super-K
detector started observation on April, 1996 achieving a 92\,kiloton-yr
(1489\,live-day) exposure to atmospheric neutrinos through July, 2001
during the Super-Kamiokande I running period.

\begin{figure}[hbt]
  \includegraphics[width=3.2in]{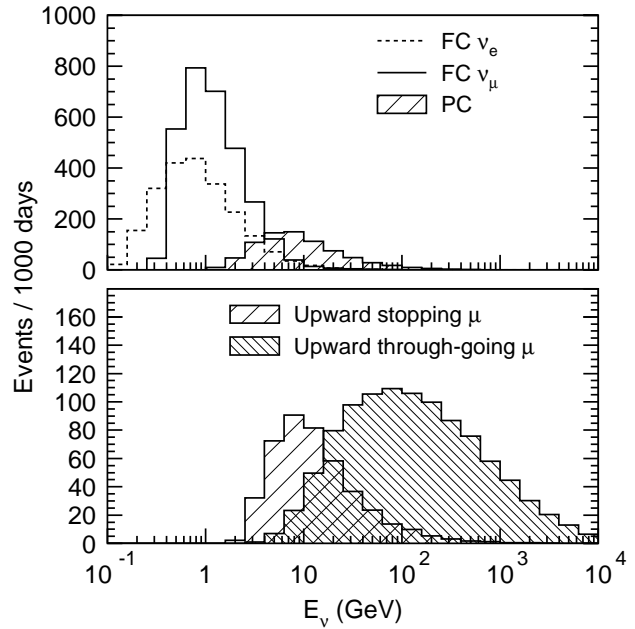} \caption{The
  parent neutrino energy distributions for the fully-contained,
  partially-contained, upward stopping-muon and upward through-going
  muons samples. Rates for the fully-contained and partially-contained
  samples are for interactions in the 22.5 kiloton fiducial
  volume. Taken together, the samples span five decades in neutrino
  energy.} \label{fig:parent_spectra}
\end{figure}

Neutrinos can also be detected by their interactions with the rock
surrounding the detector. Charged-current $\nu_\mu$ interactions with the
rock produce high energy muons which intersect the detector. While these
interactions can not be distinguished from the constant rain of cosmic ray
muons traveling in the downward direction, muons traveling in an upward
direction through the detector must be neutrino induced.  Upward-going muon
events are separated into two categories: those that come to rest in the
detector (upward stopping muons) and those that traverse the entire detector
volume (upward through-going muons).  The energies of the neutrinos which
produce stopping muons are roughly the same as for partially-contained events,
$\sim 10~$GeV. Upward through-going events, however, are significantly more
energetic; the parent neutrino energy for these events is about 100~GeV on
average.

Figure~\ref{fig:parent_spectra} shows the expected number of neutrino events
in each event category as a function of neutrino energy. The samples taken
together span nearly five decades in energy. This broad range of available
energies, in combination with the variation in neutrino travel distance,
makes the combined data sample well suited for a precise measurement of
neutrino oscillation parameters.

There have been numerous other measurements of atmospheric neutrinos.
Kamiokande~\cite{Hirata:1988uy, Hirata:1992ku},  
IMB~\cite{Casper:1991ac, Becker-Szendy:1992hq} and
Soudan~2~\cite{Allison:1996yb, Allison:1999ms} observed significantly 
smaller $\nu_\mu$ to $\nu_e$ flux
ratios of $\sim 1$~GeV atmospheric neutrinos, which were interpreted as
a signature for neutrino oscillation. The ratio was used in order to
normalize the uncertainty in the overall atmospheric neutrino flux.
Data on multi-GeV atmospheric neutrino 
events~\cite{Fukuda:1994mc} and upward-going
muons~\cite{Ambrosio:1998wu,Hatakeyama:1998ea,Ambrosio:2000ja} have also
shown a zenith-angle dependent deficit of the $\nu_\mu$ flux. The $\nu_{\mu}
\leftrightarrow \nu_{\tau}$ oscillation analyses of these various
data over various energy
ranges~\cite{Fukuda:1994mc,Hatakeyama:1998ea,Ambrosio:2000ja,
  Ambrosio:2003yz, Sanchez:2003rb,Fukuda:1998mi,Fukuda:1998ah,Fukuda:1999pp,
Ambrosio:2004ig}
indicated similar $\Delta m^2$ and $\sin ^2 2 \theta$ regions as the first
measurements from Super-K as well as those reported here. 

The K2K long baseline experiment used an accelerator beamline to
produce muon neutrinos that traveled 250 km to the Super-K detector,
as a means to study neutrino oscillation in the atmospheric neutrino
energy and distance scales. The results from
K2K~\cite{Ahn:2002up, Aliu:2004xx} are also consistent with the neutrino
oscillation parameters reported here.


\section{Atmospheric Neutrinos}
\label{sec:atmnu}
 
To carry out detailed studies of neutrino oscillations using atmospheric
neutrinos, it is important to know the expected flux without neutrino
oscillations. The difficulties and the uncertainties in the calculation of
atmospheric neutrino fluxes differ between high and low energies. For low
energy neutrinos around 1~GeV, the primary fluxes of cosmic ray components
are relatively well known. Low energy cosmic ray fluxes of less than about
10~GeV are modulated by solar activity, with the minimum flux occurring at
times of high solar activity. At these energies, the primary cosmic rays are
also affected by the geomagnetic field through a rigidity (momentum/charge)
cutoff. For high energy neutrinos, above 100~GeV, primary cosmic rays with
energies greater than 1000~GeV are relevant. At these energies, solar
activity and the rigidity cutoff do not affect the cosmic rays, but details
of the higher energy primary cosmic ray flux are not as well measured.

There are several flux calculations
~\cite{Battistoni:1999at,Honda:2001xy,Tserkovnyak:2002xx,Liu:2002sq,Battistoni:2003ju,Wentz:2003bp,Favier:2003gn, Honda:2004yz,Barr:2004br}.
Unlike older calculations~\cite{Honda:1995hz,Agrawal:1996gk}, in which the
secondary particles were assumed to travel in the direction of the primary
cosmic ray (1-dimensional calculations), the current calculations employ
three dimensional Monte Carlo methods.  We outline below the methods of the
calculation. We compared results from three atmospheric neutrino flux
calculations~\cite{Honda:2004yz,Barr:2004br,Battistoni:2003ju} which cover
the energy range relevant to the present analysis. The flux from
Honda {\it et al.}~\cite{Honda:2004yz} is used for the main numbers
and figures quoted for the Super-Kamiokande analysis.

Calculations start with primary cosmic rays based on measured fluxes, and
include solar modulation and geomagnetic field effects. The interaction of
cosmic ray particles with the air nucleus, the propagation and decay of
secondary particles are simulated.  We used a neutrino flux calculated
specifically for the Kamioka site.  According to the cosmic ray proton,
helium and neutron measurements~\cite{Shikaze:2003,NeutronMonitor}, the
cosmic ray flux was near that of solar minimum until the summer of 1999,
rapidly decreased during the next year, and was at the minimum value
consistent with solar maximum from summer of 2000 until Super-Kamiokande
stopped taking date in July 2001.  Therefore, the atmospheric neutrino Monte
Carlo is calculated for 3 years of solar minimum, 1 year of changing
activity, and 1 year of solar maximum.

\begin{figure}[thb]
  \includegraphics[width=3.1in]{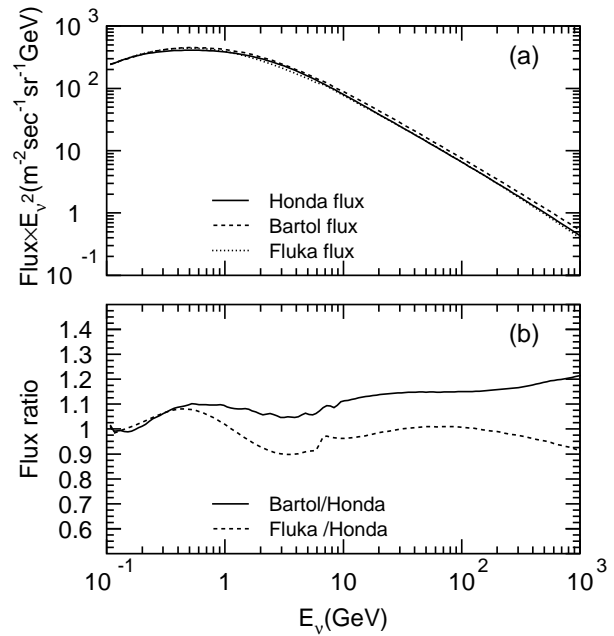}
  \caption{(a) The direction averaged atmospheric neutrino
energy spectrum for $\nu_{\mu}+\overline{\nu}_{\mu}$
calculated by several authors are shown by solid
line~\cite{Honda:2004yz}, dashed line~\cite{Barr:2004br} and dotted
line~\cite{Battistoni:2003ju}. (b) The ratio of the
calculated neutrino flux. The fluxes calculated in 
\cite{Barr:2004br} 
(solid line) and \cite{Battistoni:2003ju} 
(dashed line) are normalized by the flux in
\cite{Honda:2004yz}.  }
  \label{fig:enu_spectra}
\end{figure}

The calculated energy spectra of atmospheric neutrinos at Kamioka are shown
in Fig.~\ref{fig:enu_spectra}(a).  Also shown in
Fig.~\ref{fig:enu_spectra}(b) is the comparison of the calculated fluxes as a
function of neutrino energy.  The agreement among the calculations is about
10\,\% below 10~GeV. This can be understood because the accuracy in recent
primary cosmic ray flux measurements~\cite{Alcaraz:2000vp,Sanuki:2000wh}
below 100~GeV is about 5\,\% and because hadronic interaction models used in
each calculation are different.

However the primary cosmic ray data are much less accurate above 100~GeV.
Therefore, for neutrino energies much higher than 10~GeV, the uncertainties
in the absolute neutrino flux could be substantially larger than the
disagreement level among the calculations.  In Ref.~\cite{Gaisser:2002jj},
the authors discussed that the fit to the low energy ($<$100~GeV) proton
spectra gave a spectrum index of -2.74+-0.01.  However, this spectrum does
not fit well to the high energy data.  Therefore, authors in
Ref.~\cite{Honda:2004yz} fit the high energy data allowing a different
spectral index above 100~GeV and found the best fit value of -2.71.  There is
0.03 difference in the spectrum index for low energy ($<$100~GeV) and high
energy ($>$100~GeV) protons.  Also, it is discussed in
Ref.~\cite{Gaisser:2002jj} that the spectrum index for the He flux can be fit
by either -2.64 or -2.74.  There could be 0.10 uncertainty in the spectrum
index for He.  The spectrum indices for heavier nuclei have uncertainties
larger than 0.05~\cite{Gaisser:2002jj}.  Taking the flux weighted average of
these spectrum index uncertainties, we assign 0.03 and 0.05 for the
uncertainties in the energy spectrum index in the primary cosmic ray energy
spectrum below and above 100~GeV, respectively.
 
\begin{figure}[htb]
  \includegraphics[width=3.1in]{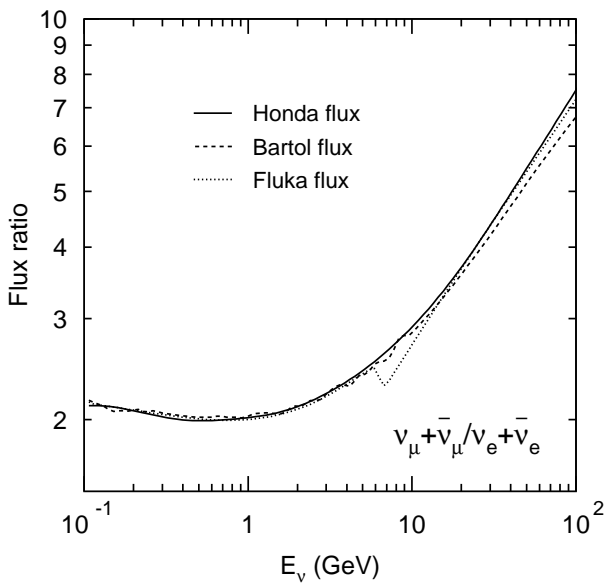}
  \caption{The flux ratio of $\nu_{\mu} + \overline{\nu}_{\mu}$ to
    $\nu_e + \overline{\nu}_e$ averaged over all zenith and azimuth angles
versus neutrino energy.
Solid, dashed and dotted lines
show the prediction by \cite{Honda:2004yz}, 
\cite{Barr:2004br} and
\cite{Battistoni:2003ju}, respectively.
}
  \label{fig:flavor_vs_enu}
\end{figure}

Figure~\ref{fig:flavor_vs_enu} shows the calculated flux ratio of $\nu_{\mu}
+ \overline{\nu}_{\mu}$ to $\nu_e + \overline{\nu}_e$ as a function of the
neutrino energy, integrated over solid angle. This ratio is essentially
independent of the primary cosmic ray spectrum. Especially in the neutrino
energy region of less than about 5~GeV, most of the neutrinos are
produced by the decay chain of pions and the uncertainty of this ratio is
about 3\,\%, which is estimated by comparing the three calculation results.
The contribution of $K$ decay in neutrino production is more important in the
higher energy region; about 10\,\% for $\nu_e + \overline{\nu}_e$ and 20\,\% for
$\nu_\mu + \overline{\nu}_\mu$ at 10~GeV. It increases to more than 30\,\% at
100~GeV for both $\nu_e + \overline{\nu}_e$ and $\nu_\mu +
\overline{\nu}_\mu$.  There, the ratio depends more on the $K$ production
cross sections and the uncertainty of the ratio is expected to be larger. A
20\,\% uncertainty in the $ K / \pi$ production 
ratio~\cite{Battistoni:2002ew,Agrawal:1996gk} 
causes at least a few percent uncertainty in the $\nu_{\mu} +
\overline{\nu}_{\mu}$ to $\nu_e + \overline{\nu}_e$ ratio in the energy range
of 10 to 100~GeV.  However, as seen from Fig.~\ref{fig:flavor_vs_enu}, the
difference in the calculated $\nu_{\mu} + \overline{\nu}_{\mu}$ to $\nu_e +
\overline{\nu}_e$ ratio is as large as 10\% at 100 GeV.
As a consequence, above 5~GeV,
we assumed that the uncertainty linearly increases with $\log E_\nu$ from 
3\,\% at 5~GeV to 10\,\% at 100~GeV.

\begin{figure}[htb]
  \includegraphics[width=3.1in]{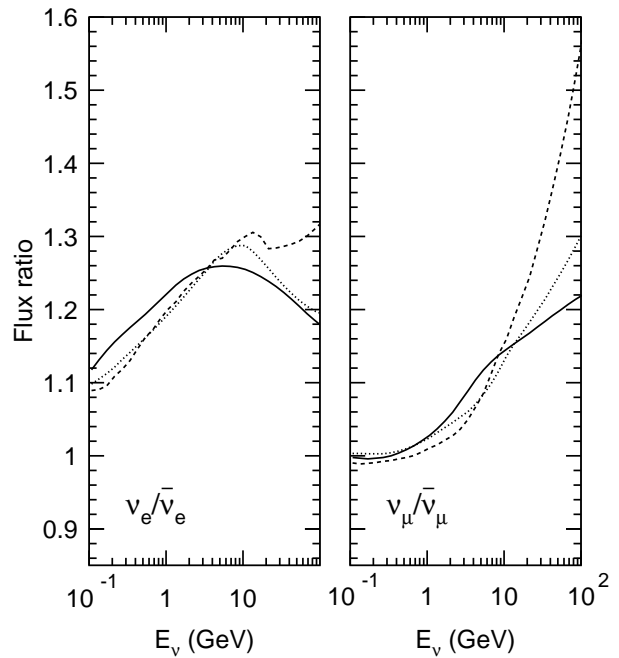}
\caption{The flux ratios of $\nu_{\mu} $ to $\overline{\nu}_{\mu}$
and $\nu_e$ to $\overline{\nu}_e$ versus neutrino energy.  Solid,
dashed and dotted lines show the prediction by
\cite{Honda:2004yz},
\cite{Barr:2004br} and
\cite{Battistoni:2003ju}, respectively 
(same key as Fig.~\ref{fig:flavor_vs_enu}).
}
        \label{fig:nu_vs_anti-nu}
\end{figure}

Figure~\ref{fig:nu_vs_anti-nu} shows the calculated flux ratios of
$\nu_{\mu} $ to $\overline{\nu}_{\mu}$ and $\nu_e$ to
$\overline{\nu}_e$.  The calculations agree to about 5\,\% for both of
these ratios below 10~GeV. However, the disagreement gets larger
above 10~GeV as a function of neutrino energy. The systematic errors
in the $\nu/\overline{\nu}$ ratio are assumed to be
5\,\% below 10~GeV and linearly increase with $\log E_\nu$ to 
10\,\% and 25\,\% at 100~GeV, for the $\nu_e$ to $\overline{\nu}_e$ 
and $\nu_{\mu} $ to $\overline{\nu}_{\mu}$ ratios, respectively.

\begin{figure}[htb]
  \includegraphics[width=3.1in]{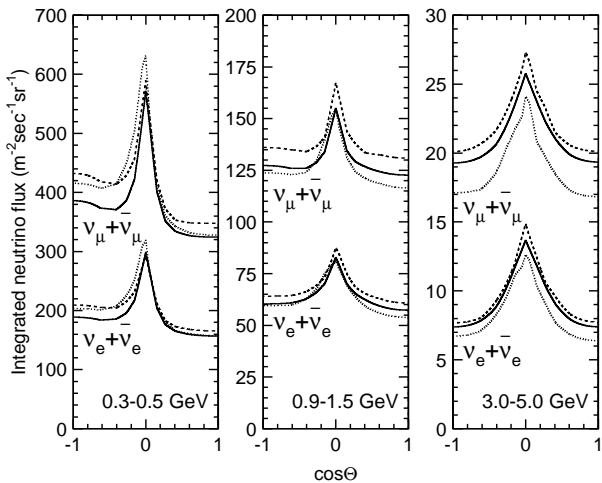}
  \caption{The flux of atmospheric neutrinos versus zenith angle.
Solid, dashed and dotted lines
show the prediction by \cite{Honda:2004yz}, 
\cite{Barr:2004br} and
\cite{Battistoni:2003ju}, respectively
(same key as Fig.~\ref{fig:flavor_vs_enu}).
}
  \label{fig:zenith_vs_enu}
\end{figure}

Figure~\ref{fig:zenith_vs_enu} shows the zenith angle dependence of
the atmospheric neutrino fluxes for several neutrino energies.  At low
energies, and at the Kamioka location, the fluxes of downward-going
neutrinos are lower than those of upward-going neutrinos.  This is due
to the deflection of primary cosmic rays by the geomagnetic field,
roughly characterized by a minimum rigidity cutoff. For neutrino
energies higher than a few GeV, the calculated fluxes are essentially
up-down symmetric, because the primary particles are more energetic
than the rigidity cutoff. 

The enhancement of the flux near horizon for low energy neutrinos is a
feature characteristic of the three dimensional nature of the
neutrino production
in cosmic ray hadronic showers.  This is properly treated in current flux
calculations~\cite{Battistoni:1999at,
  Honda:2001xy,Tserkovnyak:2002xx,Liu:2002sq,
  Battistoni:2003ju,Wentz:2003bp,Favier:2003gn, Honda:2004yz,Barr:2004br}.
However, in Super-Kamiokande, the horizontal enhancement cannot be seen in
the lepton zenith angle distribution, due to the relatively poor angular
correlation between neutrinos and leptons below 1~GeV.  The uncertainties in
the up-down and vertical-horizontal ratios of the number of events are
estimated by comparing the predicted ratios by various flux models. These
uncertainties generally depend on the energy and the neutrino flavor. The
uncertainty in the up-down event ratio is about 1 to 2\,\% in the sub-GeV
energy region and is about 1\,\% in the multi-GeV energy region.  The main
source of the uncertainty in the vertical-horizontal ratio around a GeV is
the size of the horizontal enhancement of the flux due to the three
dimensional effect; the uncertainty is estimated to be less than a few
percent.

\begin{figure}[htb]
  \includegraphics[width=3.1in]{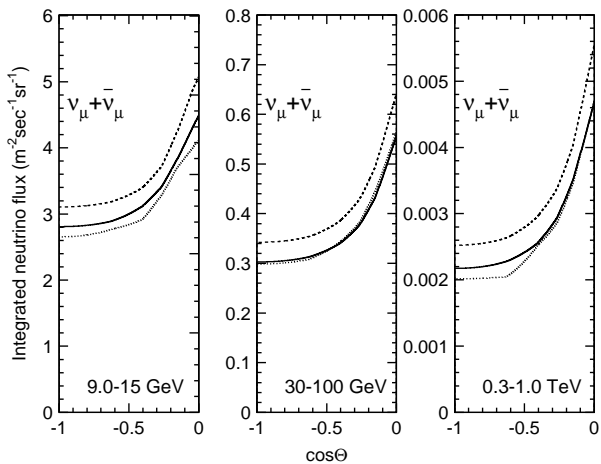}
  \caption{The flux of upward-going atmospheric neutrinos versus zenith
angle for higher energy region. Solid, dashed and dotted lines
show the prediction by \cite{Honda:2004yz}, 
\cite{Barr:2004br} and
\cite{Battistoni:2003ju}, respectively
(same key as Fig.~\ref{fig:flavor_vs_enu}).
}
  \label{fig:zenith_vs_enu-high}
\end{figure}

In the higher energy region, where upward through-going muons are relevant,
the largest source of the uncertainty in the vertical-horizontal ratio
is the $K$ production cross section. We assume that the $K / \pi$
production ratio uncertainty is 20\,\% in the whole energy 
region~\cite{Battistoni:2002ew, Agrawal:1996gk}.
The uncertainties in the zenith angle and energy distributions 
due to the  $K / \pi$ production uncertainty are included in the
systematic errors in the analysis. This error is most important
for higher energy neutrinos. For example, the vertical-horizontal
uncertainty for upward through-going muons due to 
the  $K / \pi$ production uncertainty is estimated to 
be 3\,\%~\cite{Lipari2000}. Figure~\ref{fig:zenith_vs_enu-high} shows the
zenith angle dependence of the atmospheric neutrino fluxes for higher
energy region observed as upward muons in Super-Kamiokande.

\begin{figure}[thb]
  \includegraphics[width=3.1in]{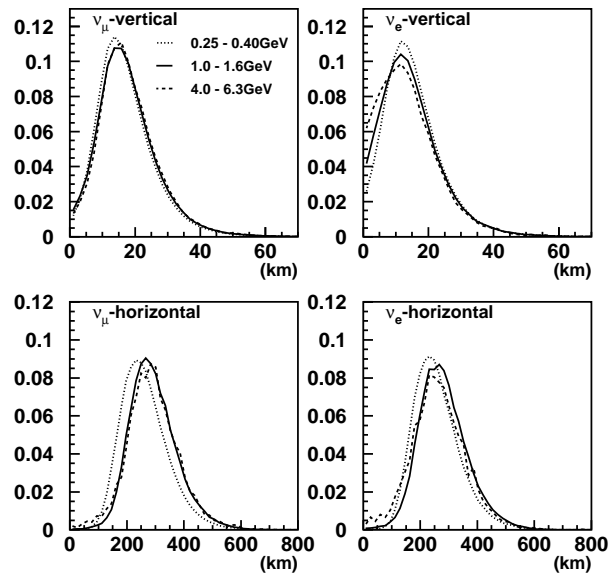} \caption{The
  calculated flight length of neutrinos for
vertically down-going ($\cos \Theta = 0.95 - 1.00$) 
and near horizontal-going ($\cos \Theta = 0.05 - 0.10$) directions.
Distributions are made for both muon-neutrinos and 
electron-neutrinos and for three energy 
intervals~\cite{Honda:2004yz}.
}
  \label{fig:flight-length}
\end{figure}

The flight length of neutrinos is an important ingredient in the
analysis of neutrino oscillation. For neutrinos passing a great
distance through the Earth, the flight length can be accurately
estimated. However, for horizontal and downward going neutrinos, the
height of production in the upper atmosphere must be
distributed by the Monte Carlo method. Figure~\ref{fig:flight-length}
shows the calculated flight length distributions for
vertically down-going and horizontal neutrinos.

 In summary of the atmospheric neutrino flux, we remark that, 
the ($\nu_{\mu} + \overline{\nu}_{\mu} $) over $(\nu_e +
\overline{\nu}_e$) flux ratio is predicted to an accuracy
of about 3\,\% in the energy region relevant to the 
data analysis discussed in this paper. 
The zenith angle dependence of the flux is well
understood, and especially, above a few GeV neutrino energies, the
flux is predicted to be up-down symmetric.

\section{The Super-Kamiokande Detector} 

\begin{figure}[htb]
  \includegraphics[width=3.1in]{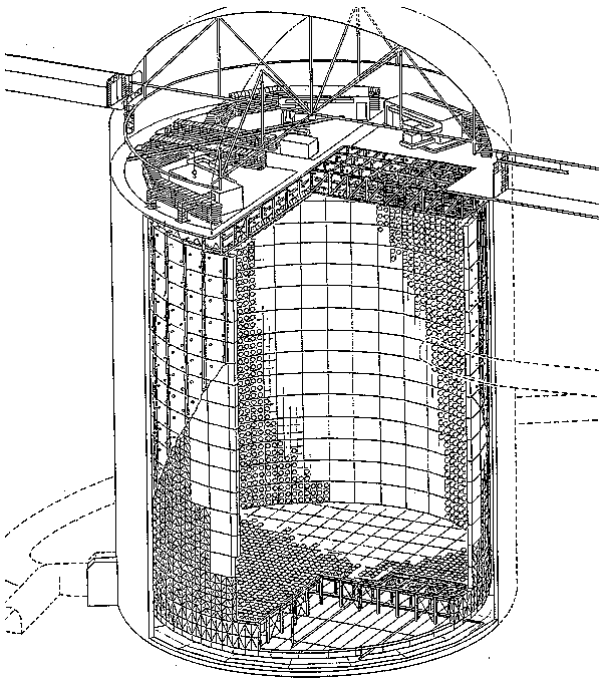}
  \caption{A drawing of the Super-Kamiokande detector. The cutaway shows the
    inside lined with photomultiplier tubes comprising a photocathode
    coverage of about 40\%. The support structure is stainless steel beams.
    The thin outer region is shown with sparser density of outward facing
    PMTs: 2 outer PMTs for every unit of $3 \times 4$ inner PMTs. The top of
    the detector, under the hemisphere, consists of electronics huts and open
    work area.}
   \label{fig:skdetector}
\end{figure}

Super-Kamiokande is a 50~kiloton water Cherenkov detector located at the
Kamioka Observatory of the Institute for Cosmic Ray Research, University of
Tokyo. Figure~\ref{fig:skdetector} shows a cut-away diagram of the
Super-Kamiokande detector. This facility is in the Mozumi mine of the Kamioka
Mining Company in Gifu prefecture, Japan, under the peak of Mt.  Ikenoyama,
providing a rock overburden of 2,700~m.w.e. Super-K consists of two
concentric, optically separated water Cherenkov detectors contained in a
stainless steel tank 42 meters high and 39.3 meters in diameter, holding a
total mass of 50,000 tons of water. The inner detector (ID) is comprised of
11,146 Hamamatsu~R3600 50~cm diameter photomultiplier tubes (PMTs), viewing a
cylindrical volume of pure water 16.9~m in radius and 36.2~m high.  The 50~cm
PMTs were specially designed~\cite{Suzuki:1993as} to have good single
photoelectron (p.e.)  response, with a timing resolution of 2.5~nsec RMS.
The ID is surrounded by the outer detector (OD), a cylindrical shell of water
2.6 to 2.75~m thick including a dead space 55~cm. The OD is optically
isolated from the ID, and is instrumented with 1,885 outward-facing Hamamatsu
R1408 20~cm PMTs, providing both a 4$\pi$ active veto and a thick passive
radioactivity shield. The information from the outer detector is used to
identify both incoming and outgoing muons.

Both ID and OD PMT signals are processed by asynchronous,
self-triggering circuits that record the time and charge of each PMT
hit over a threshold.  Each ID PMT signal is digitized with custom
Analog Timing Modules (ATMs)~\cite{Tanimori:1989qi,Ikeda:1992cy} which
provide 1.2~$\mu$sec timing range at $0.3$~nsec resolution and $550$~pC
charge range at $0.2$~pC resolution ($\sim0.1$~p.e.). The
ATM has automatically-switched dual channels to provide deadtime-free
data acquisition. The outer PMT signals are processed with custom
charge-to-time conversion modules and digitized with LeCroy~1877
multi-hit TDCs over a $-10$~$\mu$sec to $+6$~$\mu$sec window centered on
the trigger time.  More details of the Super-K detector can be found
in~\cite{Fukuda:2002uc}.

An event used in the atmospheric neutrino analysis is triggered by the
coincidence of at least 30 PMT hits in a 200~nsec window.  The hit threshold
for each individual PMT is about $1/4$~p.e. This trigger condition
corresponds to the mean number of hit PMTs for a 5.7~MeV electron. The
trigger rate is 10-12~Hz. The trigger rate due to cosmic ray muons is 2.2~Hz.
Digitized data are saved at a total rate of 12~GB per day.

The detector is simulated with a Monte Carlo program based on the
GEANT package\cite{Brun93aa}, in which the propagation of particles,
the generation and propagation of Cherenkov photons, and the response
of the PMTs is considered.
For hadronic interactions in water, the CALOR
package~\cite{Zeitnitz:1994bs} was employed in our simulation code.
This package is known to reproduce the pion interactions well down
to low momentum regions of $\sim$\,1\,GeV/$c$. For still lower 
momenta ($p_\pi$\,$\leq$\,500\,MeV/$c$), a custom program based on experimental
data from $\pi-^{16}$O scattering~\cite{Bracci:1972}
and $\pi-p$ scattering~\cite{Carroll:1976hj} was used in our simulation code.

In connection with the propagation of charged particles, Cherenkov photons
are generated. For the propagation of Cherenkov photons in water, Rayleigh
scattering, Mie scattering and absorption were considered in our simulation
code. The attenuation coefficients used were tuned to reproduce the
measurement using laser system (Section~\ref{sec:calibration}).  Light
reflection and absorption on detector material, such as the surface of PMTs
and black plastic sheets between the PMTs was simulated based on direct
measurements, using probability functions that depend on the photon incident
angle.

\subsection{Calibration of the Super-Kamiokande Detector}
\label{sec:calibration}

Water transparency was measured using a dye laser beam injected into
detector water at wavelengths of 337, 371, 400, and 420 nm.  From the
spatial and timing distribution of observed laser light, both
absorption and scattering coefficients were studied and incorporated
into our detector simulator.  The water transparency was continuously
monitored using cosmic ray muons as a calibration source.

\begin{figure}[htb]
  \includegraphics[width=3.5in]{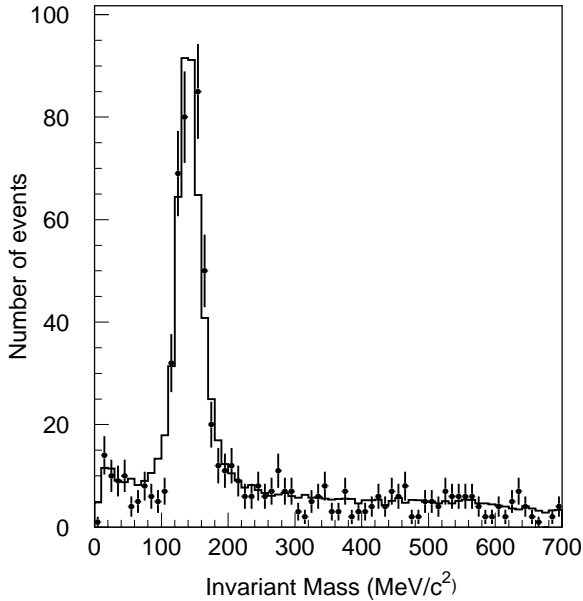}
  \caption{The invariant mass distribution of fully-contained
    events with two $e$-like rings and no muon-decay electron, for SK data
    (points) and atmospheric neutrino Monte Carlo (histogram). A peak
    from neutrino induced $\pi^0$ is clearly observed.}
   \label{fig:pizero}
\end{figure}

\begin{figure}[htb]
  \includegraphics[width=3.2in,height=2.0in]{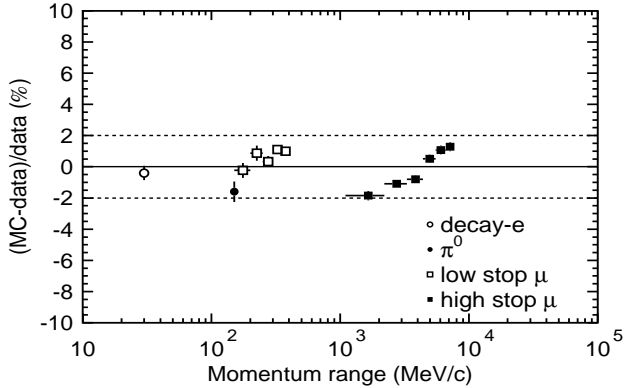}
  \caption{The determination of the absolute energy scale of 
    Super-Kamiokande based on in situ calibration with $\mu$-decay electrons,
    $\pi^0 \rightarrow \gamma \gamma$ invariant mass, and the Cherenkov
  light of stopping cosmic ray muons.}
   \label{fig:energy_scale}
\end{figure}

\begin{figure}[htb]
  \includegraphics[width=3.1in]{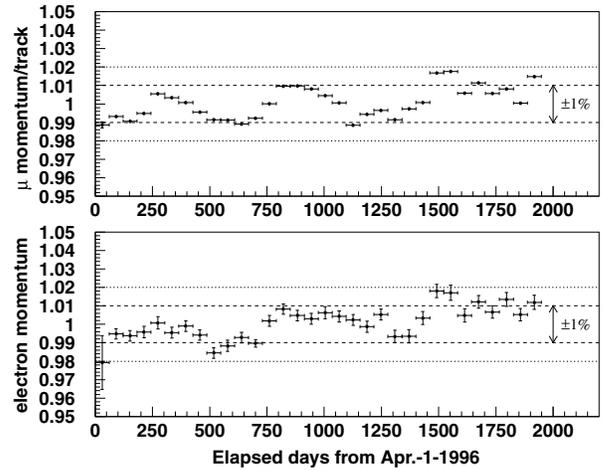}
  \caption{The mean reconstructed energy of cosmic ray stopping muons
    divided by their range (upper) and muon-decay electron (lower) as a
    function of elapsed days. Vertical axes in both figures are normalized to
    mean values and each data point corresponds to two month period.  The
    variation is within $\pm$ 2\,\%.}
   \label{fig:energy_stability}
\end{figure}

The accuracy of the absolute energy scale was estimated to be
$\pm1.8\,\%$ based on the following calibration sources: 
the total number of photo-electrons as a function of muon track
length, where the muon track length is estimated by the reconstructed
muon entrance point and the reconstructed vertex point of an
electron from the muon-decay; the total number of photo-electrons 
as a function of Cherenkov angle for low energy cosmic ray muons;   
the spectrum of muon-decay electrons; and the invariant mass of
$\pi^0$s produced by neutrino interactions (Figure~\ref{fig:pizero}).
Figure~\ref{fig:energy_scale} summarizes the absolute energy scale
calibration by these studies.  The stability of the energy scale was
also monitored continuously using stopping muons and muon-decay
electrons. Figure~\ref{fig:energy_stability} shows the time variation
of the mean reconstructed energy of stopping muons divided by muon
range and the mean reconstructed electron energy from
muon-decays. 
The R.M.S of the energy scale variation is $\pm0.9\,\%$
over the time of the experiment.  From combining the absolute
energy scale accuracy study ($\pm1.8\,\%$) and the energy scale time variation
($\pm0.9\,\%$), the total uncertainty of the energy scale 
of atmospheric neutrino detection was estimated to be $\pm2.0\,\%$.

\begin{figure}[htb]
  \includegraphics[width=3.1in]{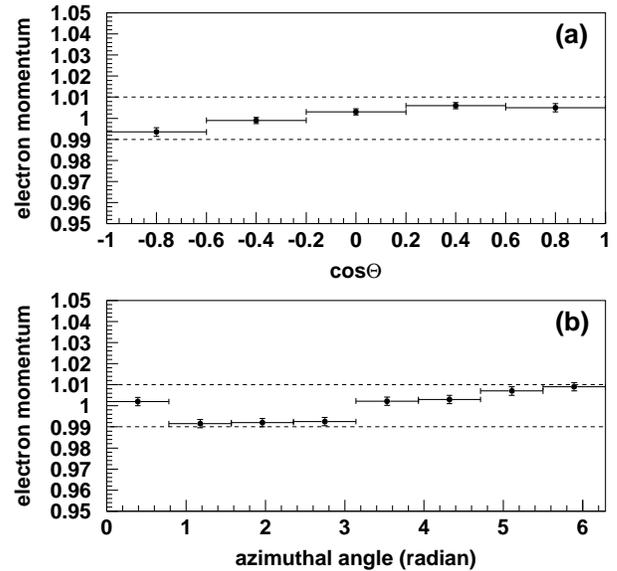}
  \caption{The gain uniformity of the Super-Kamiokande detector
    as determined by the mean value of
    the reconstructed decay electron momentum (a) as a function of zenith
    angle, and (b) as a function of azimuthal angle.
    Vertical axes in both figures are normalized to the mean values.}
  \label{fig:mue_uniformity}
\end{figure}

\begin{figure}[htb]
  \includegraphics[width=3.1in]{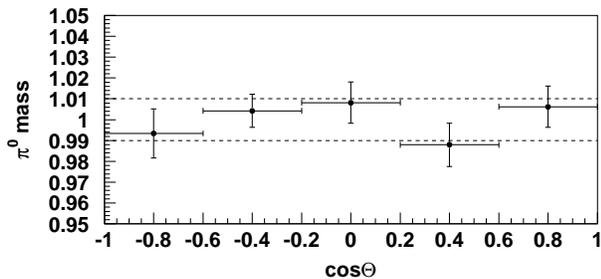}
  \caption{The gain uniformity of the Super-Kamiokande detector
   as determined by the
   fitted peak of the $\pi^0 \rightarrow \gamma \gamma$ mass
   distribution as a function of zenith angle.
   Vertical axis is normalized to the mean value.}
\label{fig:pi0_uniformity}
\end{figure}

The uniformity of the detector response was studied by decay electrons
from stopping cosmic ray muons and neutrino induced $\pi^0$.  Both are
good calibration sources because the vertex position is distributed
in the fiducial volume and the momentum distribution is nearly uniform in
all directions. To account for muon polarization in the
estimation of the zenith and azimuthal angle dependence of the
detector gain, only electrons decaying in the direction perpendicular
to the initial muon direction are used.  This condition is \(-0.25 <
\cos\Theta_{e\leftrightarrow\mu} < 0.25\) where
\(\Theta_{e\leftrightarrow\mu}\) is the opening angle between the
electron and muon directions.  Using the selected electrons, the
mean of the reconstructed momentum of the electrons are plotted as a 
function of the zenith angle of the electrons in Figure
\ref{fig:mue_uniformity}--(a).
From the figure, the detector gain was uniform
over all zenith angles within \(\pm 0.6\)\,\%.  Figure
\ref{fig:mue_uniformity}--(b) shows the azimuthal angle dependence of
the reconstructed momentum.  Again, the detector gain is uniform over
all azimuthal angles within \(\pm 1\)\,\%.  Finally, Figure
\ref{fig:pi0_uniformity} shows the zenith angle dependence of the
reconstructed $\pi^0$ mass.  This figure also suggests that the
detector gain was uniform over all zenith angles within \(\pm 1\)\,\%.

\section{Atmospheric Neutrino Monte Carlo}
\label{sec:atmnumc}

The result published in this paper relies heavily on detailed
comparison of the experimental data with the theoretical expectation.
An important element of this is to simulate the interaction of
neutrinos from 10 MeV to 100 TeV with the nuclei of water, or in the
case of upward muons, the nuclei of the rock surrounding the detector,
assumed to be ``standard rock''(Z=11, A=22).  Therefore, we have developed 
two Monte Carlo models
designed to simulate neutrino interactions with protons,
oxygen and sodium~\cite{Hayato:2002sd, Casper:2002sd}. Both models
use similar input physics models. Here, one of the 
models~\cite{Hayato:2002sd} ({\tt NEUT}) will be described. The detailed
description for the other model ({\tt NUANCE}) can be found 
elsewhere~\cite{Casper:2002sd}.

 In the simulation program, the following charged
and neutral current neutrino interactions 
are considered:
\begin{itemize}
\item (quasi-)elastic scattering, $\nu\ N \rightarrow l\ N'$,
\item single meson production, $\nu\ N \rightarrow l\ N'\ m$,
\item coherent $\pi$ production, $\nu\ ^{16}{\rm O}\ \rightarrow\ l\ \pi\  
    ^{16}{\rm O}$,
\item deep inelastic scattering, $\nu N\ \rightarrow l\ N'\ hadrons$.
\end{itemize}
Here, $N$ and $N'$ are the nucleons (proton or neutron), $l$ is the
lepton, and $m$ is the meson, 
respectively.  For single meson production, $K$ and $\eta$ production
are simulated as well as the dominant $\pi$ production processes.  If
the neutrino interaction occurred in the oxygen nuclei, generated
particles like pions and kaons interact with the
nucleus before escaping.

\subsection{Elastic and quasi-elastic scattering}
         The formalization of quasi-elastic scattering 
off a free proton, which was used in the simulation programs, was
described by Llewellyn-Smith\cite{LlewellynSmith:1972zm}.  For
scattering off nucleons in $^{16}$O, the Fermi motion of the nucleons
and Pauli Exclusion Principle were taken into account.  The nucleons
are treated as quasi-free particles using the relativistic Fermi gas
model of Smith and Moniz\cite{Smith:1972xh}.
The momentum distribution of the nucleons
were assumed to be flat up to the fixed Fermi surface momentum of 225~MeV/$c$.
This Fermi momentum distribution was also used for other
nuclear interactions.  The nuclear potential was set to 27~MeV/$c$.

\subsection{Single meson production}

Rein and Sehgal's model was used to simulate the resonance
productions of single $\pi$, $K$ and
$\eta$~\cite{Rein:1981wg,Rein:1987cb}.  
In this method, the
interaction is separated into two parts:
\begin{eqnarray}
        \nu + N &\rightarrow& l + N^*, \nonumber \\
        N^* &\rightarrow& m + N', \nonumber
\end{eqnarray}
where $m$ is a meson, $N$ and $N'$ are nucleons, and $N^*$ is a baryon
resonance.  The hadronic invariant mass, $W$, the mass of the intermediate
baryon resonance, is restricted to be less than 2~GeV/$c^2$.  In addition to
the dominant single $\pi$ production, $K$ and $\eta$ production is
considered. The production of $\eta$ is evidently much smaller than $\pi$, as
seen in Fig.~\ref{fig:pizero} where there is no evidence for a mass peak near
549 MeV/$c^2$ in data or Monte Carlo. 

To determine the angular distribution of pions in the final state, we also
use Rein and Sehgal's method for the $P_{33}$(1232) resonance.  For the other
resonances, the directional distribution of the generated pions is set to be
isotropic in the resonance rest frame.  The angular distribution of $\pi^+$
has been measured for $\nu p \rightarrow \mu^- p \pi^+$\cite{Kitagaki:1986ct}
and the results agree well with the Monte Carlo prediction.  We also consider
the Pauli blocking effect in the decay of the baryon resonance by requiring
that the momentum of nucleon should be larger than the Fermi surface
momentum.  Pion-less delta decay is also considered, where 20\,\% of the
events do not have the pion and only the lepton and nucleon are
generated~\cite{Singh:1998ha}.  

\begin{figure}
\includegraphics[width=7cm]{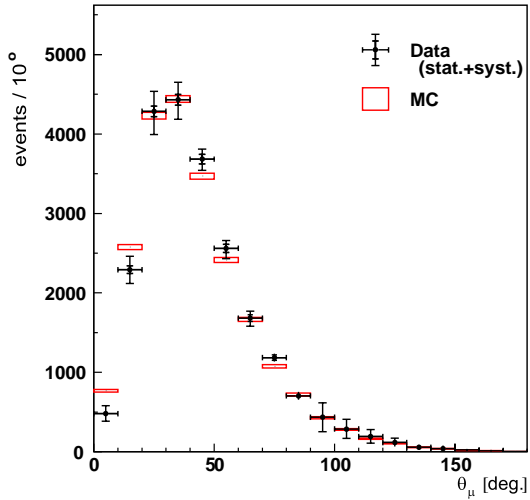}
\caption{ The scattering angle distribution by neutrino
interactions off the $H_2O$ target from the K2K experiment
(data are from Fig.1(b) of Ref.~\cite{Ahn:2002up}).  Single
Cherenkov ring events observed by the 1~kton water Cherenkov detector
are used. The histogram shows the prediction by the Monte Carlo used
in the present analysis. }
\label{fig:theta_nu-mu}
\end{figure}

The quasi-elastic and single meson production models have a parameter
(axial vector mass, $M_{A}$) that must be determined by
experiments. For larger $M_{A}$ values, interactions with higher
$Q^2$ values (and therefore larger scattering angles) are enhanced for
these channels.  The $M_{A}$ value was tuned using the
K2K~\cite{Ahn:2002up} near detector data. In our atmospheric
neutrino Monte Carlo simulation, $M_{A}$ is set to
1.1~GeV for both the quasi-elastic and single-meson production
channels, but the uncertainty of the value is estimated to be 10\,\%.
Figure~\ref{fig:theta_nu-mu} shows the K2K 1~kton water
Cherenkov data on the scattering angle for single Cherenkov ring
events~\cite{Ahn:2002up} together with the prediction by the Monte
Carlo used in this analysis. The scattering angle agrees well between
the data and Monte Carlo overall, although the suppression of events at
small angle is being studied by several groups\cite{Cavanna:2004}.
 
Coherent single-pion production, the interaction between the neutrino
and the entire oxygen nucleus, is simulated using the formalism developed
by Rein and Sehgal \cite{Rein:1983pf}.

\subsection{Deep inelastic scattering}
In order to calculate the cross-sections of deep inelastic scattering, the
GRV94~\cite{Gluck:1995uf} parton distribution function is used.  In the
calculation, the hadronic invariant mass, $W$, is required to be greater than
1.3~GeV/$c^2$.  However, the multiplicity of pions is restricted to be
greater than or equal to 2 for $1.3<W<2.0$~GeV/$c^2$, because single pion
production is separately simulated as previously described.  In order to
generate events with multi-hadron final states, two models are used. For $W$
between 1.3 and 2.0~GeV/$c^2$, a custom-made program~\cite{Nakahata:1986zp}
is used to generate the final state hadrons; only pions are considered in
this case. For $W$ larger than 2~GeV/$c^2$,
PYTHIA/JETSET~\cite{Sjostrand:1994yb} is used.

\begin{figure}
\begin{center}
\includegraphics[width=7.2cm]{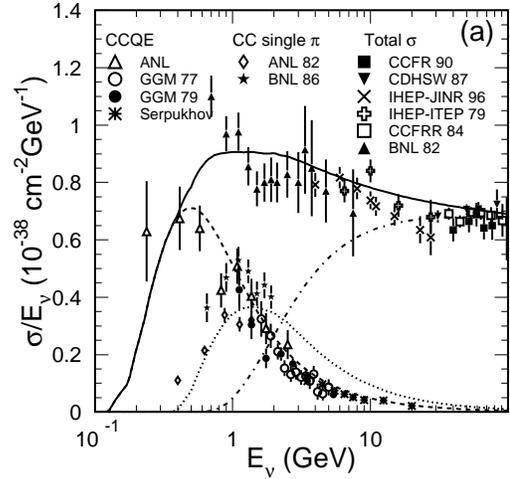}
\par\vspace{-0.25in}
\includegraphics[width=7.2cm]{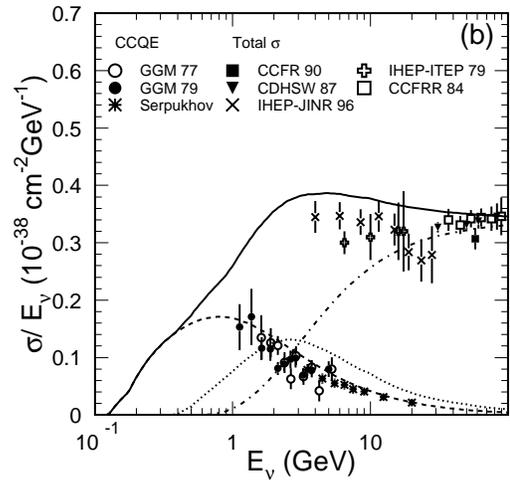}
\par\vspace{-0.15in}
\caption{ Charged current
total cross section divided by $E_\nu$ for (a) neutrino and
(b) anti-neutrino nucleon charged current interactions. Solid line 
shows the calculated total cross section. The dashed, dot and dash-dotted 
lines show the calculated quasi-elastic, single-meson
and deep-inelastic scatterings, respectively.
Data points are taken from the following
experiments:
\mbox{({$\triangle$})ANL\protect\cite{Barish:1977qk}},
\mbox{({$\bigcirc$})GGM77\protect\cite{Bonetti:1977cs}},
\mbox{({$\bullet$})GGM79(a)\protect\cite{Ciampolillo:1979wp},(b)\protect\cite{Armenise:1979zg}},
\mbox{({$\ast$})Serpukhov\protect\cite{Belikov:1985kg}},
\mbox{({$\Diamond$})ANL82\protect\cite{Radecky:1982fn}},
\mbox{({$\star$})BNL86\protect\cite{Kitagaki:1986ct}},
\mbox{({$\blacksquare$})CCFR90\protect\cite{Auchincloss:1990tu}},
\mbox{({$\blacktriangledown$})CDHSW87\protect\cite{Berge:1987zw}},
\mbox{({$\times            $})IHEP-JINR96\protect\cite{Anikeev:1996dj}},
\mbox{({$+$})IHEP-ITEP79\protect\cite{Mukhin:1979bd}},
\mbox{({$\Box$})CCFRR84\protect\cite{MacFarlane:1984ax}},
and \mbox{({$\blacktriangle$})BNL82\protect\cite{Baker:1982ty}}.
}
\label{plot_tot}
\end{center}
\end{figure}

Total charged current cross sections including quasi-elastic scattering,
single meson productions and deep inelastic scattering are shown in 
Fig.\ref{plot_tot}.

\subsection{Nuclear effects}
        The interactions of mesons within
the $^{16}$O nucleus are also important for the atmospheric neutrino
analysis. Basically, all of the interactions are treated by
using a cascade model. The interactions of pions are very important
because the cross section for pion production is quite large for
neutrino energies above 1 GeV and the interaction cross sections for
pions in nuclear matter is also large.

In our simulation program, we consider the following pion interactions 
in $^{16}$O: inelastic scattering, charge exchange and absorption. The
procedure to simulate these interactions is as follows. The
initial position of the pion generated according to the Woods-Saxon
nucleon density distribution\cite{Woods:1954}. The interaction mode is
determined from the calculated mean free path of each interaction.  To
calculate the mean free path, we adopt the model described by Salcedo 
{\it et al.}~\cite{Salcedo:1988md}. 
The calculated mean free path depends not only on the momentum of the pion but
also on the position of the pion in the nucleus. If inelastic scattering or
charge exchange occurs, the direction and momentum of the pion are determined
by using the results of a phase shift analysis obtained from $\pi-N$
scattering experiments\cite{Rowe:1978fb}. When calculating the pion
scattering amplitude, the Pauli blocking effect is also taken into account by
requiring the nucleon momentum after interaction to be larger than the Fermi
surface momentum at the interaction point. The pion interaction simulation
was checked using data for the following three interactions: $\pi ^{12}$C
scattering, $\pi ^{16}$O scattering and pion photo-production ($\gamma +
^{12}$C $\rightarrow \pi^- + X$)\cite{Ashery:1981tq}.


\section{Super-Kamiokande Data}
\label{sec:skdata}

The Super-Kamiokande data set was acquired from May, 1996
to July, 2001. Three separate data reduction paths were used to isolate
samples of fully-contained events, partially-contained events, and
upward-going muons. The fully-contained and partially-contained data sets
shared a common set of good run selection criteria, and have identical
live-time. The upward-going muon data set relies mostly on fitting long muon
track directions; it was less susceptible to detector effects, and therefore
had looser data quality cuts and somewhat higher live-time.

\begin{figure}[htb]
  \includegraphics[width=3.1in]{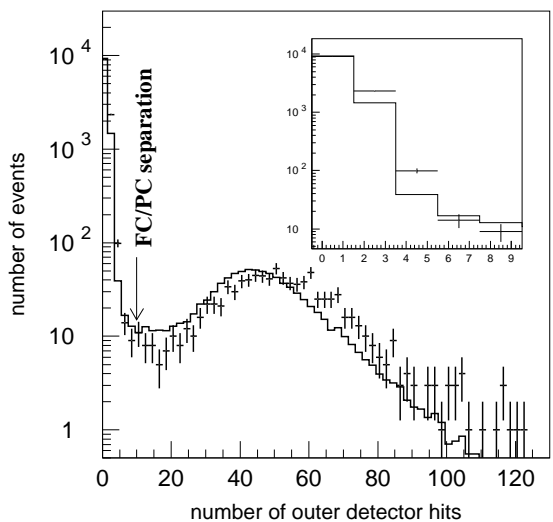} 
\caption{The number of hits in the largest outer detector cluster, 
         which is used to separate the fully-contained and
         partially-contained event samples. The histogram shows the MC
         prediction with neutrino oscillations.}  
\label{fig:od-nhit}
\end{figure}

To separate fully-contained and partially-contained events, a fast spatial
clustering algorithm was applied to the outer detector hits; if the number of
hits in the largest OD cluster was less than 10, the event was defined as
fully-contained (FC), otherwise, it was defined as partially-contained (PC).
Figure~\ref{fig:od-nhit} shows the number of the outer detector hits in the
largest OD cluster. A clear separation of FC and PC events is seen at
10~hits.  The systematic uncertainty of the FC and PC separation was
estimated by scaling the number of outer detector hits to match the
distribution among data and MC.

In the early stage of the Super-Kamiokande experiment, two data analyses
based on independent data reduction, reconstruction, and simulation were
carried out to make sure that the atmospheric neutrino results did not have
any serious mistakes~\cite{Fukuda:1998tw,Fukuda:1998ub}. After confirming
that the two analyses described atmospheric neutrino results equally well,
they were unified. This paper is based on the unified analysis plus methods
developed since that time.
 
\subsection{Fully-Contained Events} 

\subsubsection{Data Reduction}
\label{sec:FCdatareduction}

The Super-Kamiokande event sample consists mainly of downward-going cosmic
ray muons and low energy radioactivity from parents such as radon. Owing to
the double structure of the inner and outer detectors, cosmic ray muons are
easily removed with high efficiency by requiring little or no activity in the
outer detector. For atmospheric neutrino analysis, we then consider only
events with visible energy above 30 MeV, where visible energy ($E_{vis}$) is
defined as the energy of an electromagnetic shower that gives a certain
amount of Cherenkov light (for example, a muon of momentum 300 MeV/$c$ yields
a visible energy of about 110 MeV). To make the final FC data sample, five
steps of data reduction criteria were used:

(i-ii) Simple and efficient criteria were applied 
in the first and second reduction steps:
(1) the total charge collected in the inner detector within a
300~nsec time window must be greater than $200$~p.e.'s;
(2) the ratio of the maximum p.e. in any single ID PMT to the total number
of p.e.'s in the inner detector must be less than $0.5$;
(3) the number of hits in the outer detector within an 800~nsec time window
should be less than 25~hits;
(4) the time interval from the preceding event should be greater
than 100~$\mu$sec, to 
reject electrons from stopping muon decays.

(iii) More complex criteria were applied in the third reduction step
with the help of event reconstruction tools, for further rejection of 
cosmic ray muons and low energy events:
(1) no spatial cluster of more than 10 OD PMT hits is allowed within 
8~m from the entrance or exit point of a candidate muon track fit
to the inner detector light pattern,
(2) the number of ID hits in 50~nsec residual time window
should be 50~hits or more.

(iv) In the fourth reduction step, additional selection criteria
were used to eliminate spurious
events, such as those due to ``flashing'' PMTs that emit light from
internal corona discharges.
Flasher events were removed by two different methods.
(1) Typical flasher events have broader PMT timing distributions
than the neutrino events. Events with broader timing distributions
were eliminated.
(2) Since flasher events have a tendency to be
repeated with similar spatial hit distribution, the pattern
information of observed charge was used to eliminate these events. 
A correlation parameter based on the charge pattern was
calculated with other data events and a ``matched'' tag was assigned
for highly correlated events. A cut was applied based on maximum
correlation value and the number of ``matched'' with other events.

(v) Two further event types are eliminated in the fifth reduction step.
(1) Events are removed which have $\geq$~10 OD hits in 200~nsec coincidence
preceding the trigger time ( $-8900 \sim -100$~nsec );
this eliminates decay electrons 
from invisible cosmic ray muons that are below Cherenkov threshold 
in the inner detector.
(2) Cosmic ray muons are removed using a more precise fitter
and the same criteria as (1) of (iii).

(vi) Finally, the vertex was required to be within a fiducial
volume, 2 meters from the wall of the inner detector, and the
visible energy was required to be greater than 30~MeV.

Table~\ref{tab:fc-reduction} shows the number of events for each
reduction step. Also shown are the the number of Monte Carlo events
for each reduction step.

\begin{table}[htbp]
\begin{center}
  \begin{tabular}{l|c|c}
\hline\hline
Reduction step & Data & Monte Carlo \\
\hline\hline
Trigger & 1889599293 & 14013.9~(100.00\,\%) \\
First reduction & 4591659 & 14006.3~(99.95\,\%) \\
Second reduction & 301791 & 14006.1~(99.94\,\%) \\
Third reduction &  66810  & 13993.3~(99.85\,\%) \\
Fourth reduction & 26937  & 13898.1~(99.17\,\%) \\
Fifth reduction &  23984  & 13895.3~(99.15\,\%) \\
Fiducial volume and  & 12180 & 13676.7~(97.59\,\%) \\
~~visible energy cuts  &  &  \\
\hline\hline
  \end{tabular}
  \caption{Number of events after each reduction for fully-contained events
    during 1489~days of the detector live-time.
   The Monte Carlo numbers and efficiencies down to the fifth reduction
   are for events whose real vertex is
   in the fiducial volume, the number of outer detector hits fewer 
   than 10 and the visible energy larger than 30~MeV. In the last line,
   the fitted vertex is used for both data and Monte Carlo.}
  \label{tab:fc-reduction}
\end{center}
\end{table}

\subsubsection{Event Reconstruction}

\begin{figure}[htb]
  \includegraphics[width=3.1in]{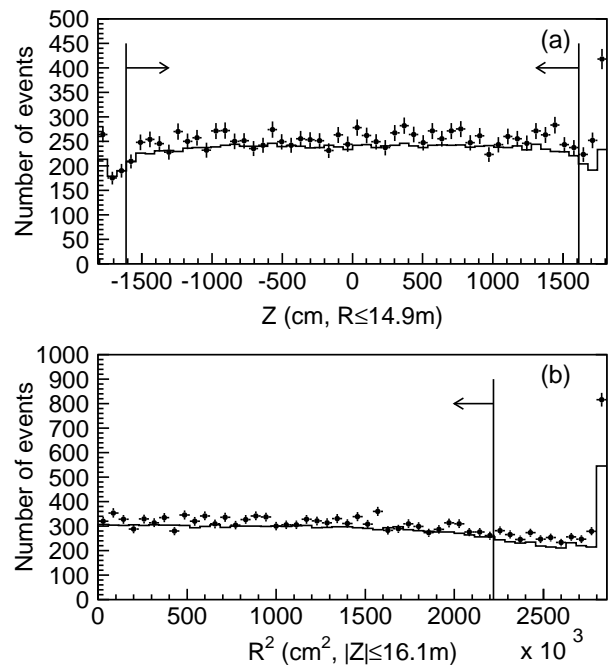}
  \caption{The distribution of fully-contained event vertices
    in the (a) $z$-coordinate and (b) $r^2$-coordinate,
    comparing SK data and atmospheric neutrino
    Monte Carlo. The points show the data and the histogram shows
    the Monte Carlo prediction. The Monte Carlo includes neutrino oscillation
    with $(\sin^22\theta = 1.00,$ $\Delta m^2 = 2.1\times10^{-3} $~eV$^2$).}
   \label{fig:fc-vtx}
\end{figure}

The fully-contained events underwent a series of reconstruction steps
in order to classify their origin and properties. First, the vertex
position of an event was determined using PMT hit times; the point
which best fit the distribution of PMT times (when adjusted for the
time of flight of the Cherenkov light) was defined as the vertex
position.  This vertex was reconstructed again after particle
identification was established, to correct for particle track length.
The vertex resolution was estimated to be 30~cm for single-ring
fully-contained events.  The distribution of vertex position for both
data and MC as a function of the $z$-coordinate and $r^2$-coordinate
are shown in Figures~\ref{fig:fc-vtx} (a) and (b),
respectively. In these and several further figures, the original 
Monte Carlo prediction is modified by the oscillation of CC $\nu_\mu$
interactions according to the best-fit parameters 
$(\sin^2 2\theta = 1.0, \Delta m^2 = 2.1 \times 10^{-3} {\rm eV^2})$,
as found in Section VI. Only a simple survival probability
suppression is applied for these comparisons, the adjusted systematic
terms that will be described in Section~\ref{sec:oscillation}.

\begin{figure}[htb]
  \includegraphics[width=2.9in]{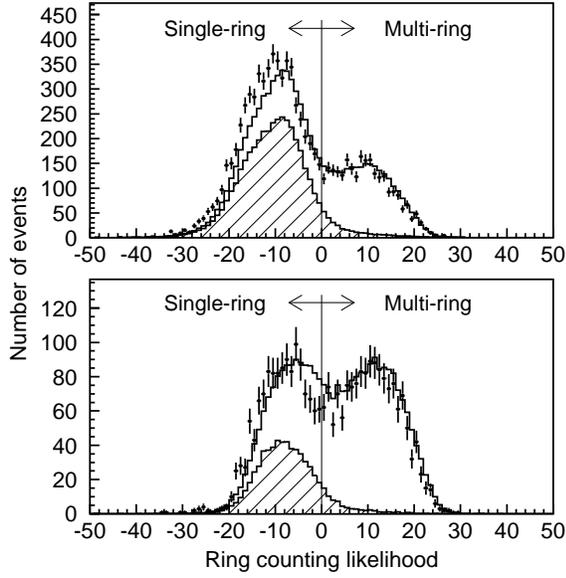}
  \caption{The distribution of the likelihood difference between
   a single-ring assumption and a multi-ring assumption for sub-GeV (top)
   and multi-GeV (bottom) FC events. The points show the data and the 
   histograms show the Monte Carlo prediction. 
    The Monte Carlo includes neutrino oscillation with
  $(\sin^22\theta = 1.00,$ $\Delta m^2 = 2.1\times10^{-3} $~eV$^2$).
 The hatched histograms show the charged current quasi-elastic interactions. }
   \label{fig:ring-counting-likelihood}
\end{figure}

\begin{figure}[htb]
  \includegraphics[width=2.8in]{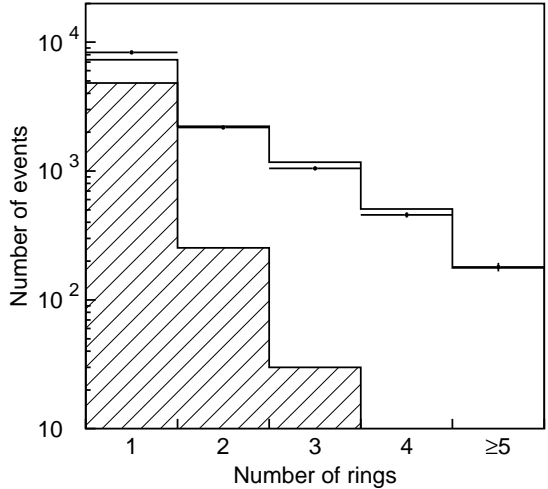}
  \caption{The distribution of the number of identified Cherenkov
    rings, comparing SK data and atmospheric neutrino Monte Carlo.
   The Monte Carlo includes neutrino oscillation with
  $(\sin^22\theta = 1.00,$ $\Delta m^2 = 2.1\times10^{-3} $~eV$^2$).
The hatched histogram shows the charged current quasi-elastic interactions.}
   \label{fig:fc-nring}
\end{figure}

After an initial ring direction and vertex were found by use of the timing
method, a Hough transform~\cite{Davies:machine-vision} based technique was
applied to automatically determine the number of Cherenkov rings in an event
and their directions.  The technique was iterative.  A second ring was
searched for by choosing possible ring directions based on the Hough map, and
a likelihood technique was used to determine if a second ring from this list
of possible rings was more consistent with the data than just one ring.  If a
second ring was found to be necessary, then this procedure was repeated as
often as needed (to a maximum of 5 found rings), each time fixing the
previously found rings, until finally no further rings were necessary to fit
the data.  Figure~\ref{fig:ring-counting-likelihood} shows the likelihood
difference between the 2-ring assumption and a 1-ring assumption.  A cut was
made at likelihood difference of 0 to separate single and multi-ring events.
The likelihood distributions, especially the one for multi-GeV energy region,
have a slight difference in the peak positions between the data and the Monte
Carlo. This difference is taken as a source of the systematic error in the
measurements of the $\nu_{\mu}$ and $\nu_e$ rates.  More details will be
discussed later.  Figure~\ref{fig:fc-nring} shows the distribution of the
number of reconstructed Cherenkov rings for both the data and MC.

The efficiency for identifying charged current (CC) quasi-elastic
$\nu_e$($\nu_\mu$) events as single-ring was 93.2 (95.8)\,\%, and the angular
resolution for these single-ring events was estimated to be
3.0$^{\circ}$ and 1.8$^{\circ}$ for single-ring $e$-like and $\mu$-like
events, respectively. 

\begin{figure}[htb]
  \includegraphics[width=3.2in]{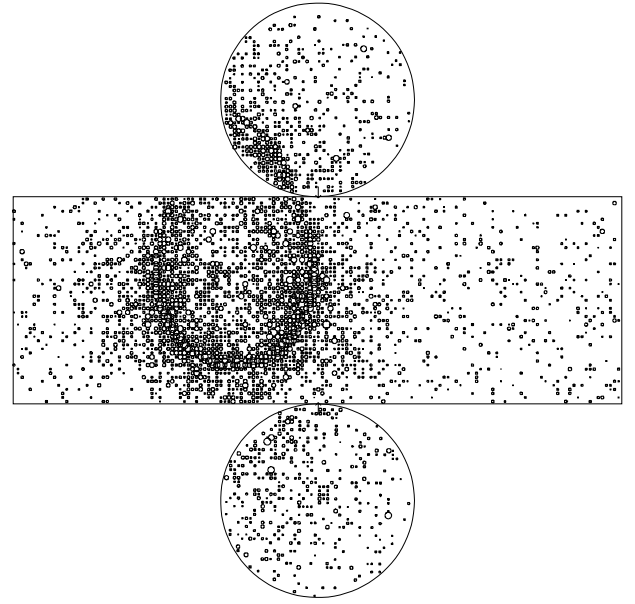}
  \caption{An example event display of a single-ring $e$-like event. 
    Each small circle represents a hit PMT and the size of the circle
    represents the number of photons to hit it. In this event the boundary of
    the Cherenkov light is smeared over many PMTs as the light comes from
    numerous positrons and electrons in the electromagnetic
    shower.}
   \label{fig:e-like}
\end{figure}

\begin{figure}[htb]
  \includegraphics[width=3.2in]{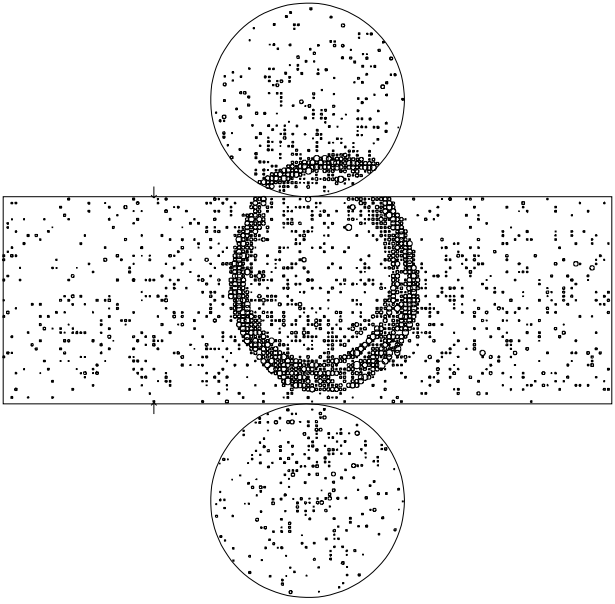}
  \caption{An example event display of a single-ring $\mu$-like event.
    In this event the boundary of the Cherenkov light is sharp as the
    muon travels relatively straight as it comes to a stop.
    Distant hit PMTs come from scattered light and Cherenkov light from
    delta-rays.}
   \label{fig:mu-like}
\end{figure}

\begin{figure}[htb]
  \includegraphics[width=3.in]{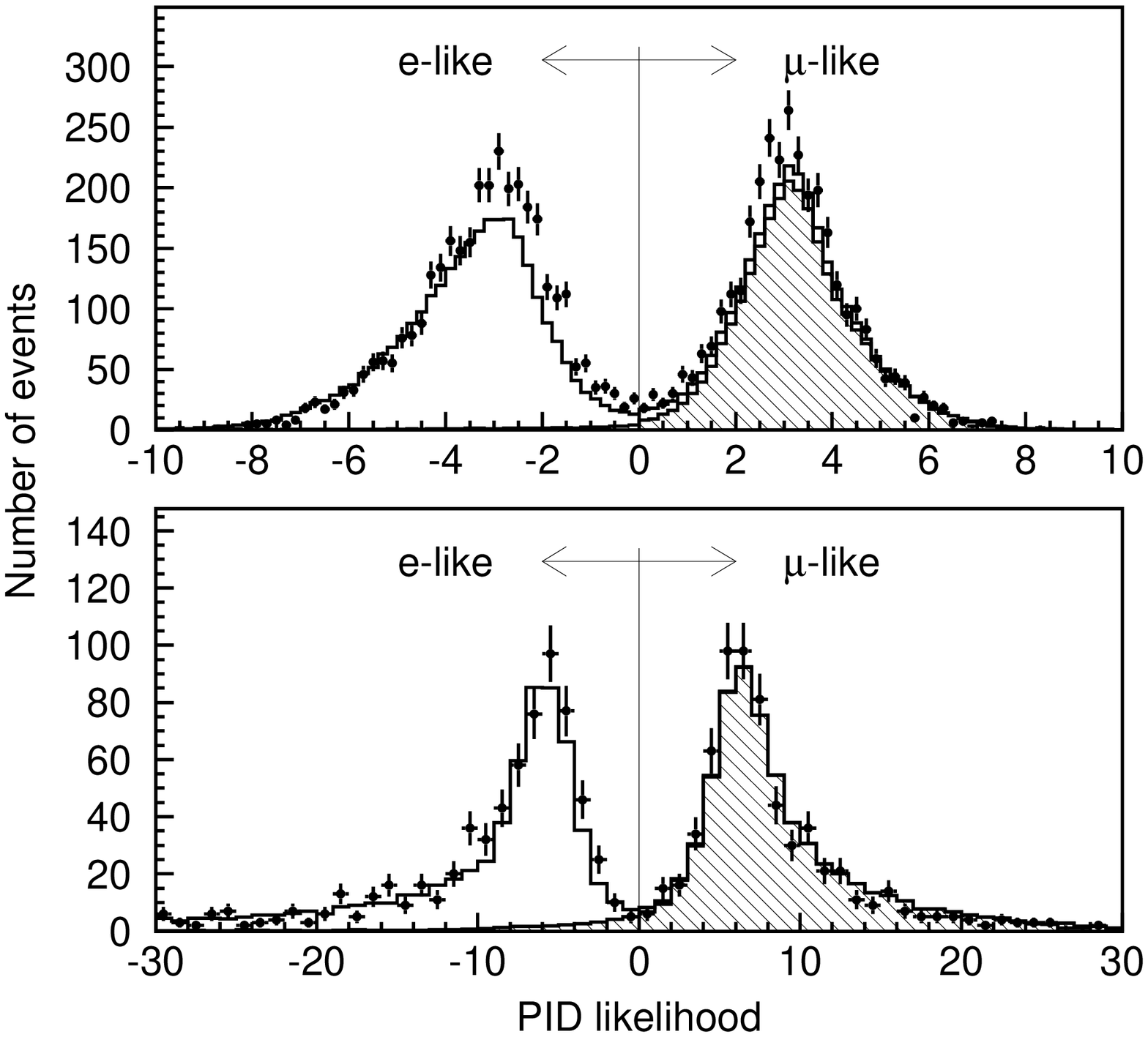}
  \caption{The distribution of particle identification likelihood
    for sub-GeV (top) and multi-GeV (bottom) FC single-ring events, 
 comparing SK data (points) and atmospheric neutrino Monte Carlo (histograms).
 The Monte Carlo includes neutrino oscillation with
 $(\sin^22\theta = 1.00,$ $\Delta m^2 = 2.1\times10^{-3} $~eV$^2$).
 The hatched histograms show the $\nu_{\mu}$ charged current interactions.}
   \label{fig:pid-single-ring}
\end{figure}

To determine the identity of the final state particles, a particle
identification algorithm was applied which exploited systematic
differences in the shape and the opening angle of Cherenkov rings
produced by electrons and muons.  Cherenkov rings from electromagnetic
cascades exhibit a more diffuse light distribution than those from
muons. Figures~\ref{fig:e-like} and \ref{fig:mu-like} show observed
single-ring $e$-like and $\mu$-like events, respectively.  The opening
angle of the Cherenkov cone, which depends on $\beta(\equiv v/c)$, was
also used to separate electrons and muons at low momenta.  The
validity of the method was confirmed by a beam test experiment at
KEK~\cite{Kasuga:1996dm}.  The misidentification probabilities for
single-ring muons and electrons were estimated to be 0.7\,\% and
0.8\,\% respectively, using simulated CC quasi-elastic neutrino
events. The distribution of the likelihood variable used to
discriminate single-ring electrons and muons are shown for both the
data and MC for the sub-GeV and multi-GeV samples in
Figure~\ref{fig:pid-single-ring}. In both of these cases there is a
clear separation of the likelihood variable. 

\begin{figure}[htb]
  \includegraphics[width=3.in]{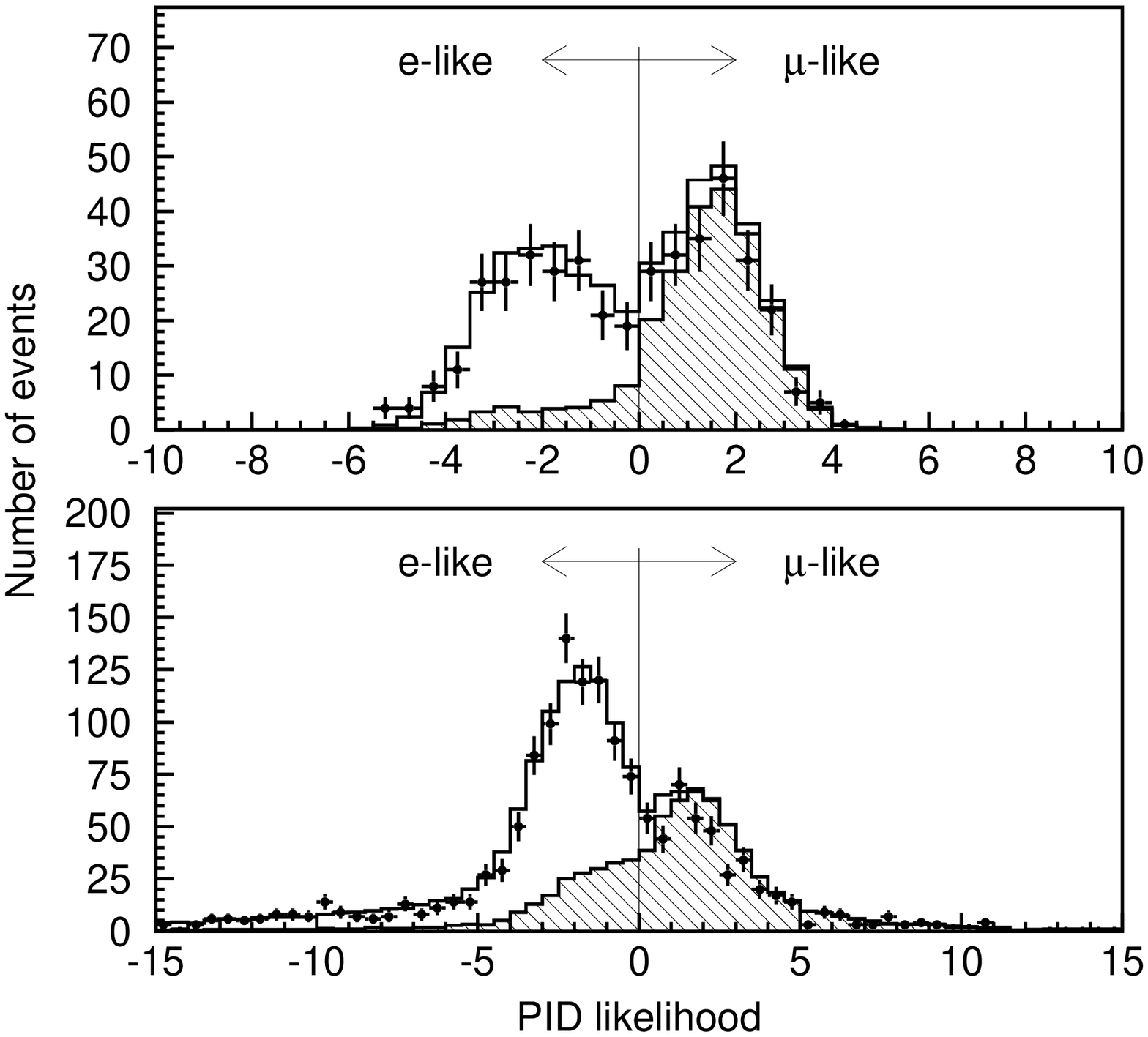}
  \caption{Particle ID likelihood distribution for sub-GeV (top) and multi-GeV (bottom) FC multi-ring events 
and the brightest ring of the multi-ring events. Points show the data and
the histograms show the Monte Carlo results. The Monte Carlo includes
neutrino oscillation with
 $(\sin^22\theta = 1.00,$ $\Delta m^2 = 2.1\times10^{-3} $~eV$^2$). 
The hatched histograms show the $\nu_{\mu}$ charged current interactions. }
   \label{fig:pid_fc-multi-ring}
\end{figure}

Figure~\ref{fig:pid_fc-multi-ring} shows the likelihood variable
distribution for the brightest ring of FC multi-ring events. Due
mostly to overlapping of Cherenkov photons from multiple particles,
the separation of the particle type for a Cherenkov ring in a
multi-ring event is not as good as that for a single-ring event.

The identification efficiency was checked using cosmic ray muons that
stop in the detector and subsequently decay to electrons.  These
events are easily selected by their timing signature.  The resulting
misidentification probabilities for stopping cosmic ray muons and
decay electron light patterns were $0.4\pm0.1\,\%$ and $1.8\pm0.5\,\%$
respectively, in good agreement with the Monte Carlo estimates. This
check was performed continuously during data-taking, and particle
identification performance remained stable despite water transparency
that varied from about 90~m to 120~m.

Next, the Cherenkov rings were re-fit taking into account the expected light
pattern given by the particle identification, and in the case of single-ring
events, a specialized event fitter was applied.  After the rings were re-fit
and the total photo-electrons in the event were apportioned between all of
the rings, each ring was assigned a momentum based on proportion of Cherenkov
photons. The momentum of a particle was determined from the total number of
p.e.'s within a 70$^{\circ}$ half-angle cone relative to the track direction,
with corrections for light attenuation and PMT angular acceptance.  The
resulting momentum resolution is estimated to be $0.6 +
2.6/\sqrt{\rm{P(GeV/c)}} \,\%$ for single-ring electrons and $1.7 +
0.7/\sqrt{\rm{P(GeV/c)}} \,\%$ for single-ring muons. A final procedure was
performed which utilized the final energy and angle information of the rings
to remove rings which were most likely not real.

Although decay electrons are not used in this oscillation analysis, they are
a useful signature in other atmospheric neutrino analyses and the search for
proton decay; therefore, we document their treatment here. Decay electrons
were identified either as: (a) PMT hits within the same time window as the
primary event trigger (up to 900 ns later) or (b) a later independent event
trigger. In the first case, a sliding search window of width 30 ns began 100
ns after the primary trigger; a decay electron was counted if 40 hits were
found in coincidence above the background level. In the second type, 60 hits
were required in a 50 ns time window, and goodness-of-fit for a Cherenkov
ring pattern is required. In both cases the vertex is known from the primary
event and is used to subtract the time-of-flight of the Cherenkov light. If
the decay occurs around 900 ns, the hits may be split between the primary
event trigger and a subsequent event trigger. In some analyses, electrons in
the time interval 800 ns to 1200 ns after the primary trigger are excluded,
owing to this splitting effect as well as a reduced efficiency due to
electrical reflection on the PMT cables. The contamination level for these
criteria is very good, with no events out of 32000 stopping cosmic ray muons
having more than one decay electron. The efficiency for fully-contained
sub-GeV neutrino interactions was estimated by Monte Carlo to be 80\% for
$\mu^+$ and 63\% for $\mu^-$, where the lower efficiency is due to $\mu^-$
capture on $^{16}$O.

\subsubsection{Background and Efficiency}

The main sources of the background 
for the FC sample are cosmic ray muons, neutrons
generated by high energy cosmic ray muons and PMT flasher events.  
The contamination of the background events was 
estimated for lower energy (visible energy 
lower than 1.33~GeV, called $sub$-GeV) and higher energy
(visible energy higher than 1.33~GeV, called $multi$-GeV) samples
separately, since the contamination could have an energy dependence.  
The cosmic ray muon background contaminations to the final FC sample were
estimated to be 0.07\,\% for sub-GeV and 0.09\,\% for multi-GeV $\mu$-like
events. They were estimated using the distribution of the distance 
of the vertex position from the inner detector wall along the particle 
direction.
The neutron background contamination was estimated to be 0.1\,\% for sub-GeV
$e$-like events and multi-GeV $e$-like events.
This was estimated using the distribution of the distance 
of the vertex position from the inner detector wall ($D_{wall}$).
The contaminations from PMT flasher events were estimated to be 0.42\,\%
for sub-GeV $e$-like events and 0.16\,\% for multi-GeV $e$-like events
using the goodness of the vertex fitting and the
$D_{wall}$ distributions.

The reduction efficiency was estimated using the atmospheric neutrino MC.
The detection efficiency for events which satisfy the data reduction
conditions (i)-(iv) was 99.15\,\% for events which have a true vertex in the
fiducial volume, $E_{vis}>30$~MeV, and less than 10 hits in the largest OD
cluster. The systematic error in the event reduction was estimated to be
0.2\,\%. The main source of the systematic error in the event reduction was
the flasher cut based on the pattern matching algorithm.  The systematic
error for this cut was estimated by mixing the different flasher samples in
the atmospheric MC and comparing the reduction efficiencies.  The
inefficiency was estimated to be 0.7\,\% $\pm$ 0.2\,\%, in which 0.2\,\% was
considered as the systematic uncertainty.  The systematic errors for other
reduction steps were negligibly small because the reduction efficiency was
almost 100\,\% and the distributions of cut parameters for data agree with
that of MC.  In the early stage of the experiment, an independent data
selection and reconstruction program was also employed; the results were
compared and event samples and classifications were found to agree by around
95\,\%\cite{Fukuda:1998tw}.


\subsection{Partially-Contained Events} 

\subsubsection{Data Reduction}
\label{sec:pcdatareduction}

The data reduction for PC events differed from the reduction for FC
events because of the presence of additional hits in the OD. Because
these extra hits result from the exiting particle (usually a muon), a
simple criterion based on the number of hit OD tubes could not be used
to reject cosmic ray background. The criteria used to identify
partially-contained events are as follows:

(i) Low energy events with fewer than 1000 total p.e.'s in the inner 
detector were removed,
corresponding to muons (electrons) with momentum less than 310 (110)
MeV/$c$. By definition, an exiting particle in the PC sample
must have reached the
OD from the inner fiducial volume, and so must have had a
minimum track length of about 2.5~m (corresponding to muons with
$\geq 700$~MeV/$c$ momentum).

(ii) Events for which the width of the time distribution of hits in the
OD exceeded 260~nsec were rejected, as well as events with two or more
spatial clusters of OD hits.  These cuts eliminated many through-going
muons, which typically left two well separated clusters in the OD.
Muons which clipped the edges of the detector were eliminated based upon
the topology of the OD cluster.  Cosmic ray muons which entered and
stopped in the inner volume of the detector were eliminated by excluding
events with a relatively small number of ID photoelectrons near the OD
cluster (1000~p.e.'s within 2~m).  This cut did not remove PC neutrino
events because PC events produced large numbers of photoelectrons
(typically 3500~p.e.'s) in the region where the particle exited.

(iii) In the next step, a simple vertex fit and p.e. weighted
direction estimation were used.  A requirement of $\leq 10$ hits in the OD
within 8~m of the back-projected entrance point was imposed. Also in this step,
flasher events were removed by using their broader timing distribution
feature.

(iv) The remaining background still had muons which left 
few or no entrance hits in the OD. These events were rejected by requiring the
angle subtended by the earliest inner detector PMT hit, the vertex, and
the back-projected entrance point be $>37^\circ$.  Remaining corner
clipping muons were rejected by requiring a fitted vertex at least 1.5~m
away from the corners of the ID volume.  A through-going muon fitter was
also applied to reject events with a well fitted muon track greater than
30~m long.

(v) In the last reduction step, various remaining background events
were eliminated by several selection criteria:
(1) Fully-contained events were eliminated by requiring that PC events 
have more than 9 hits in the most highly charged cluster in OD;
(2) A minimum requirement of 3000 total p.e.'s in the inner detector, which
corresponds to 350~MeV of visible energy, well below that of any
exiting muon, was applied to get rid of low energy background events;
(3) Clusters in the OD were searched for again with the same clustering 
algorithm used in the 2nd reduction step but with different 
clustering parameters. 
Events were eliminated if there existed two or more clusters 
with more than 10~p.e.'s and they were apart by more than 20~meters.
Some obvious through-going muons were removed by this cut;
(4) After those steps, most remaining
background events are due to the imprecision of the fast fitters used to
quickly filter the data stream.
A precise fitting algorithm was then applied to obtain more
accurate information on ring direction and vertex position.
With much more accurate information of the event, we were able to
eliminate most remaining through-going and stopping events based on
their distinct geometry and OD signatures.
(5) Some through-going muons have a very special geometry---they 
passed through the tank vertically along the wall of ID. These events 
were eliminated by counting the number of p.e.'s and hits in the OD within 
the region defined by an 8~m
sphere around the top and bottom fringes and checking the time 
interval between the average timings of those top and bottom hits.
(6) Remaining cosmic ray muon background events are those entering the
ID through relatively weak OD regions---there are four holes covered
by veto counters on the top of OD through which cables run. Events
with a veto counter hit were eliminated, as well as those satisfying
a detailed cosmic ray muon consistency requirement.

(vi) After this final reduction step, events were scanned by physicists
to check the data quality. However, no event was rejected based on the 
scanning. Finally, the vertex was required to be within a
fiducial volume, 2 meters from the wall of the inner detector.
 The final event sample is an almost 100\,\% pure $\nu$ sample.  
The background contamination has been estimated to be about 0.2\,\%.

 Table
~\ref{tab:pcdetecteff} shows the number of events after each reduction
step and the detection efficiency of PC events as a
function of reduction steps.

\begin{table}[htbp]
\begin{center}
  \begin{tabular}{l|c|c}
\hline\hline
Reduction step & Data & Monte Carlo\\
\hline\hline
Trigger & 1889599293 & 1,417.0~(100.0\,\%) \\
First reduction & 34536269 & 1,402.8~(99.0\,\%) \\
Second reduction & 5257443 & 1,334.7~(94.2\,\%) \\
Third reduction & 380053 & 1,318.7~(93.1\,\%) \\
Fourth reduction & 53825 & 1,246.2~(87.9\,\%) \\
Fifth reduction & 1483 & 1,201.0~(84.8\,\%) \\
Fiducial volume & 911 & 1,129.6~(79.7\,\%) \\

\hline\hline
  \end{tabular}
  \caption{Number of events after each reduction step for 
  partially-contained events during 1489~days of the detector
   live-time. 
   The Monte Carlo efficiencies are for events whose real vertices
   are in the fiducial volume and the number of outer detector hits more
   than 9. In the last line, we used the events whose
   fitted vertices are
   inside the fiducial volume both for data and Monte Carlo. The Monte Carlo
   does not include neutrino oscillation.}
  \label{tab:pcdetecteff}
\end{center}
\end{table}

\begin{figure}[htb]
  \includegraphics[width=3.1in]{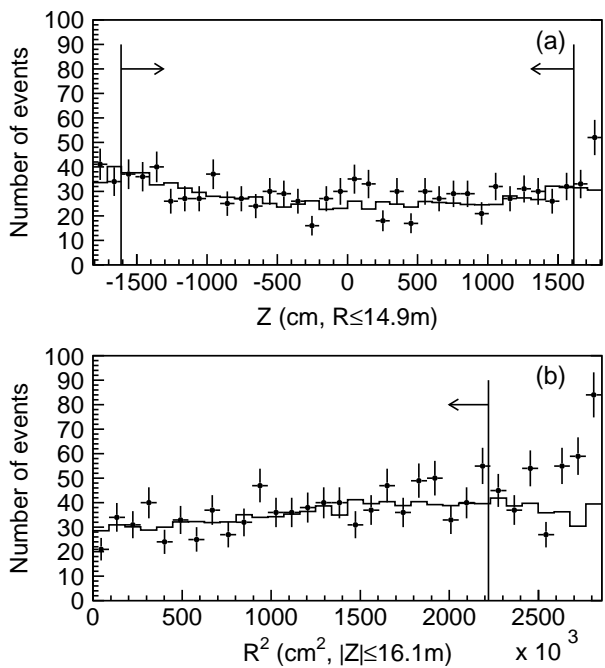}
  \caption{The distribution of partially-contained event vertices
     in the (a) $z$-coordinate and (b) $r^2$-coordinate,
     comparing SK data and atmospheric neutrino
       Monte Carlo. The Monte Carlo includes neutrino oscillation
    with $(\sin^22\theta = 1.00,$ $\Delta m^2 = 2.1\times10^{-3} $~eV$^2$).}
       \label{fig:pc-vtx}
\end{figure}

\subsubsection{Event Reconstruction}

The partially-contained events were reconstructed using inner detector PMT
information by similar vertex, direction fit and ring-counting algorithms, as
were applied to fully contained events.

For some PC events, however, the direction fit was slightly modified.  
Under some conditions, rather than using the results of the precise 
fitting algorithm for the PC event direction, outer detector spatial 
information was used instead.
In order to use the outer detector cluster for the direction, the
number of tubes in the largest OD cluster was required to be greater or equal
to 20.  If this condition was satisfied, 
and also, if the ID PMT nearest to the
projected ID exit point of the fitted track with more than 200 p.e.'s was
more than 200 cm away (i.e. no clear ID
exit point), $or$ if there was a clear exit point in the ID but yet
there were more than 800 ID PMT hits with more than 200 p.e.'s in each of
them (i.e. saturated in our electronics), then the vector from the fitted 
vertex to the largest OD cluster was used for the PC event direction.  
Otherwise, the standard direction provided by the precise fitting 
algorithm was used. 

The estimated vertex position resolution for PC events was 64~cm. The
angular resolution for the penetrating particle in a PC event was
estimated to be 2.8$^{\circ}$.

Finally, the fiducial volume cut was applied. The event rate in the
fiducial volume was 0.62~events/day. 

\subsubsection{Background and Efficiency}

The background for the PC sample originates from cosmic ray muons.
They were efficiently removed by the reduction steps mentioned above.
We estimated the contamination of non-neutrino events in the fiducial
volume by two methods. One method utilized scanned results.
After applying all the reduction steps, all the events were
scanned and the estimated fraction of background contamination
was found to be 0.2\,\% in the fiducial volume.
Another method was by examining the vertex distribution
of non-neutrino events as a function of distance from the wall.
By extrapolating the distribution from outside the fiducial volume,
we obtained that the contamination of the background was less than 0.1\,\%.
Since these two results were statistically consistent, we took the larger
number (0.2\,\%) as the contamination of background events
in the fiducial volume.
Figures~\ref{fig:pc-vtx} (a) and (b)
show the distribution of the vertex position for both data and MC
as a function of the $z$-coordinate and $r^2$-coordinate.
Some contamination of background is evident near the side and 
top PMT walls. However, no evidence for substantial background
contamination is seen in the fiducial volume.

The PC reduction efficiency was estimated based on Monte Carlo events.
The definition of partially-contained events is that (1) the
interaction point of the parent neutrino is inside the fiducial
volume, and (2) number of outer detector 
hits within 8~m around an estimated exiting point is larger than nine.
We applied those five reduction steps to the atmospheric neutrino Monte Carlo 
sample and after each reduction step, we counted the number of PC events 
left inside fiducial volume and calculated the efficiency of this particular 
reduction step. We found the overall the efficiency of the reduction for 
PC neutrino events was 79.7\,\%. 

The systematic uncertainties on the reduction efficiency from the
first to the fourth steps were estimated by two methods.
For the ID contribution, they were estimated
by comparing the distributions of cut parameters used 
in the reduction criteria for the data and Monte Carlo.
For the OD part, we created Monte Carlo samples with two different sets of 
OD-related tuning parameters. Then, the change in reduction efficiency is the 
OD-related systematic error. 
The two sets produced different amounts of light in the OD
within the limits 
of good overall agreement with the standard tuning sample (several
hundreds of well-measured stopping muons).
The estimated uncertainties were 1.5\,\% and 1.4\,\% 
for inner- and outer-detector related selection criteria, respectively.

The systematic uncertainty in the fifth reduction efficiency is mainly 
from ID variables and involves the precise fitting algorithm.
The main contributions come from the cluster cut mentioned above.
The uncertainty on the cluster cut was estimated
by examining the distributions of cut variables. The uncertainties
in other cuts were assumed to be the inefficiency in each cut,
since the inefficiency itself was small compared with other errors.
Combining these estimated uncertainties, we obtained the value of 1.6\,\%
for the fifth reduction step.

Thus the overall systematic uncertainty in the reduction of the PC events
was estimated to be 2.6\,\%. 

\subsection{Upward-going muon Data}

\subsubsection{Data Reduction}
\label{sec:upmureduction}

The upward-going muons observed in Super-K are
classified into two categories: (1)~upward stopping muon events having
only an entrance signal in the OD; and (2)~upward through-going muon
events having both entrance and exit signals in the OD.  The 
criteria used in Sec.~\ref{sec:pcdatareduction} to determine an event
entry or exit were used: 10~OD hits in-time and within 8~m of the
tracks projected entry or exit point constitutes a muon entry or exit
signal.  We required the geometrical trajectory of through-going muons
to be greater than 7~m in the inner detector, and
we imposed an equivalent 7~m path-length cut on upward stopping
muons based on the momentum reconstruction using Cherenkov light.

The purpose of the data reduction is to isolate the upward muon events
and the horizontal muon events (needed for background estimation),
provide a classification of stopping or through-going muon type, 
and to reject the background
from cosmic ray muons and noise such as flashing PMTs. Decay
electrons associated with stopping muons were also
saved. A charge cut of $8,000 \leq Q < 1,750,000$~p.e.'s in the
ID was applied.  For a muon, $Q (p.e.) \simeq 25 \cdot L (cm) $, ensuring that
that we can detect all muons with path-length $\geq $ 7~m while
eliminating events at lower energies. At very high ID charge
corresponding to $\simeq 1,750,000$~p.e.'s the ID electronics becomes 
saturated causing the muon fitters fail. 

 To isolate the rate of about one neutrino induced
upward-going muon per day from the remaining background of about $2
\times 10^5$ cosmic ray muons, we used a logic tree involving as many as
seven different muon fitters. Some of these fitters were specialized to
fit stopping muons, others were specialized for through-going muon events,
and some of them were specialized to fit muon events with Bremsstrahlung.
The main logic behind the upward-going muon reduction was that if a muon
fitter classifies an event as upward with a goodness of fit which was
above the fitter's goodness threshold then the event was automatically
saved. Conversely, if a muon fitter classified an event as downward with
a goodness of fit which was above the threshold then the event was
automatically rejected. If a muon fitter classified an event as
horizontal and with goodness above threshold, or if the fitter could not
give a good fit for the event, the event was passed to the next fitter.
This sequence continued until all the events had passed through all the
fitters or had been classified.  If no fitter was able to give a good fit then
this event was automatically rejected.  If at least one fitter classified
this event as horizontal then the event was saved.  
All events from the output of the upward muon reduction were then
passed to the precise fitter which is described in
Sec.~\ref{sec:precisionfitter}.

\subsubsection{Event Reconstruction}
\label{sec:precisionfitter}

All events from the output of the upward-going muon reduction were
passed through the precise fitter. The basic algorithm was identical
to that used for the vertex and direction determination for single
ring fully-contained and partially-contained events.  The fitter
assumes that the particle is a muon and the vertex position of the
event is at the inner detector surface. However, when the muon
produces an energetic electro-magnetic shower, the assumption of
single non-showering muon does not give an accurate direction. For
these events, the information of OD hit is used to determine the
particle direction. 
The angular resolution of the fitter was about 1.0$^\circ$ for 
both through-going and stopping muons.  Taking into account 
multiple scattering from the point of muon creation, 
68\,\% of through-going and stopping muons fit within $1.3^\circ$ 
and  2.4$^\circ$ of the 
muon's true initial direction, respectively.
The direction determined by this fitter was used in the neutrino
oscillation analysis.

\subsubsection{Background and Efficiency}

The effective detection efficiency for the data reduction process was
estimated by a Monte Carlo simulation, and was found to be 102.4\,\% for
upward stopping muons and 95.9\,\% for upward through-going muons. The
efficiency higher than 100\,\% for upward stopping muons is due to a slight
bias in the separation of stopping and through-going muons, causing a small
fraction of the more numerous through-going muons to be misidentified as
stopping muons.  The up/down symmetry of the detector geometry allows a check
of this Monte Carlo efficiency calculation using real cosmic ray induced
downward-going muons. The efficiency is approximately constant for $ -1 <\cos
\Theta< 0 $; bin-by-bin efficiencies are listed in Table~\ref{tab:upmubins}
in the Appendix.

After the reduction, the precise fitter described in
Sec.~\ref{sec:precisionfitter} was applied to determine the entry position and
the muon direction. This is the final direction used for further physics
analysis, including the determination of upward versus downward
classification. Events selected as upward by this direction were then scanned
by eye with a graphical event display program in order to reject
difficult-to-remove instances of corner clipping or bremsstrahlung cosmic ray
muons and noise events.  The event scanning only rejected events judged to be
background and did not change the direction and the vertex decided by the
precise fitter, nor the stop/through judgment made by the reduction programs
using the entrance and exit points of this fit.  The event scanning was done
independently by two physicists and testing had shown that both scanners had
never rejected the same good upward-going muon event. About 50\,\% of the
events remaining after all automated reduction steps were rejected by this
final scan.  Table~\ref{tab:upmudetecteff} summarizes the data reduction for
upward muons.

\begin{table*}[htbp]
\begin{center}
  \begin{tabular}{l|cc|cc}
\hline\hline
Reduction step & \multicolumn{2}{c|}{Data} & \multicolumn{2}{c}{Monte Carlo} \\
               & stopping  & through-going   & stopping  &  through-going \\
\hline\hline
Trigger & \multicolumn{2}{c|}{ 2129729843 } & 697.1~(100\,\%) & 1741.0~(100\,\%) \\
Reduction & \multicolumn{2}{c|}{ 89911 } & 693.9~(99.5\,\%) & 1722.3~(98.9\,\%) \\
Precise fitter ($\cos \theta \leq 0$) & \multicolumn{2}{c|}{ 4266 } & 692.4~(99.3\,\%) 
& 1721.7~(98.9\,\%) \\
Scan   & \multicolumn{2}{c|}{ 2447 } & - & - \\
Stop-through separation and $E_{\mu} \geq 1.6$GeV & 458  & 1856  & 713.5~(102.4\,\%) & 1669.5~(95.9\,\%) \\
CR BG subtraction~(subtracted ev) & 417.7~(40.3)  & 1841.6~(14.4)  & - & - \\
\hline\hline
  \end{tabular}
  \caption{Number of events after each step of the data selection for
   upward muons during 1646 days of the detector live-time.
   The Monte Carlo efficiencies are relative to the generated events
  with track length longer than 7~m (for through-going muons) or
  with energy higher than 1.6~GeV/$c$ (for stopping muons). 
  The efficiencies are for Monte Carlo upward muon events
   with track length in the inner detector longer than 7~m 
   (for through-going muons) 
   or with muon momentum at the wall of the inner 
   detector higher than 1.6~GeV/$c$ (for stopping muons).
   The Monte Carlo does not include neutrino oscillation.}
  \label{tab:upmudetecteff}
\end{center}
\end{table*}

\begin{figure}[htb]
  \includegraphics[width=3.3in]{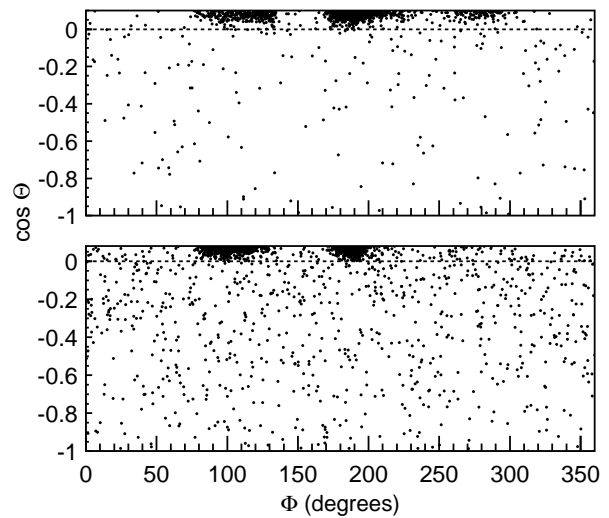}
  \caption{The zenith versus azimuth directions for a 641.4 day 
    sample of upward muons used to estimate background contamination. The
    dense regions of events near the horizon correspond to thin regions of
    the mountainous overburden. The upper panel is for upward stopping
    muons and the lower panel is for upward through-going muons.}
   \label{fig:upmu-zen-phi}
\end{figure}

\begin{figure}[htb]
\includegraphics[width=3.3in]{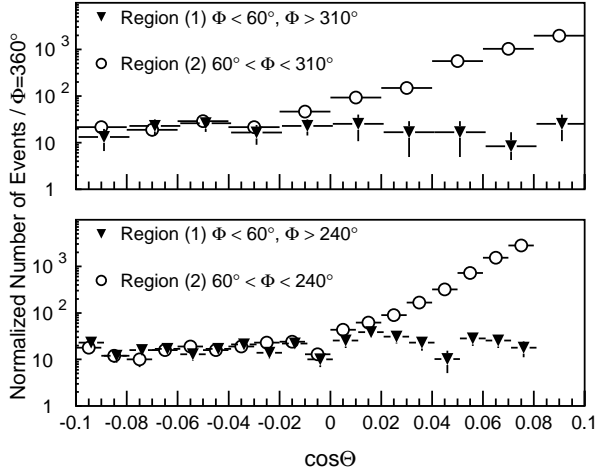}
  \caption{The zenith angle distribution of upward muon candidates
    near the horizon for two different regions in azimuth. Regions (1)
    and (2) are thin and thick overburden respectively. The live-time of upward
    and downward events is 1645.9~days and 641.4~days respectively. The
    number of downward events is normalized to be 1645.9~days live-time.
    The upper panel is for upward stopping muons and the lower panel is
    for upward through-going muons.}
   \label{fig:upmu-bg}
\end{figure}

Near the horizon, horizontal cosmic ray muons are a non-negligible source of
background for both through-going and stopping upward muons.  Because of
finite fitter resolution and multiple Coulomb scattering of muons in the
nearby rock, some downward going cosmic ray muons may appear to be coming
from $\cos\Theta< 0$.  Figure~\ref{fig:upmu-zen-phi} shows the zenith versus
azimuth directions for the upward-going muon sample. Clusters of cosmic ray
downward muons are seen for relatively thin overburden directions.
Figure~\ref{fig:upmu-bg} shows the zenith angle distribution of upward muon
candidates near the horizon for two different regions in azimuth. The thick
overburden region has negligible downward going cosmic ray muon
contamination, even above the horizon. The thin overburden region has
non-negligible contamination.  The shape of the distribution above the
horizon was extrapolated below the horizon to estimate the background
contamination in the upward muon sample.  The number of background events,
based on Fig.~\ref{fig:upmu-bg}, to the upward stopping muon signal were
estimated to be $14.4 \pm7.2 (stat) \pm6.0(sys)$ and $40.3 \pm 13.7(stat)
\pm 4.3(sys)$ events for the through-going and stopping muon samples,
respectively.  Horizontal muon background was contained in the $-0.1
<\cos\Theta< 0$ zenith angle bin, and was subtracted from this bin.  The
stopping muon contamination is larger than the through-going contamination
since lower muon energies allow larger scattering angles.

Away from the horizon, a potential source of upward particles is the
photo-production of pions by energetic muons that pass nearby the
detector~\cite{Ambrosio:1998qh}. The pathlength requirement of 7~m within the
inner detector limits the background from this source to $\simeq 0.01\,\%$ in
the upward through-going muon data sample and $\simeq 0.3\,\%$ in the upward
stopping muon data sample.


\subsubsection{Expected Upward-Going Muon Signal}

The expected upward-going muon flux was calculated using the same tools
as for the contained vertex events, extended to higher energies and to
outside the detector volume.  The input neutrino flux (see
Sec.~\ref{sec:atmnu}) in Ref.~\cite{Honda:2004yz} was used up to
1~TeV. At 1~TeV, the calculated flux in Ref.~\cite{Volkova:1980sw}
was rescaled to that in Ref.~\cite{Honda:2004yz}. Above 1~TeV, the
rescaled flux in \cite{Volkova:1980sw} was used up to 100~TeV.
The target volume for these
neutrinos is primarily the rock around the detector, parameterized as
standard rock, with Z=11, A=22 and density =
2.7~gm/cm$^3$.  However, neutrinos interacting in the water of the OD
and insensitive region can also be seen as upward-going muons (1.8\,\% of
through-going and 6.6\,\% of stopping muons), so water interactions were
also simulated.

The neutrino interactions were modeled as discussed in
Sec.~\ref{sec:atmnumc}.  The same GEANT detector simulation discussed
previously was used to track muons from the interaction vertex through
the rock into the detector itself.  The output of the detector
simulation was passed through the same reduction and fitting routines as
was the real upward-going muon data.  100 years equivalent exposure was
generated.  The results of this Monte Carlo data before and after the
reduction were used to estimate efficiencies and systematic errors as
well as provide an expected upward-going muon signal. 
Table~\ref{tab:upmudetecteff} summarizes the detection efficiency
at each step of the data reduction.


\subsection{Observations}

\subsubsection{Contained Vertex Events}
We have accumulated 1489.2 days of FC and PC data from May 17, 1996 to
July 16, 2001. These are compared with statistically larger
samples of Monte Carlo events 
based on the two neutrino interaction 
models~\cite{Hayato:2002sd, Casper:2002sd}, 
both equivalent to 100 years of the detector exposure.
However, upward-going Monte Carlo muons were 
only generated based on {\tt NEUT}. Both Monte Carlo samples were
generated based on the flux model of Ref.~\cite{Honda:2004yz}.
The Monte Carlo samples were processed by the same event-selection and
event reconstruction steps as the real events. Throughout this paper,
unless otherwise indicated, we used Monte Carlo events generated 
based on the flux model of Ref~\cite{Honda:2004yz} and the 
{\tt NEUT} neutrino interaction model~\cite{Hayato:2002sd}.

Fully-contained events were divided into two sub-samples according to
the reconstructed visible energy.  We refer to the event sample below
1.33 GeV as {\it sub-GeV}, and above 1.33 GeV as {\it
multi-GeV}. Fully-contained events were further divided into the events
with single reconstructed Cherenkov ring, {\it single-ring}, and
events with more than one rings, {\it multi-ring}. All single-ring events
were classified as either $e$-like or $\mu$-like based on the PID result.
Lower energy cuts were applied only to the single-ring sample,
$P_{e}>100$\,MeV/c for $e$-like and $P_{\mu}>200$\,MeV/c for $\mu$-like.
In addition, multi-ring events were used to study the atmospheric neutrino
flux. A simple set of criteria that the most energetic ring in a 
multi-ring event was $\mu$-like with $P_{\mu} >$600~MeV/$c$ and 
$E_{vis} > 600$~MeV selected
relatively pure CC $\nu_{\mu}$ events. The estimated 
fraction of CC $\nu_{\mu}$ 
in this sample was 90.5\,\% and 94.9\,\% for the sub- and multi-GeV energy
ranges, respectively.
On the other hand, a similar set of criteria for $e$-like 
events obtained only a 54.4\,\% pure CC $\nu_e$ sample, and therefore 
we decided to use multi-ring $\mu$-like events only.

\begin{figure}[htb]
  \includegraphics[width=3.2in]{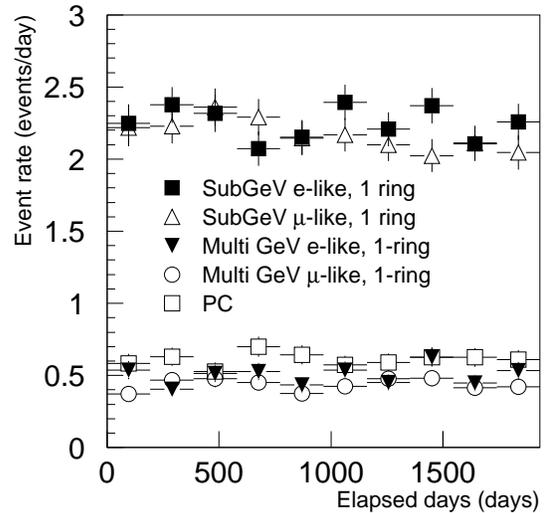}
  \caption{Event rates as a function of elapsed SK running time for
   contained events.}
   \label{fig:rate-stability}
\end{figure}

Table~\ref{tab:datasummary} summarizes the number of observed events in the
sub-GeV and multi-GeV samples as well as the expected number of events in the
absence of neutrino oscillations. The fraction of various neutrino
interaction modes, predicted by the Monte Carlo sample, are also listed.
Figure~\ref{fig:rate-stability} shows the event rates for contained events as
a function of the elapsed days. The event rate should change due to the solar
modulation.  However, the expected decrease in the event rate from minimum
solar activity (which approximately corresponds to the period when the SK-I
started taking data) to maximum solar activity (which approximately
corresponds to the period when the SK-I finished taking data) period is
6-7\,\% for sub-GeV events and 3-4\,\% for multi-GeV events. The data cannot
distinguish a constant event rate from the expected rate change due to solar
activity.

\begin{table*}
 \begin{center}
\begin{tabular}{lrr|rrr|rrr} 
  \hline \hline
  & Data & Monte Carlo &CC $\nu_{e}$ &CC 
$\nu_{\mu}$& NC & \multicolumn{3}{c}{Monte Carlo} \\
  & & ({\tt NEUT}) & & & & ({\tt NEUT}) & ({\tt NEUT}) & ({\tt NUANCE}) \\
  & & (Flux A) & & & & (Flux B) & (Flux C) & (Flux A) \\
  \hline
  {} sub-GeV & 8941 & 9884.3 \ \ \ \ &&&& 9967.8 & 10619.4 & 9074.2 \\ 
  {} single-ring   & 6580 & 7092.6 \ \ \ \ &&&&7273.2&7643.3& 6694.0 \\
  {} {} ~~~~$e$-like   & 3353 & 2879.8 \ \ \ \ &2533.9~(88.0\%)&66.3~(2.3\%)&279.6~(9.7\%) &2944.2&3069.5& 2762.3 \\
  {} {} ~~~~$\mu$-like & 3227 & 4212.8 \ \ \ \ &22.8~(0.5\%)&3979.7~(94.5\%)&210.3~(5.0\%)&4329.0&4573.9& 3931.6 \\ 
  {} multi-ring    & 2361 & 2791.7 \ \ \ \ &&&&2694.6&2976.0& 2380.2 \\
  {} {} ~~~~$\mu$-like & 208 & 322.6 \ \ \ \ &11.6~(3.6\%) &292.0~(90.5\%)&18.9~(5.9\%) &301.5&342.1& 274.0 \\
  \hline
  \multicolumn{1}{r}{$R=$} &
    \multicolumn{3}{l}{$0.658~\pm~0.016~(stat)~\pm~0.035~(sys)$} &&&0.655&0.646& 0.676 \\ 
  \hline \hline 
  {} multi-GeV & 2901 & 3472.0 \ \ \ \ &&&& 3212.6 & 3708.7 & 3179.3 \\ 
  {} single-ring   & 1397  & 1580.4 \ \ \ \ &&&&1456.8&1676.6& 1463.7 \\
  {} {} ~~~~$e$-like   & 746  &  680.5 \ \ \ \ &562.2~(82.6\%)&47.6~(7.0\%)&70.7~(10.4\%) &635.3&729.2& 635.3 \\
  {} {} ~~~~$\mu$-like & 651  &  899.9 \ \ \ \ &3.6~(0.4\%) &894.2~(99.4\%)&2.1~(0.2\%)&821.4&947.4& 828.4 \\
  {} multi-ring    & 1504 & 1891.6 \ \ \ \ &&&&1755.9&2032.1& 1715.5 \\
  {} {} ~~~$\mu$-like & 439 & 711.9 \ \ \ \ &16.6~(2.3\%)&675.8~(95.0\%)&19.4~(2.7\%)&645.9&749.1& 618.9 \\
  \hline  
  {} {partially-contained} & 911 & 1129.6 \ \ \ \ &20.8~(1.8\%)&1098.8~(97.3\%)&10.0~(0.9\%) &1065.0&1236.6& 1074.9\\ 
  \hline
  \multicolumn{1}{r}{$R_{FC+PC}=$} & 
    \multicolumn{3}{l}{$0.702~^{+0.032}_{-0.030}~(stat)~\pm~0.101~(sys)$} &&&0.705 & 0.699& 0.699 \\
  \hline \hline
  \end{tabular} 
  \end{center} 
  \caption{Summary of the sub-GeV, multi-GeV and PC event samples
  compared with the Monte Carlo prediction based on the neutrino interaction
  model of Ref.~\cite{Hayato:2002sd} ({\tt NEUT}) and neutrino flux
  calculation of Ref.~\cite{Honda:2004yz}, as well as different flux models.
  Fluxes A, B and C refers to \cite{Honda:2004yz}, \cite{Battistoni:2003ju} and
  \cite{Barr:2004br}, respectively. The Monte Carlo prediction 
  with {\tt NUANCE}~\cite{Casper:2002sd} 
  and Flux A is also shown. The Monte Carlo predictions do not include
  neutrino oscillations.}
\label{tab:datasummary}
\end{table*}

Figures~\ref{fig:fc-vtx}, \ref{fig:pc-vtx} show the reconstructed vertex
distributions for FC and PC events projected on the $z$ and $r^2 \equiv (x^2
+ y^2)$ axes.  The shape of the vertex distributions of the data and MC agree
well in the fiducial volume.  Near the fiducial volume boundary, the event
rate for FC events slightly decreases, and the event rate for PC events
increases.  This is because energetic muons produced by neutrino interactions
near the wall tend to escape and are identified as PC events.  Peaks near the
edge of the ID for both the MC and the data are caused by a constraint of the
vertex reconstruction programs: the reconstruction of the vertex is
restricted within the ID, and the events whose vertex is estimated outside of
the ID are constrained to be within the ID, where they pile up at the edge.
The peak at $z$ = 1810.0 cm in the distribution for FC data is caused by the
cosmic ray muons passing through inefficient regions of the OD. These muons
are safely rejected by the fiducial volume cut.

\begin{figure}[htb]
  \includegraphics[width=2.9in]{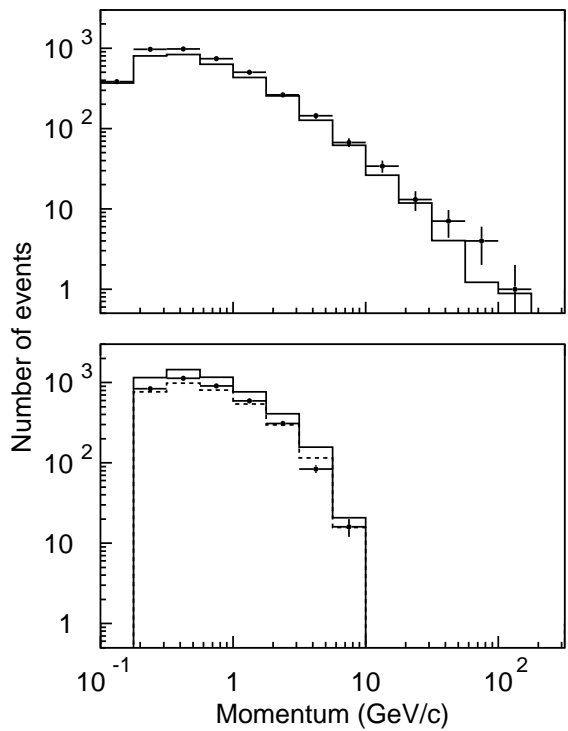}
  \caption{Momentum distribution of FC single ring $e$-like (top) and 
   $\mu$-like (bottom) events. The sharp cut in the muon momentum 
   spectrum in the high energy end is due to the requirement on
   fully containment. The points show the data, solid lines show the
   Monte Carlo prediction without neutrino oscillation and
   dashed lines show the oscillated Monte Carlo events with
   $(\sin^22\theta = 1.00,$ $\Delta m^2 = 2.1\times10^{-3} $~eV$^2$).}
   \label{fig:mom_fc}
\end{figure}

\begin{figure}[htb]
  \includegraphics[width=2.9in]{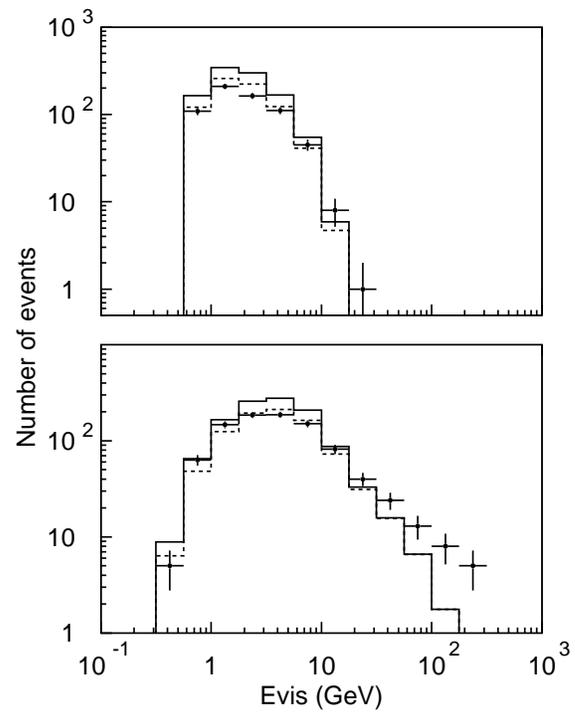}
  \caption{Visible energy distribution of multi-ring $\mu$-like events (top)
   and PC events (bottom). The points show the data, solid lines show the
   Monte Carlo prediction without neutrino oscillation and
   dashed lines show the oscillated Monte Carlo events with
   $(\sin^22\theta = 1.00,$ $\Delta m^2 = 2.1\times10^{-3} $~eV$^2$).}
   \label{fig:evis}
\end{figure}

Fig.~\ref{fig:mom_fc} shows the reconstructed momentum distributions for FC
single-ring events.  The data and MC show good agreement except for
significantly fewer numbers of FC $\mu$-like events. Fig.~\ref{fig:evis}
shows the visible energy distribution for FC multi-ring $\mu$-like and PC
events.
The PC data have more events than the Monte Carlo prediction 
at the highest energies.
This could indicate that the neutrino energy spectrum in the Monte Carlo
is too soft around 100~GeV. We note that the upward through-going data
also suggest that the neutrino energy spectrum is too soft (see 
Table~\ref{tab:upmudatasummary}.)
The neutrino energy spectrum up to 1~TeV was considered
in the Monte Carlo prediction for the FC and PC samples.
We estimated that two PC events are expected with visible energy above 
100~GeV from neutrinos above 1~TeV.

The flavor ratio of the atmospheric 
neutrino flux, $(\nu_{\mu}+\overline{\nu}_{\mu}) /
 (\nu_e+\overline{\nu}_{e})$, is predicted with 3\,\% accuracy. 
As shown in Table~\ref{tab:datasummary}, the particle identification
for FC single-ring events gives a good estimation of the flavor of
the parent neutrinos, and the ratio of the number of $e$-like events
and $\mu$-like events,$\left(\mu/e\right)$,
gives a good estimation of the flavor ratio of the atmospheric neutrinos.
We define a $(\mu / e)$  double ratio,
$R \equiv (\mu/e)_{DATA} / (\mu/e)_{MC}$.
Without neutrino oscillation, $R$ should be consistent with unity.
$R$ is measured to be:
\begin{equation}
R_{sub-GeV} = 0.658 \pm 0.016 \pm 0.035,
\end{equation}
for the sub-GeV sample.

A substantial fraction of muons in the multi-GeV energy range exit
from the inner detector and are detected as PC events.  The
partially-contained event sample is estimated to be 97\,\% pure CC
$\nu_{\mu}$ interactions, even without requiring any particle
identification or ring-number cuts. Therefore, we add FC single-ring
and PC event totals when calculating $R$ in the multi-GeV range.  We
measured $R$ in the multi-GeV energy range to be:
\begin{equation}
R_{multi-GeV+PC} = 0.702 \asymerr{0.032}{0.030} \pm 0.101.
\end{equation}
 Systematic uncertainties in the double ratio $R$ have been
discussed in detail in Refs.~\cite{Fukuda:1998tw} and 
\cite{Fukuda:1998ub}. These errors have been re-evaluated and
 are estimated to be 5.3\,\% for
 sub-GeV and 14.4\,\% for multi-GeV events. The sources of the systematic
 uncertainties in $R$ are listed in Table~\ref{table:sys_r},
 which include both theoretical and experimental errors.
Among the experimental systematic errors, the separation of single-
and multi-ring events is the largest source of the systematic 
uncertainties. As shown in Fig.~\ref{fig:ring-counting-likelihood},
the distributions of the likelihood difference between the single-ring
and multi-ring assumptions have slight shifts in the peak positions
between the data and the Monte Carlo. These differences could 
cause systematic uncertainties in the number of identified
single-ring $e$-like and $\mu$-like events (which are summarized in 
Table~\ref{table:fitsummary_fit}). Since the magnitude of the
uncertainty is different between $e$-like and $\mu$-like events,
and since the separation of single- and multi-ring is not 
applied for PC events, the uncertainty in the single- and multi-ring
separation causes the uncertainties in the $R$ measurements.

\begin{table}
\renewcommand{\arraystretch}{0.8}
\begin{center}
\begin{tabular}{lcc}
\hline\hline 
&sub-GeV(\%)&multi-GeV(\%) \\ \hline
Prediction of $\nu_{\mu}/\nu_{e}$ ratio & 2.7 &  2.2 \\
Prediction of $\nu/\overline{\nu}$ ratio & 1.6 &  0.8 \\
$K/\pi$ ratio & 0.6 & 1.9 \\
E$_{\nu}$ spectral index & 0.6 &  2.3 \\
Sample-by-sample\footnote{Different flux calculations predict different
energy dependences that cannot be explained by a simple spectral index
uncertainty. See lower part of Fig.~\ref{fig:enu_spectra}. Uncertainty of
the relative normalization of the fully-contained multi-GeV and
partially-contained sample gives the systematic error in double ratio $R$.}
 & -- & 2.9 \\
$\nu$ interaction &  &  \\
~~~~~~quasi-elastic scattering & 1.4 & 1.0 \\
~~~~~~single-meson production   & $<$0.1 & 0.3 \\
~~~~~~deep-inelastic scattering& 0.2 & 0.5 \\
~~~~~~coherent-pion production & 0.4 & 0.2 \\
~~~~~~NC/CC ratio              & 0.5 & 2.0 \\
~~~~~~nuclear effects\footnote{The mean free path of hadrons in $^{16} O$ was changed by 30\,\%. Also the uncertainty in the pion energy spectrum produced by neutrino interactions, defined to be the difference between the interaction models A and B, is taken into account.}        
                               & 1.3 & 0.8 \\
Hadron simulation              & 0.7 & $<$0.1 \\
FC reduction                   & 0.1 & 0.1 \\
PC reduction                   & -- & 1.5 \\
Non-$\nu$ background           & $<$0.5& $<$0.3 \\
$\mu$/$e$ separation           & 1.3 & 0.6 \\
Single-ring/multi-ring separation & 3.2 & 13.2 \\
Energy calibration             & 0.6 & 1.2 \\
MC statistics                  & 0.5 & 0.9 \\
\hline
Total                           & 5.3 &  14.4  \\ \hline\hline
\end{tabular}
\caption{Sources of the systematic errors in double ratio 
$R ( \equiv (\mu/e)_{DATA} / (\mu/e)_{MC} ) $ 
for the sub-GeV and multi-GeV samples. Estimated uncertainty in each
source of the systematic error is described in 
Tables~\ref{table:fitsummary_flux}, \ref{table:fitsummary_nuint},
\ref{table:fitsummary_event} and \ref{table:fitsummary_fit}.}
\label{table:sys_r}
\end{center}
\end{table}

\begin{figure}
  \includegraphics[width=3.2in]{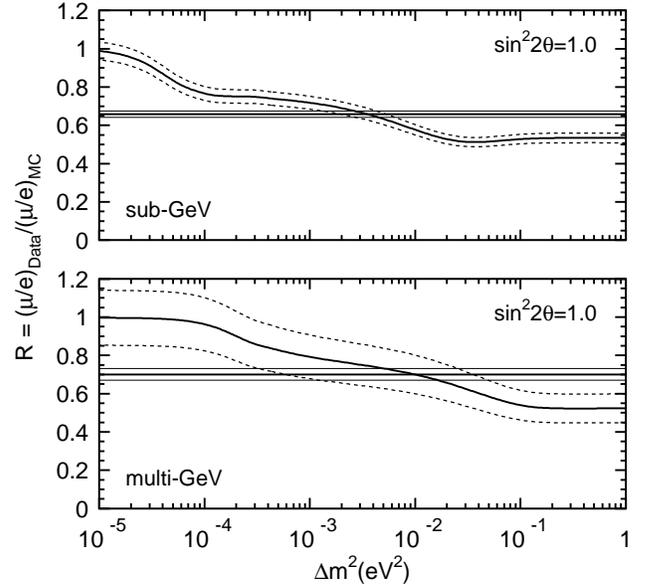}
  \caption{ Expected $(\mu / e)_{Data} / (\mu /e)_{MC}$ for singe-ring
   sub- and multi-GeV + PC samples as a 
  function of $\Delta{m}^2$ for full $\nu_{\mu} \leftrightarrow 
\nu_{\tau}$ mixing. The values 
  for the data together with $\pm 1 \sigma$
 statistical errors are shown by the horizontal lines.
 The systematic errors are shown by the band in the expectation.}
   \label{fig:ratio}
\end{figure}

Figure~\ref{fig:ratio} shows the expected
$(\mu / e)_{Data} / (\mu /e)_{MC}$ in the presence of neutrino oscillation
for sub- and multi-GeV samples
as a function of $\Delta{m}^2$. Data are consistently
explained by neutrino oscillations for $\Delta m^2$ in the range of
$10^{-3}$ to $10^{-2}$eV$^2$. 

\begin{figure*}[htb]
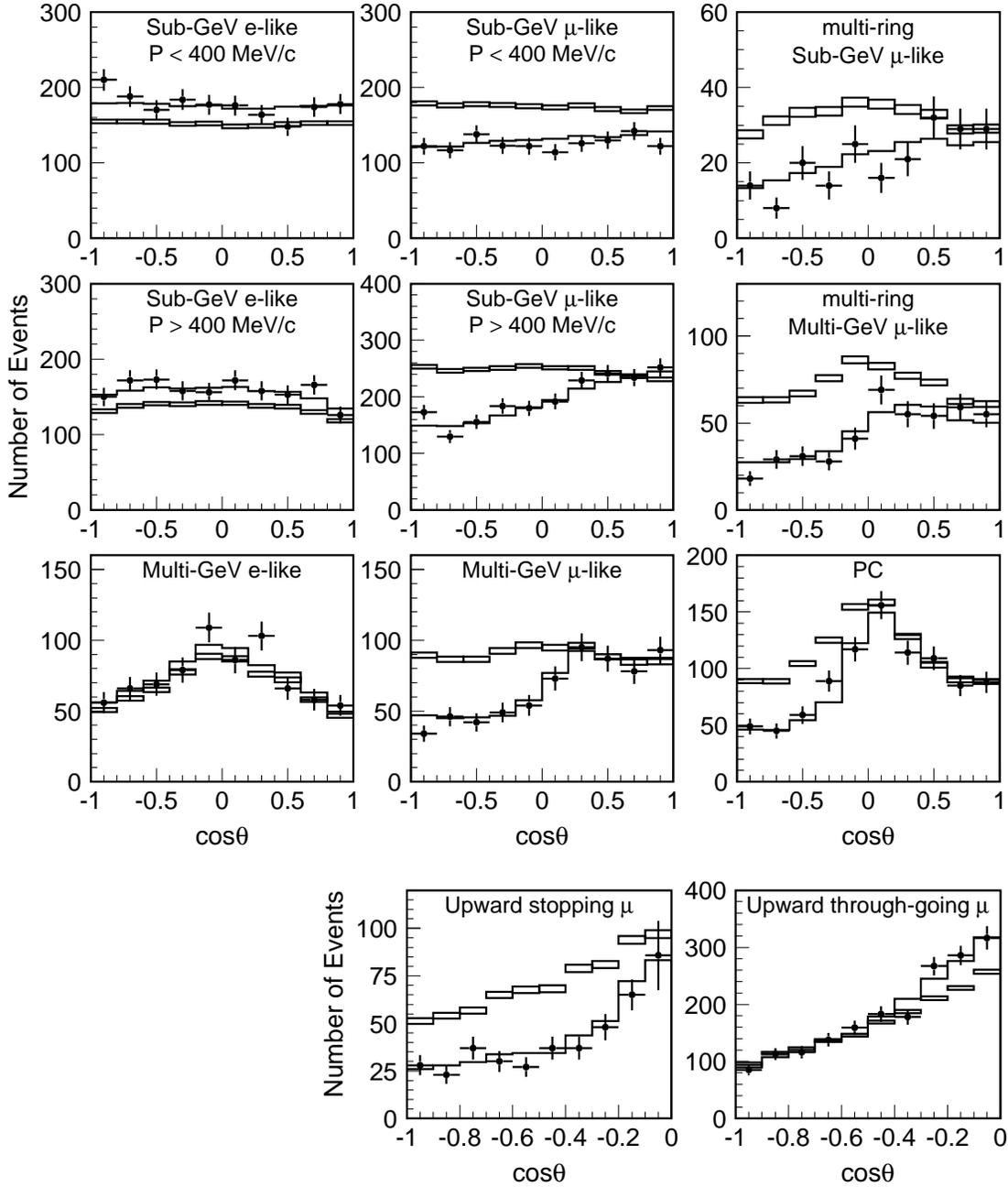

  \includegraphics[height=4.8in]{zenith_1a.epsi}
  \vspace{0.2in}
  \includegraphics[height=4.8in]{zenith_2a.epsi} \\
~~~~~~~~~~~~~~~~~~~~~~~~~~~~~~~~~~~~~~~~~~~~~~~~~~~~~~~~~~
  \includegraphics[height=1.71in]{zenith_3a.epsi}
  \caption{The zenith angle distribution for fully-contained 1-ring events, 
    multi-ring events, partially-contained events and upward muons.
    The points show the data, box histograms show the non-oscillated Monte
    Carlo events and the lines show the best-fit expectations 
    for $\nu_\mu
    \leftrightarrow \nu_\tau$ oscillations with $\sin^2 2 \theta = 1.00$
    and $\Delta m^2 = 2.1\times 10^{-3}$~eV$^2$. 
    The best-fit expectation is corrected 
    by the 39 systematic error terms, while the correction is
    not made for the non-oscillated Monte Carlo events.
    The height of the boxes shows the statistical error of the Monte Carlo. }
   \label{fig:zenith_plot}
\end{figure*}

\begin{figure}[htb]
  \includegraphics[width=3.1in]{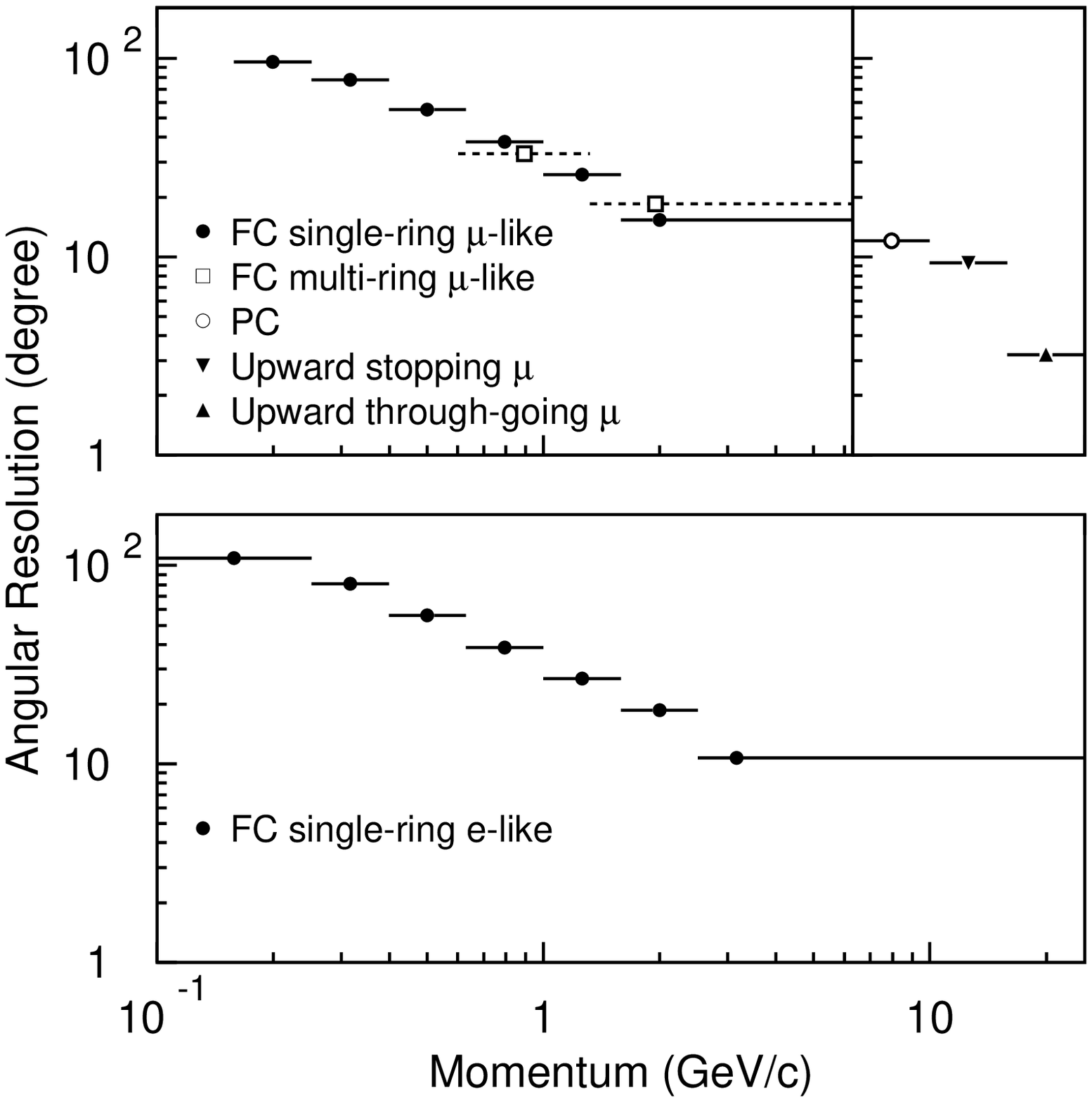}
  \caption{Angular resolution of the neutrino direction as a function of the 
  outgoing charged lepton momentum.
  The angular resolution is defined as the angular difference between the 
  parent neutrino and the reconstructed direction for which 68\,\% of 
  the events are included. }
   \label{fig:angle_cor}
\end{figure}

Figure~\ref{fig:zenith_plot} shows the measured and
expected numbers of FC and PC events as a function of the cosine of the
zenith angle ($\cos \Theta$), $\cos\Theta=-1$ refers to upward-going
and $\cos\Theta=1$ refers to downward-going.  Single-ring sub-GeV events are
separately shown in two parts, $p_l\le$ 400 MeV/$c$ and $p_l > 400$
MeV/$c$, where $p_l$ is the lepton momentum. 
In the momentum range below 400 MeV/$c$, the angular correlation
between the neutrino and outgoing lepton is very poor, the shape of
the atmospheric neutrino flux is largely washed out, and the zenith
angle distributions for the charged leptons should be approximately flat.
Fig.~\ref{fig:angle_cor} shows the angular resolution of the neutrino
directions as a function of the momentum. The angular resolution is
defined as the angular difference between the parent neutrinos and the
reconstructed directions in which 68\% of the events are included.

\begin{figure}[htb]
  \includegraphics[width=3.2in]{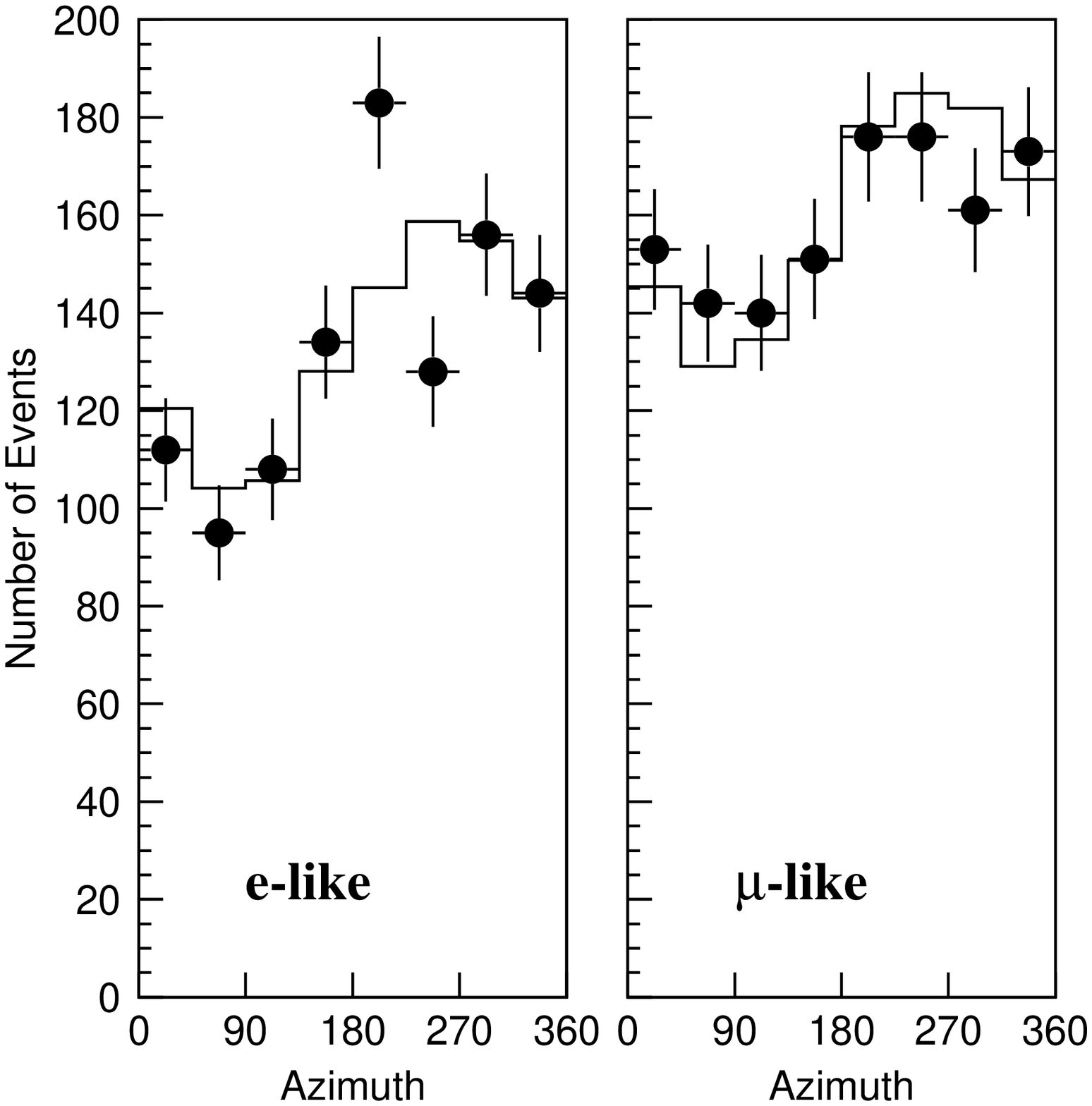}
  \caption{The azimuth distribution of the sample of events selected
   for analysis of the east-west effect. The Monte Carlo histogram
   is normalized to the total number of the real data. 
   0, 90, 180 and 270 degree azimuthal angles show particles going to 
   north, west, south and east, respectively. }
   \label{fig:east-west}
\end{figure}

We have also studied the azimuthal dependence of the atmospheric neutrino
data. This is a sensible consistency check, as neutrino oscillation should
not cause any azimuthally dependent deficit since all path lengths at a given
zenith angle are equal. In a well-selected data sample, the azimuth rates
exhibit the famous east-west effect, which was used in the 1930's to
demonstrate that cosmic rays were positively charged. The effect is caused by
the deflection of primary protons in the earth's magnetic field, where
trajectories from the east are blocked by the bulk of the earth. This also
results in a deficit of atmospheric neutrinos arriving from the east, with
the strongest effect at the lowest neutrino energy. We maximized our
sensitivity to this effect by selecting lepton momenta between 400 and 3000
MeV/$c$ in the zenith angle range $|\cos \Theta| < 0.5$. The low momentum cut
ensures good pointing resolution, and the high momentum cut diminishes the
contribution from high energy primary protons that are insufficiently
deflected. The zenith requirement enhances the statistical sensitivity as the
depletion only occurs near the horizon.  Figure~\ref{fig:east-west} updates
our previous result\cite{Futagami:1999wz} to the final data reported here.
That our data reproduces the prediction for this effect implies that the
model for geomagnetic cutoff in the flux prediction is accurately accounted
for, and checks the basic features of neutrino production and scattering.

While the measured shape of zenith angle distributions for $e$-like
events is consistent with expectations, both the FC $\mu$-like and PC
samples exhibit significant zenith-angle dependent deficits.
The up-down ratio, where $U$ is the number of upward-going
events ($-1 < \cos \Theta < - 0.2$) and $D$ is the number of
downward-going events ($0.2 < \cos\Theta < 1$), is measured to be: $U/D
= 1.133 \asymerr{0.062}{0.059} \pm 0.009$ for  single-ring sub-GeV $e$-like 
events in the momentum range
 below 400 MeV/$c$, $U/D= 1.082 \asymerr{0.063}{0.060} \pm 0.024$ above 400 MeV/$c$,
 $U/D = 0.964 \asymerr{0.062}{0.058} \pm 0.008$ for single-ring sub-GeV  
$\mu$-like events below 400 MeV/$c$,
 $U/D = 0.670 \asymerr{0.035}{0.034} \pm 0.012$  above 400 MeV/$c$,
 $U/D = 0.961 \asymerr{0.086}{0.079} \pm 0.016$ for the multi-GeV 
$e$-like events
and $U/D = 0.551 \asymerr{0.035}{0.033} \pm 0.004$ for the single-ring
multi-GeV $\mu$-like plus PC events.

\begin{figure}
  \includegraphics[width=3.2in]{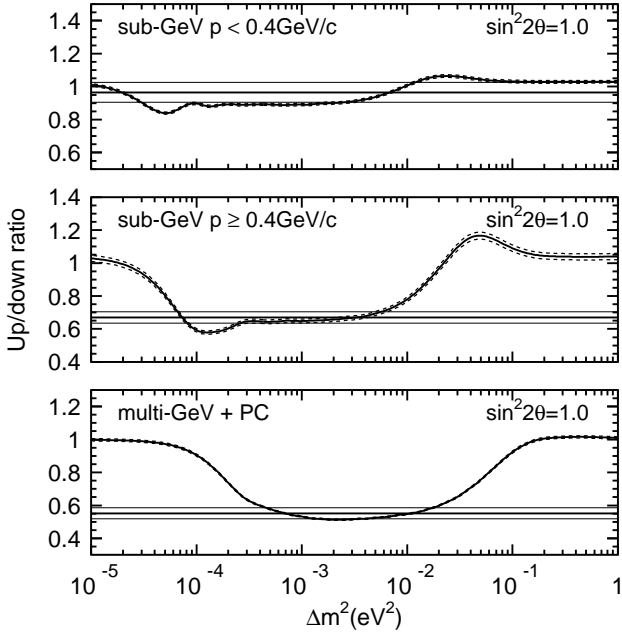}
  \caption{Expected $U/D$ ratio for FC single-ring $\mu$-like + PC events 
as a function of $\Delta{m}^2$ for full $\nu_{\mu} \leftrightarrow
\nu_{\tau}$ mixing. The data sample is 
divided into FC sub-GeV with $P_{\mu} < 400~$MeV/$c$, FC
sub-GeV with $P_{\mu} > 400~$MeV/$c$, and FC multi-GeV + PC
events. The ratio for the data together with the $\pm 1 \sigma$
statistical error are shown by the horizontal lines.
The systematic errors are shown by the band in the expectation.
}
   \label{fig:updown}
\end{figure}

Many systematic uncertainties are canceled for the up-down ratio and the
remaining sources of the uncertainties are: uncertainty in the flux
calculation, 0.5\,\% and 0.8\,\% for sub-GeV $e$-like and $\mu$-like events
in the momentum range below 400 MeV/$c$, 2.1\,\% and 1.8\,\% for 
$e$-like and $\mu$-like events above 400 MeV/$c$,
and 1.5\,\% and 0.6\,\% for multi-GeV $e$-like events and multi-GeV
$\mu$-like events plus PC events; uncertainty in the
angular dependence of absolute
energy calibration, 0.5\,\% and 0.2\,\% for sub-GeV $e$-like and $\mu$-like events
in the momentum range below 400 MeV/$c$, 0.4\,\% and 0.4\,\% for 
$e$-like and $\mu$-like events above 400 MeV/$c$,
and 0.8\,\% and 0.4\,\% for multi-GeV $e$-like events and multi-GeV
$\mu$-like events plus PC events; and uncertainty in the
non-neutrino background such as cosmic ray muons, $<$0.4\,\%, $<$0.1\,\%,
$<$0.2\,\%, $<$0.2\,\% for sub-GeV $e$-like, $\mu$-like, multi-GeV
$e$-like, $\mu$-like plus PC events, respectively.
In total, the systematic uncertainty for $U/D$
is 0.8\,\% and 0.8\,\% for sub-GeV $e$-like and $\mu$-like events
in the momentum range below 400 MeV/$c$, 2.2\,\% and 1.8\,\% above 400 MeV/$c$,
1.7\,\% and 0.7\,\% for multi-GeV $e$-like and $\mu$-like plus PC events,
respectively. While the ratio for $e$-like events is consistent with 1,
the $\mu$-like up-down ratio for the multi-GeV data differs from 1 by 
more than 12 standard deviations.
Figure~\ref{fig:updown} shows the expected $U/D$ ratios
as a function of $\Delta{m}^2$. Data are consistently
explained by neutrino oscillations with $\Delta m^2$ in the range of
$10^{-3.5}$ to $10^{-2}$eV$^2$. 

\subsubsection{Upward Muon Events}

\begin{figure}[htb]
  \includegraphics[width=3.0in]{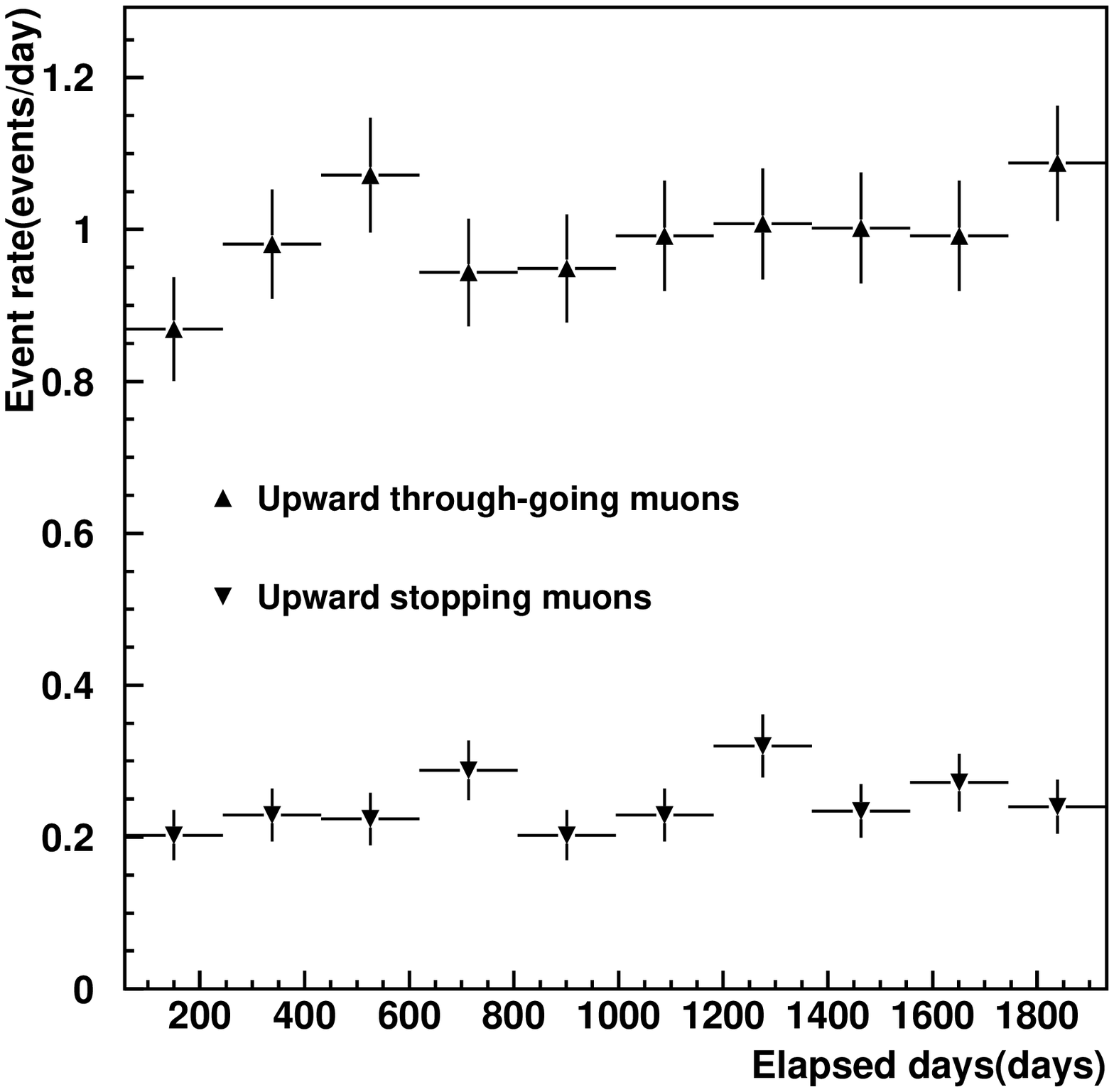}
  \caption{Event rates as a function of elapsed SK running time for 
   upward-going muon data.}
   \label{fig:rate-stability-upmu}
\end{figure}

The upward-going muon data used in this analysis were taken from
May~1996 to July~2001. The detector live-time was 1645.9~days.  Though
spanning the same period of calendar time, this live-time was larger
than that of the contained vertex events because the reconstruction of
long path length muons is less sensitive to detector conditions, allowing
looser run selection criteria. 
Fig.~\ref{fig:rate-stability-upmu} shows the event rates 
as a function of the elapsed days for upward
going muons. The event rates for these samples are stable.
Table~\ref{tab:upmudatasummary} summarizes the number of observed events
in the upward-going muon data sample and the corresponding flux and
expected flux.

  The systematic errors on the observed number
of events compared to the Monte Carlo predictions are: 
the 2\,\% energy scale uncertainty leads to a
$^{+0.9}_{-1.1}$\,\% error in the stopping muons due to the 1.6~GeV/$c$ cut;
the reduction efficiency for stopping (through-going) muons has 
an uncertainty of
$^{+0.34}_{-1.25}$\,\% ($^{+0.32}_{-0.54}$\,\%); and stopping/through-going
separation $^{+0.29}_{-0.38}$\,\% (where ``+'' means through-going muons
misidentified as stopping).  As in the contained event analysis,
comparison of data and expectations is done between observed number of
events and the live-time-scaled MC number of events.  However, to
facilitate comparisons with other experiments, these numbers are also
presented in units of flux as described
in~\cite{Fukuda:1998ah,Fukuda:1999pp}.  The additional systematic
uncertainty in the observed through-going (stopping) flux comes from
effective area of 0.3\,\% and the live-time
calculation (0.1\,\%). The absolute expected flux has 
theoretical uncertainties of at least 20\,\%
in the normalization for high energy ($> 100 $~GeV) neutrinos and 5 to
10\,\% from interaction model differences.

\begin{table*}[htbp]
  \begin{center}
    \begin{tabular}[t]{c|c|c|c|c|cc}
\hline\hline
    event class & \# events & \# expected & flux & expected flux &  \multicolumn{2}{c}{\# expected}   \\
     &  & (Flux A) & ($\times 10^{-13} cm^{-2}s^{-1}sr^{-1}$) & 
 ($\times 10^{-13}cm^{-2}s^{-1}sr^{-1}$) & (Flux B) &(Flux C)  \\
\hline
    $\Phi_{stop}$               & 417.7 & 713.5 &   $0.381 \pm0.024 ^{+0.005}_{-0.007}$ & $0.648\pm0.145$ &  681.5 &  790.5 \\
    $\Phi_{thru}$               & 1841.6 & 1669.5 & $1.661 \pm0.040 ^{+0.011}_{-0.013}$ & $1.506\pm0.337$ & 1644.3 & 1974.9 \\
    ${\mathcal{R}}=\Phi_{stop}/\Phi_{thru}$ & 0.227 & 0.427 & $0.229 \pm0.015 \pm0.003$ & $0.430\pm0.058$ &  0.414 &  0.400 \\
    $\Phi_{stop} + \Phi_{thru}$ & 2259.3 & 2382.9 & $2.042 \pm0.046 ^{+0.012}_{-0.015}$ & $2.154\pm0.482$ & 2325.8 & 2765.4 \\
\hline\hline
    \end{tabular}
    \caption{Summary of observed and expected results for upward-going 
    muons during 1645.9 live-days. 
    The first and the second errors in the observed flux show statistical
    and systematic errors, respectively. Expected event rates based on
    different flux models are also shown.
    Fluxes A, B and C refer \cite{Honda:2004yz}, \cite{Battistoni:2003ju} and
    \cite{Barr:2004br}, respectively.
    }
    \label{tab:upmudatasummary}
  \end{center}
\end{table*}

\begin{figure}[hptb]
  \includegraphics[width=3.0in]{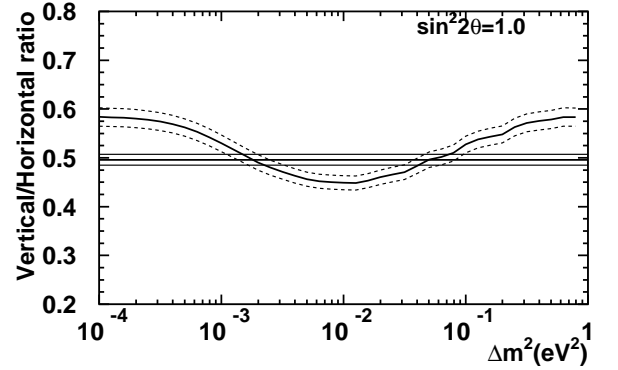}
  \caption{ Expected ratio 
   of vertical to horizontal upward through-going muons
as a function of $\Delta{m}^2$ for full $\nu_{\mu} \leftrightarrow
\nu_{\tau}$ mixing. Vertical and horizontal  
are defined to be the number of events in $-1 < \cos \Theta 
< -0.5$ and $-0.5 < \cos \Theta < 0.0$, respectively.
  The ratio for the data together with $\pm 1 \sigma$ combined statistical 
  and systematic error 
  is also shown by the horizontal lines. The systematic error is shown
  by the band in the expectation. }
   \label{fig:hvratio}
\end{figure}

The zenith angle distributions of the upward through-going and
stopping muons are shown in Fig.~\ref{fig:zenith_plot}.
The shape of the zenith angle distribution is predicted accurately.
Therefore, the vertical to horizontal ratio was taken to
study the effects of neutrino oscillations, where $V$ 
and $H$ represent the number of through-going events
with $-1 < cos \Theta < 0.5 $ and  $-0.5 < cos \Theta < 0 $,
respectively. The  $V/H$ ratio for the data
was $ 0.497 \pm 0.022 (stat) \pm 0.003 (sys)$, while
the no-oscillation prediction was
  $ 0.586 \pm 0.019 (sys)$.
Taking into account statistical, systematic, and theoretical
uncertainties (using the methods discussed in
Section~\ref{sec:oscillation}), the  $V/H$ ratio
 of the upward
through-going muon sample was smaller than the prediction 
by 3 standard deviations.
The observed flux falls off much more rapidly than
predicted as the zenith angle approaches the nadir.  
Fig.~\ref{fig:hvratio} shows the expected and observed
$V/H$ ratio of the upward through-going muon events. 
The observed ratio suggests that  $\Delta$m$^2$
lies in the range of either $(1 - 3)\times$10$^{-3}$ or 
(5 - 10)$\times$10$^{-2}$~eV$^2$.

\begin{figure}[hptb]
  \includegraphics[width=3.0in]{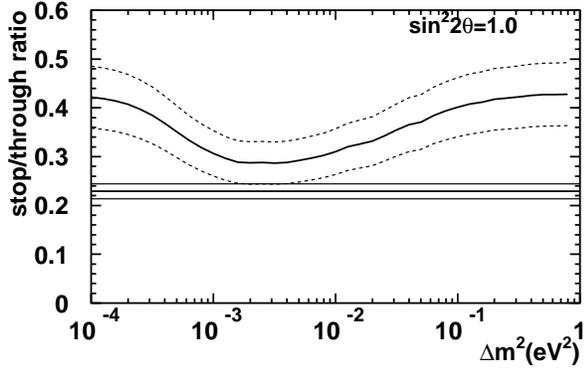}
  \caption{ Expected 
   ratio of stopping to through-going upward muons as a 
   function 
   of $\Delta{m}^2$ for full $\nu_{\mu} \leftrightarrow
   \nu_{\tau}$ mixing. The ratio for the data together with 
    $\pm 1 \sigma$ combined statistical and systematic error is also 
    shown by the horizontal lines. The systematic error is shown 
   by the band in the expectation. }
   \label{fig:stratio}
\end{figure}

The large uncertainty in the absolute flux normalization can be
canceled by taking the stopping to through-going muon ratio.   
Fewer upward stopping muons were observed than predicted, while the 
observed number of upward through-going muons was consistent with
the theoretical prediction within the errors.
The observed ratio of stopping to through-going muons was
$0.229 \pm 0.015(stat) \pm 0.003(sys)$, while the expected ratio was 
$0.430 \pm 0.065$. 
The expected ratio has theoretical uncertainties from cross sections
($\pm4.7$\,\%), the cosmic ray spectral index ($\pm12.5$\,\%),
and the flux model dependence ($\pm7.1$\,\%).  
Fig.~\ref{fig:stratio} shows the expected
ratio of stopping to through-going upward muon events as a function of
$\Delta{m}^2$ along with the measured ratio,
which was smaller than the prediction by more than 3 standard deviations. 
The observed value can be
explained assuming neutrino oscillations with $\Delta$m$^2$
in the range of 10$^{-3}$ to 10$^{-2}$~eV$^2$.
This stopping to through-going ratio
is no longer explicitly used in the oscillation fits, but is presented
for comparison to older work~\cite{Fukuda:1999pp}.

\section{Oscillation analysis}
\label{sec:oscillation}

The observed deficits of muon neutrino interactions are in strong
disagreement with the expectation in the absence of neutrino
oscillations. Oscillation between electron neutrinos and muon
neutrinos cannot explain the data, as no surplus of upward-going
electron neutrinos is observed in the multi-GeV data sample; an
attempt at a two-flavor $\nu_{\mu} \leftrightarrow \nu_e$ fit results
in a generally poor fit, with $\chi^2$ difference of more than 100
with respect to the $\nu_\mu \leftrightarrow \nu_\tau$ analysis
described below. A variety of exotic alternatives such as neutrino
decay were considered, however, none fit the data as well as the
$\nu_\mu \leftrightarrow \nu_\tau$ scenario analyzed below.
Atmospheric $\nu_\mu$ oscillation
into $\nu_\tau$ is mostly characterized by $\nu_\mu$ disappearance, as
the majority of the flux is below the 3.5~GeV neutrino energy
threshold for charged current $\tau$ production. We carefully studied
the alternative that $\nu_\mu$ could oscillate to a sterile neutrino
state\cite{Fukuda:2000np}, which would also result in $\nu_\mu$
disappearance. However, the lack of matter-induced suppression of
oscillation and the
relative up-down symmetry of the multi-ring sample with considerable
neutral current fraction eliminated this hypothesis from serious
considerations. The final Super-Kamiokande statistical analysis of
these alternative scenarios, as well as the standard three flavor 
oscillation analysis, will appear in other publications.
In this paper, we therefore establish the best-fit parameters of $\nu_\mu
\leftrightarrow \nu_\tau$ oscillation.  

The analysis is based on a comparison between data and Monte Carlo, suitably
binned to convey information about neutrino type, neutrino energy, and flight
distance. The neutrino type, $\nu_e$ or $\nu_\mu$ is classified by the
identification of the main Cherenkov pattern as showering or non-showering
respectively. Penetrating particles such as upward-going muons and
partially-contained events are assumed to arise from $\nu_\mu$ interactions.
The neutrino energy is correlated with the outgoing lepton momentum using the
interaction models described in Section IV. The flight distance is correlated
with the zenith angle as described by Figs.~\ref{fig:flight-length} and
\ref{fig:angle_cor}. To study neutrino oscillation using
Eq.~\ref{eqn:oscillation}, we reweight each simulated event using the Monte
Carlo ``truth'' information of $E_\nu$ and $L$ and bin the reweighted events
for comparison with the detected data. Unlike our analysis using the ratio
$L/E$~\cite{Ashie:2004mr}, we make no attempt to estimate $L$ or $E_\nu$ on
an event-by-event basis.''

We used all of the data samples with a well-identified CC $\nu_\mu$
component, namely: FC single-ring $\mu$-like, PC, multi-ring $\mu$-like,
upward stopping muons, and upward through-going muons. Because the flux of
electron neutrinos provides a powerful constraint through the accurately
predicted $\nu_\mu/\nu_e$ ratio, the single-ring $e$-like events were
included in the fit. The FC single ring $\mu$-like and $e$-like samples were
divided in logarithmically-spaced momentum bins.  All samples were divided in
10 zenith angle bins. In total 180 bins were used in the analysis: 150 for
the FC sample, 10 for the PC sample, 10 for the upward stopping muon sample,
and 10 for the upward through-going muon sample.  The number of observed and
expected events for each bin are summarized in the Appendix.

A $\chi^2$ statistic is defined by the following sum: 
%
\begin{equation}
\chi^2 =  \sum_{i=1}^{180}
\frac{\left(N_{i}^{\rm obs} - N_{i}^{\rm exp} (1+\sum_{j=1}^{39}f_{j}^{i}\cdot\epsilon_{j}) \right)^2
}
{ \sigma^2_{i} } + \sum_{j=2}^{39} \left(\frac{\epsilon_j}{\sigma_j}\right)^2
\vspace{-5mm}
\label{equation:chi2def}
\end{equation}
\begin{equation}
 N_{i}^{\rm exp} = N_{i}^{\rm 0} \cdot
   P(\nu_\alpha \rightarrow \nu_\beta)~~.
\label{equation:chi2def2}
\end{equation} \noindent
%
In the first sum, $N^{\rm obs}_i$ is the number of observed events in the
$i^{\rm th}$ bin and $N^{\rm exp}_i$ is the expected number of events based
on a Monte Carlo simulation and $\sigma_i$ combines the statistical
uncertainties in the data and Monte Carlo simulation. During the fit, the
values of $N^{\rm exp}_i$ are recalculated to account for neutrino
oscillations and systematic variations in the predicted rates due to
uncertainties in the neutrino flux model, neutrino cross-section model, and
detector response. $N^{\rm 0}_i$ is the number of events predicted from the
MC without neutrino oscillation for the $i^{\rm th}$ bin. The appearance of
$\nu_\tau$ as a result of oscillations is taken into account by adding into
the Monte Carlo distributions simulated $\nu_\tau$ interactions which pass
all cuts. These events show up mainly in the multi-GeV $e$-like sample, but
are not easily distinguished on an event-by-event basis. We are undertaking a
separate analysis, to be published later, which will study $\nu_\tau$
appearance in the atmospheric neutrino flux.

The systematic uncertainties are represented by 39 parameters $\epsilon_j$.
During the fit, these 39 $\epsilon_j$ are varied to minimize $\chi^2$ for
each choice of oscillation parameters $\sin^2 2\theta$ and $\Delta m^2$.
Among these, only 38 contribute to the $\chi^2$, since the absolute
normalization is allowed to be free. The factor $f^i_j$ represents the
fractional change in the predicted event rate in the $i^{\rm th}$ bin due to
a variation of the parameter $\epsilon_j$. The second sum in the $\chi^2$
definition collects the contributions from the systematic uncertainties in
the expected neutrino rates. The $\epsilon_j$ are listed in
Tables~\ref{table:fitsummary_flux}, \ref{table:fitsummary_nuint},
\ref{table:fitsummary_event}, and \ref{table:fitsummary_fit} with their
estimated uncertainties and the resulting best-fit values. Entries of the
same number are treated as fully correlated although the effect of the
uncertainty varies in size depending on its relative importance to the energy
bin of certain sub-samples. For example, the source of the up/down
uncertainty (No. 8) is due to the uncertainty in the geomagnetic field
effect, especially above the Super-Kamiokande detector. The uncertainty is
large for low energy neutrinos coming from primary cosmic rays below the
geomagnetic cutoff, but the effect of the uncertainty is decreased due to the
large scattering angle in the neutrino interactions. As a result, events in
the middle energy range are the most influenced by this particular systematic
uncertainty. Refer to the footnotes in the tables for more detail.

\begin{table}
\begin{center}
\begin{center}
\renewcommand{\thefootnote}{\alph{footnote}}
\begin{tabular}{lllccc}
 \hline \hline
 & & & $\sigma$\,(\%) & best-fit & {\it No.}\\
 \hline
 \multicolumn{6}{l}{\bf (A) Systematic uncertainties in neutrino flux} \\
 \multicolumn{2}{l}{Absolute normalization}&              & free & 11.9 & {\it 1}\\
  $(\nu_\mu + \overline{\nu}_\mu )/ (\nu_e + \overline{\nu}_e )$ \footnote[1]{A positive number means the number of MC $\nu_\mu + \overline{\nu}_\mu$ events is increased.} & \multicolumn{2}{l}{$E_{\nu}<5$\,GeV} & 3.0    & -2.4 & {\it 2} \\
                                & \multicolumn{2}{l}{$E_{\nu}>5 $\,GeV} & 
3.0\footnote[2]{Error linearly increases with $\log E_{\nu}$ from 3\,\% 
(5\,GeV) to 10\,\%(100\,GeV).}     & 0.1 & {\it 3} \\
  $\nu_{e}/\overline{\nu}_e$\footnote[3]{A positive number means the number of MC $\nu_e$ ($\nu_{\mu}$) events is increased.}             & \multicolumn{2}{l}{$E_{\nu}<10$\,GeV} & 5.0  & 1.5 & {\it 4} \\
                                         & \multicolumn{2}{l}{$E_{\nu}>10$\,GeV} & 
5.0\footnote[4]{Error linearly increases with $\log E_{\nu}$ from 
5\,\%(10\,GeV) to 10\,\%(100\,GeV).}   & 0.0 & {\it 5} \\
  $\nu_{\mu}/\overline{\nu}_{\mu}$\footnotemark[3]       & \multicolumn{2}{l}{$E_{\nu}<10$\,GeV} & 5.0  & -1.3 & {\it 6} \\
                                         & \multicolumn{2}{l}{$E_{\nu}>10$\,GeV} & 
5.0\footnote[5]{Error linearly increases with $\log E_{\nu}$ from 
5\,\%(10\,GeV) to 25\,\%(100\,GeV).} & 0.9 & {\it 7} \\
 Up/down\footnote[6]{Up/down (horizontal/vertical) uncertainty in neutrino flux is assumed to be 
fully correlated. All of the samples listed are simultaneously varied
according to the systematic uncertainty factors. A positive number means the number of MC upward (horizontally-going) events is increased.}
                        & $< 400$\,MeV          &$e$-like         & 0.5 & 0.2 & {\it 8} \\
                                &                       &$\mu$-like       & 0.8 & 0.3 & {\it 8} \\
                                & $> 400$\,MeV          &$e$-like         & 2.1 & 0.9 & {\it 8} \\
                                &                       &$\mu$-like       & 1.8 & 0.8 & {\it 8} \\
                                & Multi-GeV             &$e$-like         & 1.5 & 0.7 & {\it 8} \\
                                &                       &$\mu$-like       & 0.8 & 0.3 & {\it 8} \\
                                & PC                    &                 & 0.4 & 0.2 & {\it 8} \\
                                & \multicolumn{2}{l}{Sub-GeV multi-ring $\mu$} & 0.8    & 0.3 & {\it 8} \\
                                & \multicolumn{2}{l}{Multi-GeV multi-ring $\mu$} & 0.7  & 0.3 & {\it 8} \\
 Horizontal/vertical\footnotemark[6]                & $< 400$\,MeV          &$e$-like         & 0.3 & 0.0 & {\it 9} \\
                                &                       &$\mu$-like       & 0.3 & 0.0 & {\it 9} \\
                                & $> 400$\,MeV          &$e$-like         & 1.2 & 0.1 & {\it 9} \\
                                &                       &$\mu$-like       & 1.2 & 0.1 & {\it 9} \\
                                & Multi-GeV             &$e$-like         & 2.8 & 0.2 & {\it 9} \\
                                &                       &$\mu$-like       & 1.9 & 0.1 & {\it 9} \\
                                & PC                    &                 & 1.4 & 0.1 & {\it 9} \\
                                & \multicolumn{2}{l}{Sub-GeV multi-ring $\mu$} & 1.5    & 0.1 & {\it 9} \\
                                & \multicolumn{2}{l}{Multi-GeV multi-ring $\mu$} & 1.3  & 0.1 & {\it 9} \\ 
 $K/\pi$ ratio~\footnote[7]{20\,\% uncertainty in $K/\pi$ 
production ratio in cosmic ray interactions in the atmosphere. 
A positive number means that the fraction of $K$ is increased.}                  &                       &                 & 20.0 & -6.3 & {\it 10} \\
 \multicolumn{2}{l}{L$_{\nu}$ (production height)}      &                 & 10.0\footnote[8]{10\,\% uncertainty in the atmospheric density structure. A positive number means a more compressed atmospheric density structure.}    & -0.6 & {\it 11} \\
 Energy spectrum\footnote[9]{0.03 and 0.05 uncertainties in the spectral index of the primary cosmic rays below and above 100~GeV. Spectral index uncertainties below and above 100\,GeV are assumed to be correlated. A positive number means that the spectrum is harder. The predicted flux
was changed around an arbitrary reference energy of 10 GeV. }    &  \multicolumn{2}{l}{$E_{k}<100 $\,GeV}               & 0.03       & 0.031 & {\it 12} \\
  &  \multicolumn{2}{l}{$E_{k}>100 $\,GeV}    & 0.05       & 0.052 & {\it 12} \\
 Sample-by-sample\footnote[10]{Different flux calculations predict different energy dependences that cannot be explained by a simple spectral index uncertainty. See the lower panel of 
Fig.~\ref{fig:enu_spectra}. From a comparison of the predicted number of events based on different flux models, 5\,\% is assigned as the relative normalization uncertainty for these samples.}        & FC Multi-GeV          &                 & 5.0  & -5.2 & {\it 13} \\
                                & \multicolumn{2}{l}{PC\,$+$\,upward stopping $\mu$} & 5.0 & -3.9 & {\it 14} \\
 \hline \hline
\end{tabular}
\caption{Summary of systematic uncertainties in the prediction of the atmospheric neutrino flux.
Estimated uncertainty and the best-fit value are listed for each error.
The last column shows the error parameter numbers ($j$), which appeared in
Eqs.\ref{equation:chi2def} and \ref{equation:chi2min}. }
\label{table:fitsummary_flux}
\end{center}
\end{center}
\end{table}

\begin{table}
\begin{center}
\begin{center}
\renewcommand{\thefootnote}{\alph{footnote}}
\begin{tabular}{lllccc}
 \hline \hline
 & & & $\sigma$\,(\%) & best-fit & {\it No.}\\
 \hline
 \multicolumn{6}{l}{\bf (B) Systematic uncertainties in neutrino interaction} \\
 \multicolumn{3}{l}{$M_{A}$ in quasi-elastic and single-$\pi$}& 10.0\footnote{10\,\% uncertainty in the axial vector mass, $M_{A}$ (See Sec.~\ref{sec:atmnumc}), value.}          & 0.5 & {\it 15} \\
 \multicolumn{3}{l}{ Quasi-elastic scattering (model dependence)}& 1.0\footnote{Difference from the model in Ref.~\cite{Singh:1993rg} is set to 1.0\,.} & -0.95 & {\it 16} \\
 \multicolumn{3}{l}{ Quasi-elastic scattering (cross-section)}& 10.0    & 5.6 & {\it 17} \\
 \multicolumn{3}{l}{ Single-meson production (cross-section)}& 10.0     & -4.7 & {\it 18} \\
 \multicolumn{3}{l}{Multi-pion production (model dependence)} & 1.0\footnote{Difference from the model in Ref.~\cite{Bodek:2002vp} is set to 1.0\,.} & 1.47 & {\it 19} \\
 \multicolumn{3}{l}{ Multi-pion production (total cross-section)}& 5.0  & -0.2 & {\it 20} \\
 \multicolumn{3}{l}{ Coherent pion production (total cross-section)}& 30.0      & 0.4 & {\it 21} \\
 \multicolumn{3}{l}{ NC/CC ratio~\footnote{A positive number means more NC events in the Monte Carlo.}}                       & 20.0  & 2.9 & {\it 22} \\
 \multicolumn{3}{l}{ Nuclear effect in $^{16}$O~\footnote{30\,\% uncertainty in the mean free path of hadrons in the $^{16}$O nucleus. A positive number means stronger nuclear effect in $^{16}$O.}}                & 30.0   & -7.2 & {\it 23} \\
 \multicolumn{3}{l}{ Energy spectrum of pions}          & 1.0\footnote{The difference in the predicted pion energy spectrum by {\tt NEUT} and {\tt NUANCE} interaction models is taken as 1 standard deviation, and is set to 1.0.}                & 0.50 & {\it 24} \\
 \multicolumn{3}{l}{ CC $\nu_\tau$ interaction cross section}& 30.0    & 0.2 & {\it 25}  \\
 \hline \hline
\end{tabular}
\caption{Summary of systematic uncertainties in neutrino interactions.
Estimated uncertainty and the best-fit value are listed for each error.
The last column shows the error parameter numbers ($j$), which appeared in
Eqs.\ref{equation:chi2def} and \ref{equation:chi2min}.}
\label{table:fitsummary_nuint}
\end{center}
\end{center}
\end{table}

\begin{table}
\begin{center}
\begin{center}
\renewcommand{\thefootnote}{\alph{footnote}}
\begin{tabular}{lllccc}
 \hline \hline
 & & & $\sigma$\,(\%) & best-fit & {\it No.}\\
 \hline
 \multicolumn{6}{l}{\bf (C) Systematic uncertainties in event selection} \\
 \multicolumn{3}{l}{ Reduction for fully-contained event}     & 0.2      & 0.0 & {\it 26} \\
 \multicolumn{3}{l}{Reduction for partially-contained event}    & 2.6    & 0.3 & {\it 27} \\
Detection efficiency\footnote{Goodness of upward-going $\mu$ fit is used to select the upward-going $\mu$ sample. The difference of the goodness between the data and MC is considered as the source of the uncertainty in the detection efficiency. Uncertainties for upward stopping $\mu$ and upward through-going $\mu$ are assumed to be correlated.}&  \multicolumn{2}{l}{upward stopping $\mu$}& 1.3   & -0.2 & {\it 28} \\
 & \multicolumn{2}{l}{upward through-going $\mu$}& 0.5 & -0.1 & {\it 28} \\
 \multicolumn{3}{l}{FC/PC separation\footnote{The number of hits in the 
OD cluster is used to separate the FC and PC events. 
See Fig.~\ref{fig:od-nhit}. The systematic uncertainty in the number of 
hits in the OD cluster causes 0.9\,\% uncertainty in the 
number of the PC events. The number of FC events changes anti-correlated with 
the change in the number of PC events. A positive number means that the number of MC FC events is increased.}}   & 0.9    & -0.3 & {\it 29} \\
 \multicolumn{3}{l}{Hadron simulation}  & 1.0\footnote{Difference from the FLUKA model. A positive number means more hadrons, mostly pions, in neutral current interactions are identified as $\mu$-like.}  & -0.24 & {\it 30} \\
 Non-$\nu$ BG\footnote{The background sources are flasher PMTs and neutron interactions for $e$-like events and cosmic ray muons for $\mu$-like events. It is assumed that the background sources are un-correlated between $e$-like and $\mu$-like events. The background for sub- and multi-GeV samples in the  $e$-like and $\mu$-like events are assumed to be correlated. The background for the PC sample is also assumed to be correlated with the FC $\mu$-like samples. Only positive numbers are allowed for the background.}& Sub-GeV         &$e$-like       & 0.4  & 0.1 & {\it 31} \\
                                        &                &$\mu$-like     & 0.1  & 0.0 & {\it 32} \\
                                        & Multi-GeV      &$e$-like       & 0.2  & 0.0 & {\it 31} \\
                                        &                &$\mu$-like     & 0.1  & 0.0 & {\it 32} \\
                                        & PC             &               & 0.2  & 0.0 & {\it 32} \\
 \multicolumn{3}{l}{Upward stopping/through-going $\mu$ separation \footnote{The number of hits in the OD cluster
 at the exit point of a muon is used to separate the upward stopping and 
through-going muon events.
The uncertainty in the number of hits in the OD cluster causes 
0.4\,\% uncertainty in the stopping/through-going ratio. 
A positive number means that the number of MC stopping muons is increased.}}& 0.4 &  0.0 & {\it 33} \\
 \hline \hline
\end{tabular}
\caption{Summary of systematic uncertainties in event selection.
Estimated uncertainty and the best-fit value are listed for each error.
The last column shows the error parameter numbers ($j$), which appeared in
Eqs.\ref{equation:chi2def} and \ref{equation:chi2min}.}
\label{table:fitsummary_event}
\end{center}
\end{center}
\end{table}

\begin{table}
\begin{center}
\begin{center}
\renewcommand{\thefootnote}{\alph{footnote}}
\begin{tabular}{lllccc}
 \hline \hline
  & & & $\sigma$\,(\%) & best-fit & {\it No.} \\
 \hline
 \multicolumn{6}{l}{\bf (D) Systematic uncertainties in event reconstruction} \\
 Ring separation\footnote{Ring separation uncertainty is assumed to be 
fully correlated. Namely, if the number of single-ring sub-GeV $e$-like 
events have to be increased, the number of single-ring multi-GeV $e$-like 
events and single-ring sub- and multi-GeV $\mu$-like events have to be 
increased according to the systematic uncertainty factors. On the other 
hand, in this case, the number of multi-ring $\mu$-like events have to 
be decreased. A positive number means the number of MC events for 
the corresponding sample is increased.}                         & $< 400$\,MeV   &$e$-like         & 6.3 & 2.6 & {\it 34} \\
                                       &                &$\mu$-like       & 2.4 & 1.0 & {\it 34} \\
                                       & $> 400$\,MeV   &$e$-like         & 3.4 & 1.4 & {\it 34} \\
                                       &                &$\mu$-like       & 1.3 & 0.5 & {\it 34} \\
                                       & Multi-GeV      &$e$-like         & 15.9& 6.5 & {\it 34} \\
                                       &                &$\mu$-like       & 6.2 & 2.5 & {\it 34} \\
                                       & \multicolumn{2}{l}{Sub-GeV multi-ring $\mu$} & 3.7     & -1.5 & {\it 34} \\
                                       & \multicolumn{2}{l}{Multi-GeV multi-ring $\mu$} & 7.2   & -2.9 & {\it 34} \\
 Particle identification\footnote{The particle 
identification uncertainty is anti-correlated between $e$-like and $\mu$-like
events. It is assumed that the particle identification uncertainty
 is correlated between sub- and multi-GeV energy regions. 
However, it is assumed that it
is not correlated between single- and multi-ring events. 
A positive number means the number of MC events for the corresponding sample 
is increased.}                         & Sub-GeV        &$e$-like         & 0.6 & 0.2 & {\it 35} \\
                                       &                &$\mu$-like       & 0.6 & -0.2 & {\it 35} \\
                                       & Multi-GeV      &$e$-like         & 0.4 & 0.1 & {\it 35} \\
                                       &                &$\mu$-like       & 0.4 & -0.1 & {\it 35} \\
                                       & \multicolumn{2}{l}{Sub-GeV multi-ring $\mu$} & 3.4     & -0.9 & {\it 36} \\
                                       & \multicolumn{2}{l}{Multi-GeV multi-ring $\mu$} & 4.7   & -1.2 & {\it 36} \\
 \multicolumn{2}{l}{Energy calibration for FC event~\footnote{2\,\% uncertainty in the absolute energy scale of the detector. A positive number corresponds to increasing the visible energy of MC events.} }    &                 & 2.0               & 0.4 & {\it 37} \\
 \multicolumn{2}{l}{Energy cut for upward stopping muon}&                 & 1.1 & -0.2 & {\it 38} \\
 \multicolumn{2}{l}{Up/down symmetry of energy calibration~\footnote{A positive number means that the energy of MC events is increased for upward-going direction. }}&    & 0.6 &  0.0 & {\it 39} \\
 \hline \hline
\end{tabular}
  \caption{Summary of systematic uncertainties in event reconstruction.
Estimated uncertainty and the best-fit value are listed for each error.
The last column shows the error parameter numbers ($j$), which appeared in
Eqs.\ref{equation:chi2def} and \ref{equation:chi2min}.}
\label{table:fitsummary_fit}
\end{center}
\end{center}
\end{table}

A global scan was made on a $(\sin^22\theta, \log \Delta m^2)$ grid
minimizing $\chi^2$ at each point with respect to 39 parameters listed in
Tables~\ref{table:fitsummary_flux},~\ref{table:fitsummary_nuint},~\ref{table:fitsummary_event} and \ref{table:fitsummary_fit}. 
 At each grid point, the
local minimum of $\chi^2$ are derived by assuming a
linear dependence of $N^{\rm exp}_i$ on each of the parameters. At
the minimum $\chi^2$ location, 
$\partial \chi^2/\partial \epsilon_j=0$
for each of the parameters $\epsilon_j$. As a result,
the minimization of $\chi^2$ in Eqn.~\ref{equation:chi2def} is
equivalent to solving the following system of $k=1,39$ linear 
equations~\cite{Fogli:2002pt}:
%
\begin{eqnarray}
&&
\sum_{j=1}^{39}
\left[
  \frac{1}{\sigma_j^2}
  \delta_{jk}+\sum_{i=1}^{180}
  \left(
     \frac{N^{\rm exp}_{i} \cdot N^{\rm exp}_{i} \cdot f^i_j \cdot f^i_k}
          {\sigma_i^2}
  \right)
\right]
\cdot \epsilon_j \nonumber \\
&&{\hspace{1.2cm}}
= 
\sum_{i=1}^{180}
\frac{(N^{\rm obs}_i-N^{\rm exp}_i) \cdot N^{\rm exp}_i \cdot f^i_k}
     {\sigma_i^2}
\label{equation:chi2min}
\end{eqnarray}
where $\sigma_j$ is the estimated uncertainty in the parameter
$\epsilon_j$. One of $\sigma_j$ corresponds to the absolute 
normalization uncertainty. In this case, $ 1 / \sigma_j^2 $
is set to 0, since the absolute normalization is a free parameter 
in our analysis.
%

\begin{figure}
  \includegraphics[width=3.3in]{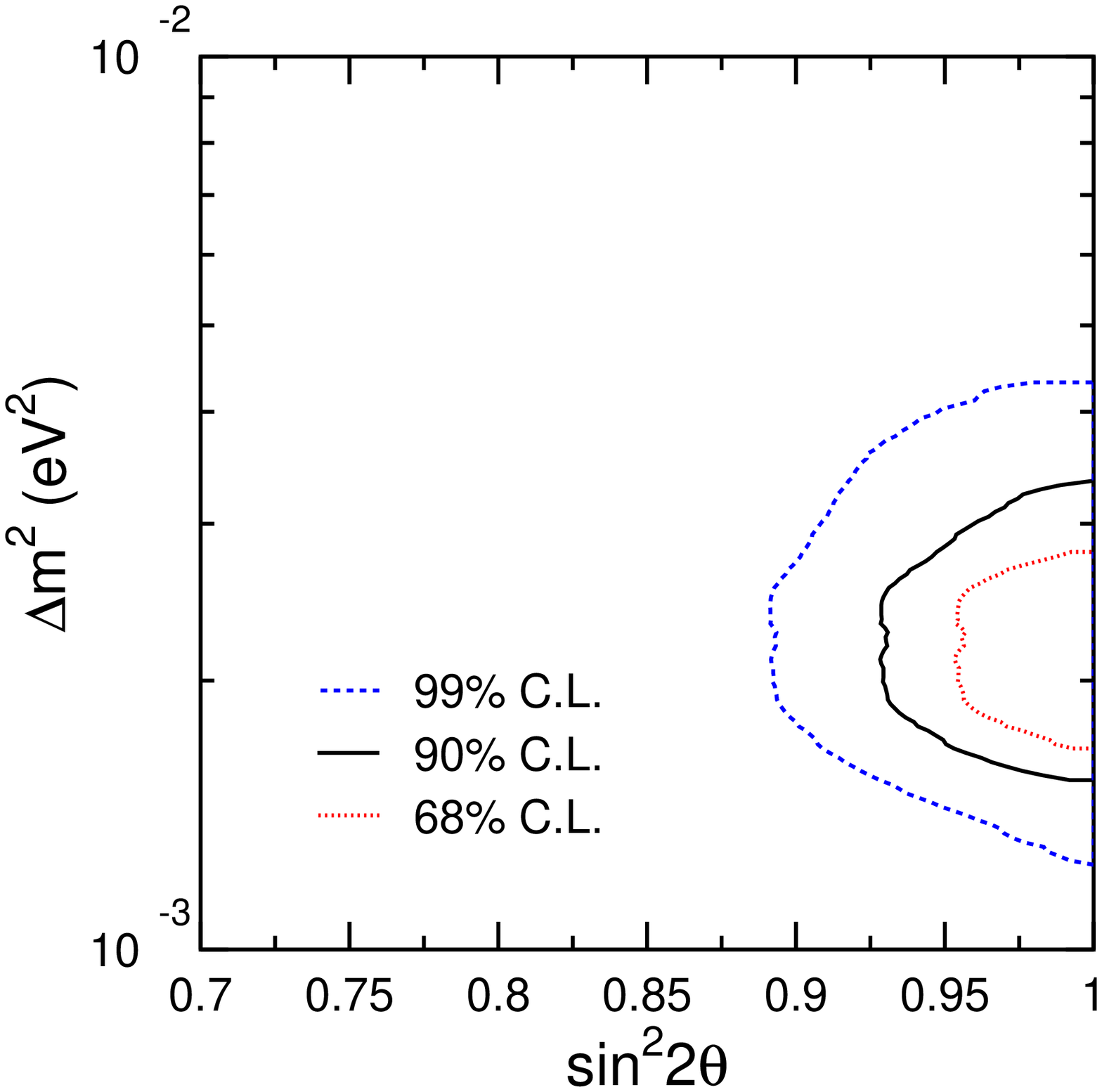}
  \caption{Allowed oscillation parameters for 
$\nu_\mu \leftrightarrow \nu_\tau$ oscillations.
Three contours correspond to the 68\% (dotted line), 90\% (solid
line) and 99\% (dashed line) C.L. allowed regions.}
   \label{fig:allowed}
\end{figure}

\begin{figure}
  \includegraphics[width=3.7in]{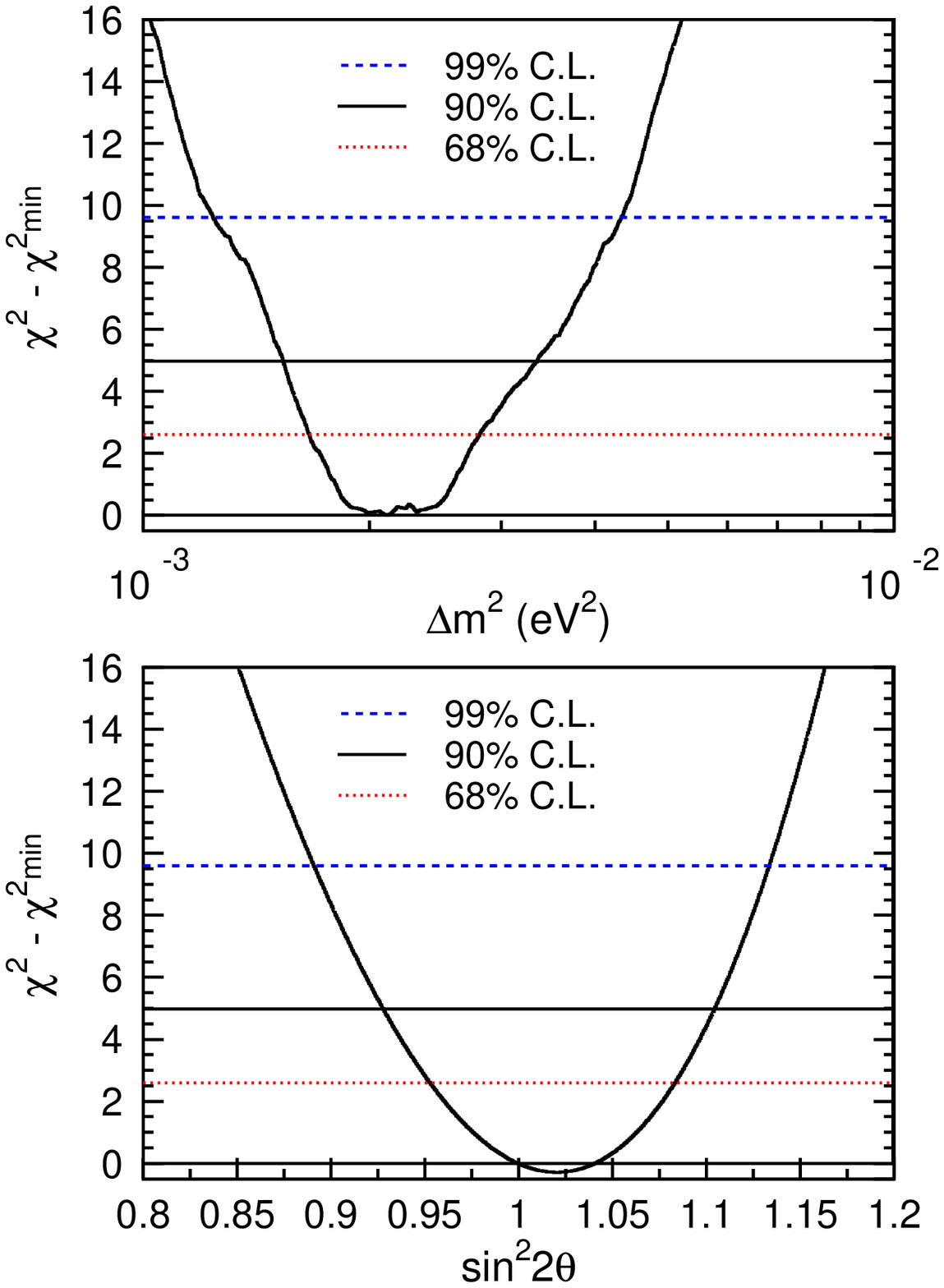}
\caption{$\chi^2 - \chi^2_{min}$ projected 
  onto the $\sin^{2}2\theta$ and $\Delta m^2$ axes. The minimum
  value at each $\sin^{2}2\theta$ and $\Delta m^2$ is
  plotted.}
   \label{fig:chi2dist}
\end{figure}

The minimum $\chi^2$ value, $\chi^2_{min} = 174.8 / 177 {\rm ~DOF}$, is
located at $(\sin^22\theta = 1.00,$ $\Delta m^2 = 2.1\times10^{-3} $~eV$^2$).
The number of DOF is found by 180 terms in the $\chi^2$ sum plus
38 systamtic constraints in the $\chi^2$ sum minus
39 minimized parameters minus the two physics parameters of
$\sin^2 2\theta$ and $\Delta m^2$. The overall normalization
is not used as a constraint to $\chi^2$.
The best-fit values of the parameters $\epsilon_j$ obtained
at the global minimum are summarized in 
Tables~\ref{table:fitsummary_flux},~\ref{table:fitsummary_nuint}
,~\ref{table:fitsummary_event} and \ref{table:fitsummary_fit}.  For
the most part, the parameters $\epsilon_j$ are fit within their
estimated $1~\sigma$ errors. 
Including the unphysical region ($\sin^22\theta > 1$)
 in the scan, the minimum $\chi^2$ value is obtained at
 $(\sin^22\theta = 1.02, \Delta m^2 = 2.1\times10^{-3} $~eV$^2$).
The minimum $\chi^2$ value, $\chi^2_{min} = 174.5 / 177 {\rm ~DOF}$, in the 
unphysical region is lower than that in the physical region by 0.29\,.
Contours corresponding to the 68\%, 90\%
and 99\% confidence intervals are located at $\chi^2_{min} +$ 2.60, 4.98,
and 9.60 respectively, where $\chi^2_{min}$ is 
the minimum $\chi^2$ value in the physical region
  and are shown in Fig.~\ref{fig:allowed}.
These intervals are derived based on a two dimensional extension of
the method described in Ref.~\cite{Barnett:1996hr}.
Figure~\ref{fig:chi2dist} shows the $\chi^2 - \chi^2_{min}$ distributions projected to
 $\sin^{2}2\theta$ and $\Delta m^2$ axes, in which the minimum $\chi^2  - \chi^2_{min}$
 values for each $\sin^{2}2\theta$ and $\Delta m^2$ are plotted.
The  $\chi^2 - \chi^2_{min}$ distribution is rather flat between
 $\Delta m^2$~=~2.0$\times 10^{-3}$eV$^2$ and 2.5$\times 10^{-3}$eV$^2$.
Any $\Delta m^2$ in this range fits the data nearly as well as the best-fit
point.

Assuming no oscillation, ($\sin^22\theta = 0$, $\Delta m^2 = 0$), we found a
$\chi^2$ value of 478.7 for 179~DOF, where only the overall normalization is
a free parameter. We allowed all systematic uncertainty terms to be
minimized, yet the fit was greatly inferior to the best-fit including
neutrino oscillations.

\renewcommand{\topfraction}{1.}
\renewcommand{\bottomfraction}{1.}
\renewcommand{\textfraction}{0.}
\renewcommand{\floatpagefraction}{1.}

\begin{figure}
  \includegraphics[width=3.3in]{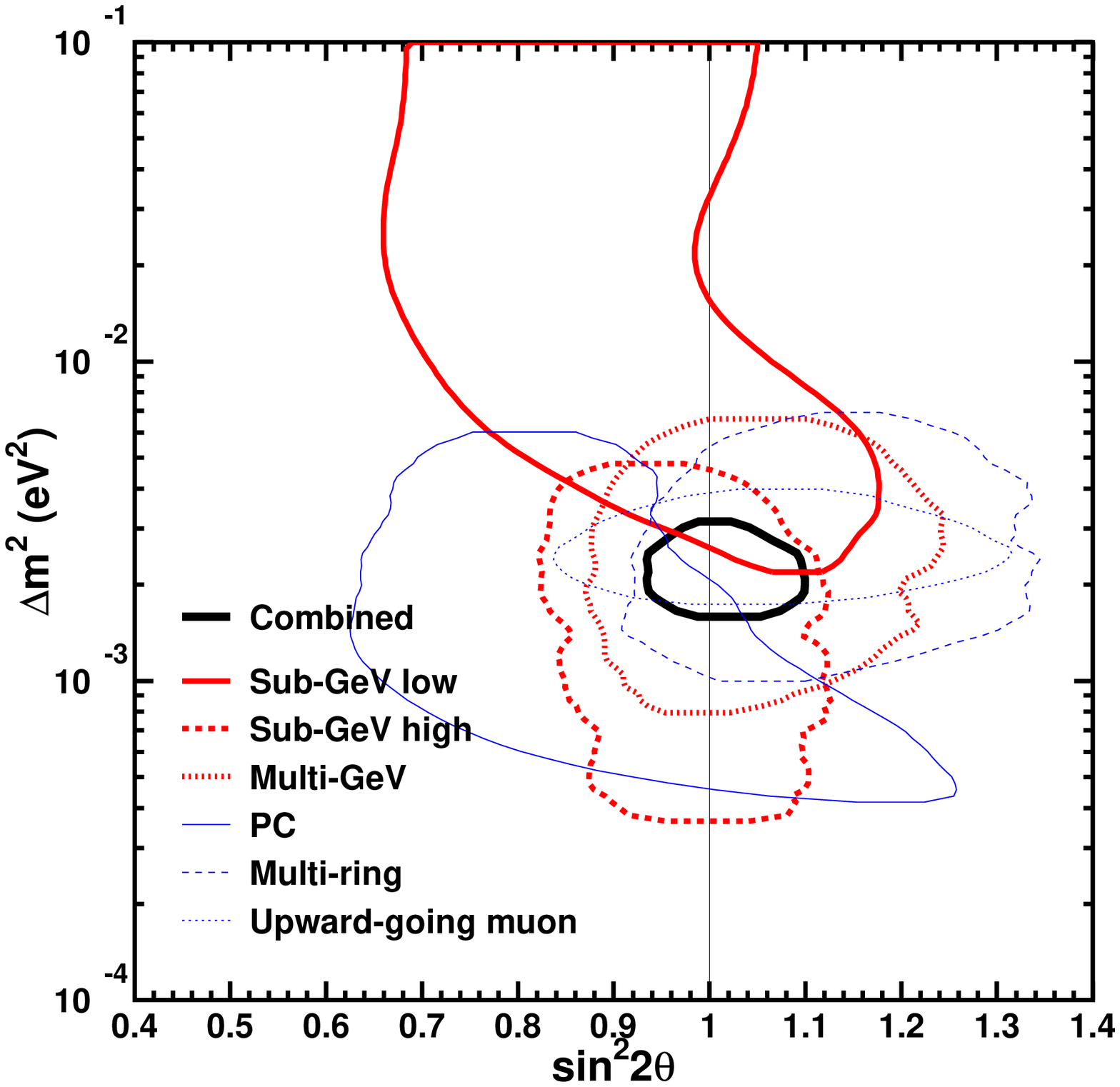}
  \caption{90\,\% confidence level allowed oscillation parameter regions for $\nu_\mu \leftrightarrow \nu_\tau$
 oscillations from six sub-samples. In this plot, 90\,\% confidence interval is defined to be $\chi^2 = \chi^2_{min}
+ 4.61$, where $\chi^2_{min}$ is the minimum $\chi^2$ value including the unphysical parameter region.}
   \label{fig:allowed_sub}
\end{figure}

We have also estimated the allowed neutrino oscillation parameters by
performing the same fitting procedure using independent subsamples of the
data: FC single-ring sub-GeV events below 400~MeV/$c$, FC single-ring sub-GeV
events above 400~MeV/$c$, FC single-ring multi-GeV events, PC events, FC
multi-ring events, and upward-going muon events. In each independent fit,
only the relevant parameters out of the set of 39 were minimized. The results
are shown in Fig.~\ref{fig:allowed_sub}. The allowed region contours found by
fitting these six subsamples are consistent with each other and with the
combined fit to all events.

\begin{figure}
  \includegraphics[width=3.3in]{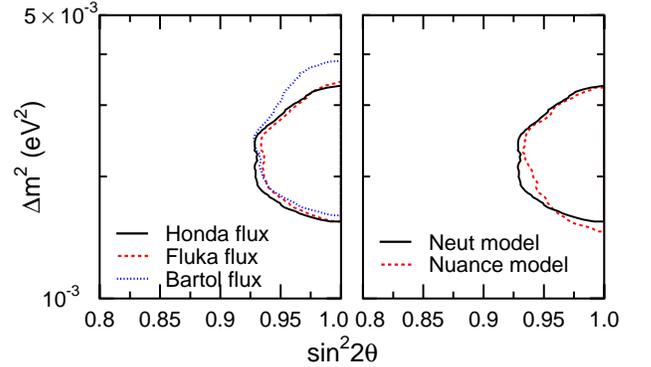}
  \caption{Left: 90\,\% confidence level allowed oscillation parameter regions 
for $\nu_\mu \leftrightarrow \nu_\tau$ oscillations, based on the 
{\tt NEUT} neutrino interaction model, from different flux models (solid
line; \cite{Honda:2004yz}, dashed line; \cite{Battistoni:2003ju}, dotted line;
\cite{Barr:2004br}).  Right: The 90\,\% C.L. allowed regions based
on a different neutrino interaction model ({\tt NUANCE}~\cite{Casper:2002sd}) 
for FC+PC events with the flux model of Ref.~\cite{Honda:2004yz} 
(dashed line) is compared with that based on {\tt NEUT} with the 
same flux.  In this plot, Monte Carlo events from {\tt NEUT} were used 
for upward-going muons.}
   \label{fig:allowed-regions-different-flux}
\end{figure}

In addition, the same oscillation analyses were repeated using different flux
models (but with the same neutrino interaction Monte Carlo program) and
different neutrino interaction Monte Carlo program (but with the same flux
model).  The 90\,\%~C.L. allowed parameter regions are compared in
Fig.~\ref{fig:allowed-regions-different-flux}. The allowed regions from these
analyses overlap well, demonstrating that the measured parameters do not
strongly depend on the choice of flux or interaction model from which we
start the fitting procedure. However, the allowed region obtained based on
the flux model of Ref.~\cite{Barr:2004br} allows for slightly higher $\Delta
m^2$. We studied the reason for this difference in detail, and found that the
main reason was the slightly harder energy spectrum in the upward-going muon
energy range (Fig.~\ref{fig:enu_spectra}).

\begin{figure}
  \includegraphics[width=3.7in]{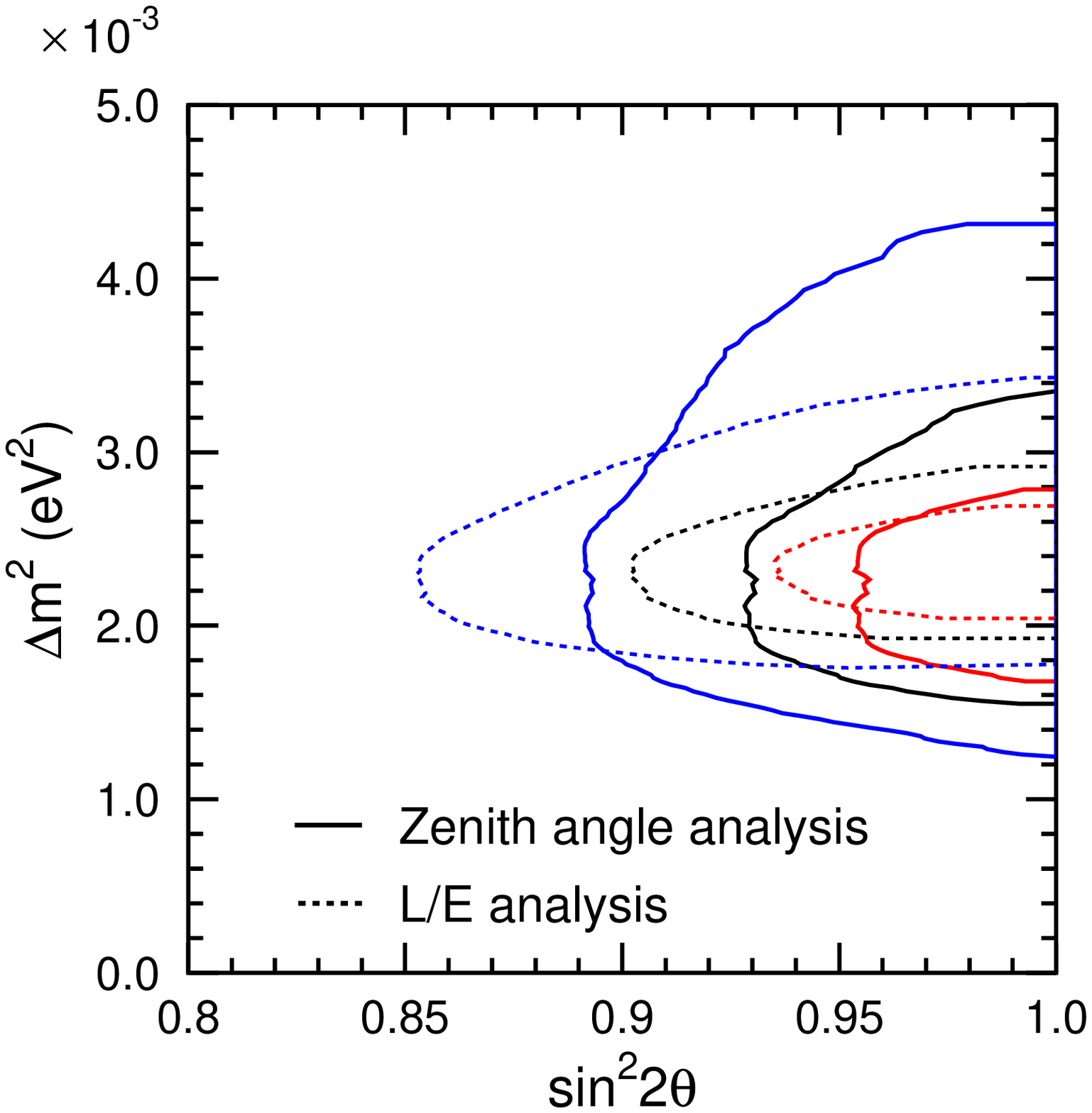}
  \caption{ The 68, 90 and 99\,\% confidence level allowed 
oscillation parameter
regions obtained by an $L/E$ analysis~\cite{Ashie:2004mr} and by the
present analysis are compared.}
   \label{fig:allowed-regions-different-int}
\end{figure}

Finally, we point out that a separate $L/E$ analysis of the same
running period~\cite{Ashie:2004mr}, using only selected high
resolution FC and PC events, gave an allowed oscillation parameter
region consistent with this result. This is shown in
Fig.~\ref{fig:allowed-regions-different-int}, with a magnified view of
the region and a linear scale in $\Delta m^2$. The $L/E$ analysis
provided a slightly better constraint in $\Delta m^2$ due to locating
the oscillatory dip; the present analysis constrains $\sin^2 2\theta$
better due to high statistics in the up-down asymmetry.

\section{Conclusion}

\label{sec:conclusion}

Super-Kamiokande has observed more than 15,000 atmospheric neutrino events
during the first data taking period between 1996 and 2001.  Atmospheric
neutrino events observed in Super-Kamiokande have an energy range from about
100~MeV to 10~TeV, and a neutrino flight-length from about 10~km to
13,000~km.  These wide energy and flight-length ranges together with high
statistics made it possible to study neutrino oscillations.  Especially, the
predicted up-down asymmetry of the atmospheric neutrino flux enabled us to
accurately estimate the mixing parameter $\sin ^2 2\theta$. The observed muon
neutrino events showed a clear zenith angle and energy dependent deficit of
events, while the electron neutrino events were in good agreement with the
prediction. A detailed neutrino oscillation analysis confirmed that the full
data set was explained well by $\nu_{\mu} \leftrightarrow \nu_{\tau}$
oscillations. Various systematic effects were included in the oscillation
analysis.  The measured neutrino oscillation parameters were 
$\sin ^2 2\theta > 0.92$ and 
$1.5 \times 10^{-3} < \Delta m^2 < 3.4 \times 10^{-3}$ eV$^2$ 
at 90\,\%~C.L. This result gives the most accurate determination of
$\sin^2 2\theta$ and is consistent with the somewhat more accurate
measurement of $\Delta m^2$ determined by the independent study of high
resolution $L/E$ events\cite{Ashie:2004mr}.

We gratefully acknowledge the cooperation of the Kamioka Mining and Smelting
Company. The Super-Kamiokande experiment has been built and operated from
funding by the Japanese Ministry of Education, Culture, Sports, Science and
Technology, the United States Department of Energy, and the U.S. National
Science Foundation. Some of us have been supported by funds from the Korean
Research Foundation (BK21) and the Korea Science and Engineering Foundation,
the State Committee for Scientific Research in Poland (grants 1P03B08227 and
1P03B03826), Japan Society for the Promotion of Science, and Research
Corporation's Cottrell College Science Award.





\subsection{Appendix}
\label{binned:data}

Table~\ref{tb:eventnumber1} summarizes the number of observed and 
expected FC and PC events for each bin. 
The Monte Carlo prediction does not include neutrino oscillations.
Table~\ref{tab:upmubins} summarizes those for upward-going muons.
These binned data are used in the oscillation analysis.
Table~\ref{tab:mcinfo} summarizes neutrino energy at which 50\% 
of events are accumulated
 for each energy bin in the absence of neutrino oscillations. The fraction of 
various neutrino interaction modes are also listed.

\begin{table*}
 \begin{center}
\begin{tabular}{|l|r|r|r|r|r|r|r|r|r|r|}
  \multicolumn{11}{c}{FC single-ring e-like}  \\
 \hline
 log($P_{lep}$) &I &II &III &IV &V & VI&VII &IIX &IX &X\\
 \hline
 1 &114(79.29) &95(83.33) &74(81.41) &94(82.04) &88(83.99) &91(79.78) &79(79.49) &74(84.19) &91(81.50) &100(82.93) \\
 2 &96(75.62) &93(71.70) &96(73.22) &90(69.44) &89(68.36) &85(68.75) &85(69.49) &74(67.18) &83(71.14) &78(69.68) \\
 3 &76(64.16) &80(66.93) &80(65.78) &69(63.60) &72(64.57) &60(64.06) &69(62.39) &71(61.66) &85(59.72) &63(57.48) \\
 4 &48(45.35) &57(47.92) &62(50.12) &52(50.91) &60(51.62) &74(51.60) &55(50.75) &58(49.13) &60(46.51) &43(42.45) \\
 5 &26(21.68) &35(23.19) &31(25.13) &37(25.84) &24(25.55) &38(25.93) &34(24.96) &24(26.14) &21(23.63) &20(18.51) \\
 6 &33(29.29) &35(33.22) &41(34.92) &37(39.72) &46(42.84) &49(43.86) &49(40.65) &32(39.50) &36(32.03) &36(27.30) \\
 7 &23(21.36) &31(25.66) &28(30.10) &42(37.93) &63(45.70) &37(42.99) &54(35.42) &34(32.54) &22(26.00) &18(19.48) \\
 \hline
 \multicolumn{11}{c}{FC single-ring $\mu$-like }  \\
 \hline
 1 & 36(54.73) &40(53.66) &39(54.39) &37(55.09) &35(55.75) &34(53.79) &35(53.46) &45(53.57) &48(52.63) &46(52.10) \\
 2 & 86(124.32) &77(122.76) &99(123.28) &86(121.54) &87(119.12) &80(119.67) &91(122.64) &85(117.84) &94(115.67) &76(120.59) \\
 3 & 94(118.74) &60(112.36) &81(113.06) &94(115.77) &87(112.80) &84(112.77) &116(113.40) &119(111.91) &97(108.09) &118(104.54) \\
 4 & 52(91.07) &48(87.99) &53(90.52) &53(91.04) &68(94.68) &68(91.11) &72(89.57) &81(88.15) &91(84.45) &86(82.90) \\
 5 & 27(43.35) &22(45.89) &22(44.91) &37(44.48) &25(47.03) &40(47.52) &41(47.91) &41(42.59) &46(44.04) &48(43.60) \\
 6+7 &34(89.46) &46(86.45) &42(86.48) &49(92.54) &54(96.41) &73(94.68) &95(96.23) &87(88.39) &78(84.46) &93(84.82) \\
 \hline
 \multicolumn{11}{c}{FC multi-ring $\mu$-like }  \\
 \hline
sub-GeV  &14(27.57) &8(31.15) &20(33.42) &14(33.68) &25(36.13) &16(35.59) &21(34.12) &32(32.94) &29(28.85) &29(29.12) \\
multi-GeV  &18(63.27) &29(63.34) &31(66.89) &28(75.69) &41(86.27) &69(82.70) &55(77.10) &54(73.27) &59(62.44) &55(60.87) \\
 \hline
 \multicolumn{11}{c}{PC }  \\
 \hline
  &49(88.97) &45(88.81) &59(104.46) &89(125.12) &117(154.61) &156(158.34) &114(128.39) &109(103.00) &85(89.70) &88(88.18) \\
 \hline
  \end{tabular} 
  \end{center} 
  \caption{Summary of the number of observed (MC expected) FC and PC events  
    for each bin. Neutrino oscillation is not included 
    in the Monte Carlo prediction. Roman numbers represent zenith angle 
    regions equally spaced between $\cos\Theta = -1$ and $\cos\Theta = 1$.
    The numbers in the $log(P_{lep})$ column show the momentum ranges. 
    The momentum ranges
    are $<$250, 250-400, 400-630, 630-1000 and $>$1000~MeV/c in sub-GeV samples for momentum range numbers 1 to 5 and $<$2500 and 
    $>$2500~MeV/c in multi-GeV samples for momentum range numbers 6 to 7.
    }
\label{tb:eventnumber1} 
\end{table*}

\begin{table*}
 \begin{center}
\begin{tabular}{|r|r|r|r|r|r|r|r|r|r|r|}
  \multicolumn{11}{c}{upward through-going muon}  \\
 \hline
   & I &II &III &IV &V & VI&VII &IIX &IX &X\\
 \hline
\# Events& 85 & 113 & 116 & 138 & 159 & 183 & 178 & 267 & 286 & 316.6 \\
\# Expected & 96.13 & 114.85 & 122.48 & 136.88 & 145.79 & 169.17 & 187.32 & 210.92 & 228.76 & 257.16 \\
 Efficiency & 95.2\% & 93.9\% & 92.4\% & 95.9\% & 94.0\% & 97.2\% & 96.9\% & 99.0\% & 96.2\% & 95.5\%\\\hline
 Flux       & $0.862$    &$1.060$   &$1.045$   &   $1.216$&   $1.388$&   $1.589$ &   $1.557$ &   $2.365$&   $2.574$&   $2.953$ \\
 Stat. err. & $\pm0.093$ &$\pm0.100$&$\pm0.097$&$\pm0.104$&$\pm0.110$&$\pm0.117$ &$\pm0.117$ &$\pm0.145$&$\pm0.152$&$\pm0.191$ \\
 Expected   &    $0.976$ &   $1.078$&   $1.103$&   $1.206$&   $1.274$&   $1.469$ &   $1.638$ &   $1.868$&   $2.060$&   $2.393$ \\
 Stat. err. & $\pm0.021$ &$\pm0.021$&$\pm0.021$&$\pm0.022$&$\pm0.022$&$\pm0.024$ &$\pm0.025$ &$\pm0.027$&$\pm0.029$&$\pm0.032$ \\
 \hline 

 \multicolumn{11}{c}{upward stopping muon }  \\
\hline
\# Events& 28 & 23 & 37 & 30 & 27 & 37 & 37 & 48 & 65 & 85.7 \\
\# Expected & 51.24 & 54.06 & 56.69 & 64.96 & 67.60 & 68.21 & 78.85 & 80.96 & 94.00 & 96.90 \\
Efficiency  & 99.8\% & 99.1\% & 99.8\% & 108.1\% & 102.1\% & 101.2\% & 103.5\% & 103.1\% & 105.0\% & 100.2\% \\\hline
 Flux       &    $0.286$ &   $0.217$&   $0.337$&   $0.265$&   $0.236$&   $0.322$ &   $0.323$ &   $0.425$&   $0.589$&   $0.807$ \\
 Stat. err  & $\pm0.054$ &$\pm0.045$&$\pm0.055$&$\pm0.048$&$\pm0.045$&$\pm0.053$ &$\pm0.053$ &$\pm0.061$&$\pm0.073$&$\pm0.172$ \\
 Expected   &    $0.523$ &   $0.509$   &$0.517$   &$0.574$   &$0.591$   &$0.594$    &$0.689$    &$0.715$   &$0.851$   &$0.913$ \\
 Stat. err. & $\pm0.015$ &$\pm0.015$&$\pm0.015$&$\pm0.015$&$\pm0.015$&$\pm0.015$ &$\pm0.016$ &$\pm0.017$&$\pm0.019$&$\pm0.020$ \\


 \hline
 \hline
  \end{tabular} 
  \end{center} 
  \caption{Summary of the number of observed and expected 
    upward-going muons for each bin, efficiencies, and the corresponding
    flux. Neutrino oscillation is not included in the Monte Carlo prediction.
     The errors on the observed fluxes are statistical, the units
    of flux $\times10^{-13} {\rm{cm^{-2}s^{-1}sr^{-1}}}$. The Roman
    numerals refer to zenith angle regions equally spaced between
    $\cos\Theta = -1$ and $\cos\Theta = 0$. }
\label{tab:upmubins} 
\end{table*}

\begin{table*}
 \begin{center}
\begin{tabular}{|l|ll|r|rrr|}
  \multicolumn{7}{c}{FC single-ring e-like}  \\
 \hline
 \multicolumn{1}{|c}{ $\log(P_{lep})$ } & & & ~$E_{\nu}$(GeV) & ~~CC $\nu_{e} (\%) $ & ~~CC $\nu_{\mu}$ (\%) & ~~NC (\%) \\
 \hline
1 & sub-GeV   & 100-250  & 0.31 & 87.6 & 2.2 & 10.1 \\
2 &           &250-400   & 0.48 & 89.1 & 1.5 & 9.5 \\
3 &           &400-630   & 0.72 & 88.7 & 1.8 & 9.5 \\
4 &           &630-1000  & 1.1  & 86.8 & 3.2 & 10.1 \\
5 &           &$>$1000   & 1.5  & 86.5 & 4.5 & 8.9 \\
6 & multi-GeV &$<$2500   & 2.3  & 85.7 & 5.8 & 8.5 \\
7 &           &$>$2500   & 5.4  & 79.0 & 8.4 & 12.6 \\
 \hline
 \multicolumn{7}{c}{FC single-ring $\mu$-like }  \\
 \hline
1 & sub-GeV   & 200-250 & 0.50 & 0.3 & 90.9 & 8.8 \\
2 &           &250-400  & 0.68 & 1.2 & 95.6 & 3.3 \\
3 &           &400-630  & 0.86 & 0.7 & 97.7 & 1.6 \\
4 &           &630-1000 & 1.2  & 0.5 & 99.0 & 0.4 \\
5 &           &$>$1000  & 1.5  & 0.4 & 99.3 & 0.2 \\
6+7 & multi-GeV &       & 2.6  & 0.4 & 99.4 & 0.2 \\
 \hline
 \multicolumn{7}{c}{FC multi-ring $\mu$-like }  \\
 \hline
\multicolumn{1}{|c}{ } & \ sub-GeV & & 1.9 & 3.6 & 90.5 & 5.9 \\
\multicolumn{1}{|c}{ } & multi-GeV & & 3.6 & 2.3 & 94.9 & 2.7 \\
 \hline
 \multicolumn{7}{c}{PC }  \\
 \hline
 \multicolumn{1}{|c}{ }& & & 7.9 & 1.8 & 97.3 & 0.9\\
 \hline
 \multicolumn{7}{c}{Upward-going muons }  \\
 \hline
\multicolumn{1}{|c}{ } & \multicolumn{2}{l|}{upward stopping muon}      & 11.1  & 1.0 & 98.6 & 0.4 \\
\multicolumn{1}{|c}{ } & \multicolumn{2}{l|}{upward through-going muon} & 113.5 & 0.2 & 99.7 & 0.1 \\
 \hline
  \end{tabular} 
  \end{center} 
  \caption{ Description of momentum bins used for this analysis, 
  corresponding to raws of Tables~\ref{tb:eventnumber1} and \ref{tab:upmubins}.
  Also tabulated are the medium parent neutrino energy, and the relative
  fractions of CC $\nu_e$, CC $\nu_\mu$ and NC interactions in the 
   absence of neutrino oscillations as estimated
  by the Monte Carlo program.}
\label{tab:mcinfo} 
\end{table*}


\bibliography{combined}
\end{document}